\documentclass[pdflatex,sn-mathphys-ay,iicol]{sn-jnl}

\usepackage[utf8]{inputenc}
\usepackage{graphicx}%
\usepackage{multirow}%
\usepackage{amsmath,amssymb,amsfonts}%
\usepackage{amsthm}%
\usepackage{mathrsfs}%
\usepackage[title]{appendix}%
\usepackage{xcolor}%
\usepackage{textcomp}%
\usepackage{manyfoot}%
\usepackage{booktabs}%
\usepackage{algorithm}
\usepackage{algpseudocode}
\usepackage{listings}%
\usepackage{dsfont}
\usepackage{mathtools}
\usepackage{float}     
\usepackage{flafter}   
\usepackage[section]{placeins} 
\usepackage{stfloats}
\usepackage{url}
\usepackage{mathabx}
\usepackage{subcaption}
\usepackage{kantlipsum,cuted}
\usepackage{tikz}
\usetikzlibrary{positioning, arrows.meta}
\usetikzlibrary{calc}
\usepackage{accents}
\usepackage[font=small]{caption}
\usepackage{siunitx} 
\usepackage{rotating}
\theoremstyle{thmstyleone}%

\newcommand{\BF}[1]{\mathbf{#1}}
\newcommand{\BS}[1]{\boldsymbol{#1}}
\newcommand{\bX}{\BF{X}}
\newcommand{\bZ}{\BF{Z}}
\newcommand{\bY}{\BF{Y}}
\newcommand{\bSigma}{\BS{\Sigma}}

\newcommand{\bE}{\BF{E}}

\newcommand{\bA}{\BF{A}}
\newcommand{\bC}{\BF{C}}
\newcommand{\bV}{\BF{V}}

\newcommand{\bphi}{\BS{\varphi}}

\newcommand{\btheta}{\BS{\theta}}

\newcommand{\bmu}{\BS{\mu}}

\newcommand{\softmax}[1]{\sigma (#1)}
\newcommand{\rmX}{\mathrm{X}}
\newcommand{\rmZ}{\mathrm{Z}}

\newcommand{\RR}{\mathds{R}}

\newcommand{\multinomial}[2]{\mathcal{M}\left(#1, #2\right)}

\newcommand{\gaussian}[2]{\mathcal{N}\left(#1, #2\right)}

\newcommand{\poisson}[1]{\mathcal{P}(#1)}

\newcommand{\eqsp}{\;}

\newcommand{\child}[3]{\widecheck{\BF{#1}}{}^{#3}_{#2}}
\newcommand{\childindex}[2]{\mathcal{C}_{#2}^{#1}}

\newcommand{\kk}[1]{k}
\newcommand{\pkk}[1]{k}
\newcommand{\jj}[1]{j}
\newcommand{\vv}[1]{v}

\newcommand{\vamp}{\bar p_{\bphi}}

\newcounter{hypH}

\newcounter{notation}

\newcounter{properties}

\begin{document}

\title[Article Title]{TaxaPLN: a taxonomy-aware augmentation strategy for microbiome-trait classification including metadata}


\author*[1]{\fnm{Alexandre} \sur{Chaussard}}\email{alexandre.chaussard@sorbonne-universite.fr}

\author[1]{\fnm{Anna} \sur{Bonnet}}\email{anna.bonnet@sorbonne-universite.fr}

\author[1]{\fnm{Sylvain} \sur{Le Corff}}\email{sylvain.le\_corff@sorbonne-universite.fr}

\author[2]{\fnm{Harry} \sur{Sokol}}\email{harry.sokol@gmail.com}

\affil*[1]{\orgdiv{CNRS, Laboratoire de Probabilit\'es, Statistique et Mod\'elisation, LPSM}, \orgname{Sorbonne Universit\'e, F-75005 Paris, France}}

\affil[2]{\orgdiv{Centre de Recherche Saint-Antoine, CRSA, AP-HP}, \orgname{Sorbonne Universit\'e, INSERM UMRS-938}, \orgaddress{\postcode{F-75012}, \state{Paris}, \country{France}}}





\abstract{
\textbf{Motivation:} The gut microbiome plays a crucial role in human health, making it a corner stone of modern biomedical research. To study its structure and dynamics, machine learning models are increasingly used to identify key microbial patterns associated with disease and environmental factors. However, microbiome data present unique challenges due to their compositionality, high-dimensionality, sparsity, and high variability, which can obscure meaningful signals. Besides, the effectiveness of machine learning models is often constrained by limited sample sizes, as microbiome data collection remains costly and time consuming. In this context, data augmentation has emerged as a promising strategy to enhance model robustness and predictive performance by generating artificial microbiome data.
\vspace{0.1cm}

\textbf{Results:} The aim of this study is to improve predictive modeling from microbiome data by introducing a model-based data augmentation approach that incorporates both taxonomic relationships and covariate information. To that end, we propose TaxaPLN, a data augmentation method built on PLN-Tree generative models, which leverages the taxonomy and a data-driven sampler to generate realistic synthetic microbiome compositions. We further introduce a conditional extension based on feature-wise linear modulation, enabling covariate-aware generation. Experiments on high-quality curated microbiome datasets show that TaxaPLN preserves ecological properties and generally improves or maintains predictive performances, particularly with non-linear classifiers, outperforming state-of-the-art baselines. Besides, TaxaPLN conditional augmentation establishes a novel benchmark for covariate-aware microbiome augmentation.
\vspace{0.1cm}

\textbf{Availability and implementation:} The MIT-licensed source code is available at \url{https://github.com/AlexandreChaussard/PLNTree-package} along with the datasets used in our experiments.
}

\keywords{Data augmentation, Microbiology, Generative model, Variational inference 
\vspace{-0.6cm}}


\maketitle
\section{Introduction}
The human gut microbiome is a highly complex and diverse ecosystem composed of bacteria, fungi, and viruses living in our gastrointestinal tract. Its composition is determined via high-throughput sequencing technologies \citep{knight2018best}, and varies depending on numerous factors such as host genetics, diet, environmental exposure, lifestyle habits, and health condition \citep{gilbert2018current}. Despite the unique signature of each individual's microbiome, growing evidence suggests that microbial composition can serve as a biomarker for various diseases, including Inflammatory Bowel Disease (IBD), Type 1 Diabetes (T1D), colorectal cancer (CRC), or psychiatric disorders such as schizophrenia \citep{lloyd2019multi,kostic2015dynamics,zeller2014potential,yu2017metagenomic,zhu2020metagenome}. These insights have led to the integration of gut microbiome data into machine learning models for disease classification, prognosis, biomarker discovery, and bacterial interaction network inference \citep{marcos2021applications,chiquet2019variational}.
However, leveraging microbiome data for predictive modeling remains challenging due to host-specific variability, high dimensionality, sparsity, and compositional constraints which raise challenges for model generalization \citep{gloor2017microbiome,quinn2018understanding}. Additionally, thousands of microbial species inhabit the gut microbiome \citep{rosenberg2024diversity}, often distributed in vastly different proportions across individuals. This inherent complexity aggravates the risk of overfitting in machine learning models, particularly when the cohort size is limited, yielding poor generalization. 

To mitigate these challenges, data augmentation has emerged as a promising approach to enrich training sets with synthetic samples \citep{mumuni2022data}. Adapting traditional techniques such as SMOTE \citep{chawla2002smote}, Mixup \citep{zhang2017mixup}, and generative adversarial networks (GAN) \citep{goodfellow2014generative}, recent methods have focused on microbiome-specific augmentation, notably MB-GAN \citep{rong2021mb}, Compositional CutMix, and Aitchison Mixup \citep{gordon2022data} which tackle the compositional nature of microbiome data. Lately, approaches such as TADA \citep{sayyari2019tada}, PhyloMix \citep{jiang2025phylomix}, and PhylaGAN \citep{sharma2024phylagan} have incorporated biological structures such as the taxonomy and phylogeny to guide the generation of synthetic samples, yielding improved predictive performances over earlier methods.

Following these advances, we introduce TaxaPLN, a novel data augmentation strategy tailored to microbiome data that aims at improving predictive analysis. In contrast to prior approaches, TaxaPLN is a deep variational generative model based on the Poisson Log-Normal Tree (PLN-Tree) framework \citep{chaussard2025tree}. By incorporating the microbial taxonomy as a structured prior on taxa distribution, TaxaPLN generates biologically faithful synthetic microbiome profiles using a variational mixture of posteriors (VAMP) sampler \citep{tomczak2018vae}. Additionally, we extend PLN-Tree with a conditional architecture based on Feature-wise Linear Modulation (FiLM) \citep{perez2018film}, enabling TaxaPLN to generate samples conditionally on exogenous metadata, thus allowing covariate-aware data augmentation in microbiome studies.

We conduct a comprehensive evaluation of TaxaPLN on nine high-quality curated datasets from the \texttt{curatedMetagenomicData} database \citep{curated}. First, we evaluate the fidelity of TaxaPLN-generated samples regarding the original cohorts using standard ecology diversity metrics, asserting valid downstream analysis. We then assess the practical utility of TaxaPLN in supervised learning tasks against several baseline augmentation methods. Our benchmark includes four widely used classifiers in microbiome research: logistic regression, random forest, XGBoost, and deep neural networks. Overall, we find that TaxaPLN preserves key biological properties of the cohorts while enhancing predictive performance, with notable gains on non-linear classifiers against baseline methods. In particular, TaxaPLN outperforms the standard PLN baseline, showing the interest of the taxonomy in microbiome modeling. Furthermore, performance increases in covariate-aware data augmentation establish a first baseline in conditional microbiome predictive analysis.



\section{Material and methods}
\label{sec:methods}

\subsection{TaxaPLN augmentation}

\subsubsection{Framework}
\label{sec:framework}
Consider a supervised learning setting with $n$ microbiome samples. Let $\bX\in\RR^{n\times K_L}_+$ the abundances matrix for the $K_L$ taxa observed at the finest taxonomic level $L$, such that the $i$-th row $\bX_i \in \RR^{K_L}_+$ indicates the abundance of each taxon in the $i$-th sample. Each sample is paired with a binary or multiclass label $\bY\in\{0,\dots,M\}^{n}$ describing the host trait of interest (e.g.\ disease status), and when available, the abundance can be joined with exogenous covariates $\bC\in\RR^{n\times d}$ (e.g.\ age, body-mass index, country) to be used as predictors.

Upon sequencing microbial data, genome-resolved features are aggregated along a curated taxonomic hierarchy up to a level $L$ of interest, typically the \emph{genus} in 16S rRNA or the \emph{species} in shotgun sequencing, offering an increasing resolution on the microbiome features as $L$ grows. Thus, the obtained features are related through the taxonomy to form taxa-abundance data, establishing hierarchical compositional relationships along the taxonomy, as illustrated in Figure \ref{fig:curated_samples}. Formally, at level $\ell \in \{1, \dots, L-1\}$ of the hierarchy, denoting by $\rmX_{ik}^\ell \in \RR_+$ the abundance of taxa $k$ at level $\ell$ of the $i$-th sample, the taxa-abundance compositional property yields 
$$
\rmX_{ik}^\ell = \sum_{j \in \childindex{\ell}{k}} \rmX_{ij}^{\ell+1} \eqsp,
$$
where $\childindex{\ell}{k}$ is the set of indices in $\{1, \dots, K_{\ell+1}\}$  related to children taxa of taxa $k$ at layer $\ell$. 
Although sequencing pipelines output integer count, the abundance is often normalized to assert the compositionality property of microbial samples \citep{gloor2017microbiome}, from which counts can be recovered through multinomial sampling \citep{chaussard2025tree,holmes2012dirichlet}.

Given the microbial samples $\bX$ with corresponding traits labels $\bY$, the objective of data augmentation is to generate $\tilde n$ synthetic microbiomes $\tilde\bX \in \RR^{\tilde n \times K_L}$ and labels $\tilde\bY \in \{0, \dots, M\}^{\tilde n}$ such that $\tilde n = \beta n$, where $ \beta > 0$ is the augmentation ratio. Then, defining the augmented training dataset $\{\bX,\tilde \bX\} \in \RR^{(n+\tilde n) \times K_L}$ with labels $\{\bY, \tilde \bY\} \in \{0,\dots, M\}^{n+\tilde n}$, we aim to improve the performances of a classifier by using the augmented training set over the raw data, evaluating on real microbiomes only. Similarly, conditional data augmentation involves generating covariates $\tilde \bC \in \RR^{\tilde n \times d}$, yielding the augmented training set $\{\bX,\tilde\bX,\bC,\tilde\bC\}$ with labels $\{\bY,\tilde \bY\}$.

\subsubsection{PLN-Tree models}
\label{sec:plntree}
\bmhead{Model definition} The Poisson Log-Normal (PLN) model is a standard approach for count data modeling and has seen widespread applications in ecology and microbiome research \citep{aitchison1989multivariate,chiquet_pln}. In this framework, observed abundances at the deepest taxonomic level $L$ are modeled via conditionally independent Poisson distributions whose parameters are given by a latent multivariate log-normal variable: for all $i\in\{1,\ldots,n\}$,
\begin{align*}
    &\bZ_i \sim \gaussian{\bmu}{\bSigma} \eqsp, \\
    \forall k \in \{1, \dots, K_L\}, \quad &\rmX_{ik} \sim \poisson{\exp{(\rmZ_{ik})}} \eqsp,
\end{align*}
where $\bmu \in \RR^{K_L}, \bSigma \in \mathcal{S}_+(\RR^{K_L})$, and conditionally on $\rmZ_{ik}$, $\rmX_{ik}$ is independent of all $\rmX_{ij}$ for $j \neq k$.
Although PLN models have been successfully applied to microbiome network inference \citep{chiquet2019variational,qian2024splang} and feature selection \citep{chiquet_plnpca}, they do not account for taxonomic relationships, and their potential for data augmentation remains unexplored. Besides, generative assessments on microbiome data conducted in \cite{chaussard2025tree}, reveal significant discrepancies between synthetic samples from PLN models and real microbial samples, thereby challenging the effective interpretability of PLN for predictive analysis with microbiome data. 

To address this gap, a tree-based extension known as PLN-Tree \citep{chaussard2025tree} has been developed. By integrating hierarchical dependencies, such as taxonomic relationships, PLN-Tree offers a biologically grounded framework that captures diversity metrics of microbial communities. This model employs a hierarchical stick-breaking process to model the counts, where the highest taxonomic level is modeled with a standard PLN, while lower levels are recursively generated via grouped PLN models under a parent-sum constraint, thus respecting the compositional structure along the taxonomy inherent to taxa-abundance data. The propagation of microbial abundances across the hierarchy through these constrained PLN blocks are parameterized by a latent Gaussian Markov process independent of the tree, affording increased flexibility over the hierarchical structure by modeling all taxa interactions independently of their taxonomic relationships. The random variables $\{(\bZ_i,\bX_i)\}_{1\leq i \leq n}$ are independent and are distributed according to the following model. 
\begin{itemize}
    \item The latent Gaussian Markov process
    \begin{align*}
        &\bZ^1_i \sim \gaussian{\bmu_1}{\bSigma_1} \eqsp, \\
        \forall \ell < L, \quad &\bZ^{\ell+1}_i \mid \bZ^{\ell}_i \sim \gaussian{\bmu_{\btheta_{\ell+1}}(\bZ^\ell_i)}{\bSigma_{\btheta_{\ell+1}}(\bZ^\ell_i)} \eqsp,
    \end{align*}
    is such that $\bZ^\ell_i \in \RR^{K_\ell}$ where $K_\ell$ is the number of taxa at the level $\ell$ of the taxonomy, and $\bmu_{\btheta_{\ell}}, \bSigma_{\btheta_\ell}$ are functions parameterized by $\btheta_\ell$ outputting respectively the mean and covariance matrix of a Gaussian distribution of dimension $K_{\ell}$, the number of taxa at level $\ell$ of the taxonomy.
    \item Conditionally on the latent process, the count emission process is defined as
    \begin{align*}
        &\rmX^1_{ik} \mid \rmZ^1_{ik} \sim \poisson{\exp{(\rmZ^1_{ik})}} \eqsp, \\
        \forall \ell < L, k \leq K_\ell, \quad &\child{\bX}{ik}{\ell} \mid \rmX_{ik}^\ell, \child{\bZ}{ik}{\ell} \sim \multinomial{\rmX_{ik}^\ell}{\softmax{\child{\bZ}{ik}{\ell}}} \eqsp,
    \end{align*}
    such that $\softmax{x} = \{\exp{(x_k)} / \sum_{j=1}^d \exp{(x_j)}\}_{1 \leq k \leq d}$ is the softmax transform, $\multinomial{N}{p}$ is the multinomial distribution with total count $N$ and probabilities $p$, and for any random variable $\bV$, $\child{\bV}{ik}{\ell}$ refers to the vector of children variables of node $k$ at layer $\ell$ for sample $i$ along the taxonomy.
\end{itemize}

Building on prior work for structured latent variable models \citep{halva2021disentangling,gassiat2020identifiability,zhou2020learning}, \cite{chaussard2025tree} establish a class of identifiability for PLN-Tree models, enabling meaningful interpretation of its parameters for medical applications. In the same study, PLN-Tree also exhibits substantial generative improvements over classical PLN models by more accurately reproducing diversity metrics observed in microbial ecosystems. As such, this model demonstrates the value of incorporating taxonomic structure into microbial modeling, while providing a probabilistic framework for interpreting taxa dependencies.

\bmhead{Conditional model} A theoretical extension of PLN-Tree to incorporate exogenous information has been proposed in \cite{chaussard2025tree}, but no practical architecture has been benchmarked. Thus, we propose a Feature-wise Linear Modulation (FiLM) architecture \citep{perez2018film} that can incorporate covariates in both the latent dynamic and the variational approximation. 
For a given sample $i$ and taxonomic level $\ell$, the FiLM block takes the covariate vector $\bC_i\in\RR^{d}$ as input to two neural networks, $\alpha^\ell, \gamma^\ell$, that returns two vectors with dimensionality matching the regular input $\mathbf{A}_i$ in the PLN-Tree architecture. The FiLM block then applies a feature-wise affine transformation
$$
\mathbf{A}^{\mathrm{FiLM}}_i = \alpha^\ell(\bC_i) \odot \mathbf{A}_i + \gamma^\ell(\bC_i) \eqsp,
$$
where  $\odot$ is the Hadamard product. Appendix \ref{fig:prior_architecture_film} and \ref{fig:res_backward_architecture_film} depict the FiLM insertions into the PLN-Tree architecture. 
Empirically, FiLM has shown comparable performances to attention mechanisms \citep{vaswani2017attention,perez2018film,dumoulin2018feature} while being much less parameter-heavy, which can be critical in microbiome studies where $n$ is often small compared to the amount of taxa.

\bmhead{Package availability} To facilitate research and encourage the use of PLN-Tree and TaxaPLN, we release a user-friendly implementation as an open-source Python package \texttt{plntree} available on PyPI. Full documentation and source code are hosted on GitHub\footnote{\url{https://github.com/AlexandreChaussard/PLNTree}}, where all the experiments performed in this paper can be found and reproduced. The framework is developed using PyTorch \citep{paszke2019pytorch} library and allows for batch training on GPU using CUDA for scalability to large datasets.

\subsubsection{Generating microbiomes}
\label{sec:generation_microbiomes}
Motivated by non-parametric methods that condition synthetic sample generation on observed data \citep{chawla2002smote,jiang2025phylomix,gordon2022data}, we introduce a data augmentation strategy that employs the Variational Mixture of Posteriors (VAMP) prior \citep{tomczak2018vae} as a post-training sampler for PLN-Tree. The VAMP prior defines a data-driven latent distribution by forming a uniform mixture of the variational posteriors using microbiome samples from the train set. Formally, if the variational approximation of the posterior is $q_{\bphi}(\bZ | \bX, \bC)$,  parameterized by $\bphi$, the VAMP distribution is given by
\[
\vamp(\bZ) = \frac{1}{n} \sum_{i=1}^n q_{\bphi}(\bZ \mid \bX_i, \bC_i) \eqsp.
\]
This methodology enables the generation of synthetic samples that are generally more representative of the training data than samples from the prior \citep{tomczak2018vae,chadebec2022data}, thereby emphasizing on the biological integrity characterizing the cohort in microbiome analysis. Following the VAMP prior definition, the generative procedure consists in drawing uniformly a microbiome sample from the training set, then encoding it using the variational sampler to obtain its latent representation, before decoding the latent representation into a new sample using the emission distribution. Applied to a PLN-Tree model trained with the taxonomy as a hierarchical prior, this sampling strategy defines the TaxaPLN augmentation strategy.

To assert the practical benefit of our approach, we compare synthetic microbiome data generated with the VAMP prior to data generated with the canonical prior employed in \citep{chaussard2025tree}, following the evaluation protocol of Section~\ref{sec:tasks}. The results provided in Appendix~\ref{app:compare_taxapln_prior} highlight the faithfulness of VAMP prior generated microbiomes to the original data compared to traditional prior sampling, aligning with previous findings in other areas of research \citep{tomczak2018vae,norouzi2020examplar,ai2021bype,chadebec2022data}.

\subsection{Metagenomic datasets}
\label{sec:datasets}
The \texttt{curatedMetagenomicData} package is an extensive collection of human microbiome shotgun metagenomic data from different body sites that has been standardized and curated from 93 published studies \citep{curated}, thus facilitating comparative metagenomics research. Focusing on stool samples, it includes over $15,000$ taxa-abundance samples in relative form sequenced with MetaPhlAn3 \citep{metaphlan} and annotated up to the species level, as well as rich annotations such as the age category, the sex, the geographical location, and the health status of the host.
This vast database integrates studies covering a variety of conditions, including inflammatory bowel disease (IBD), type 1 diabetes (T1D), adenoma, schizophrenia, colorectal cancer (CRC), and other microbiome-related disorders.

\begin{figure}
    \centering
    \includegraphics[width=0.8\linewidth]{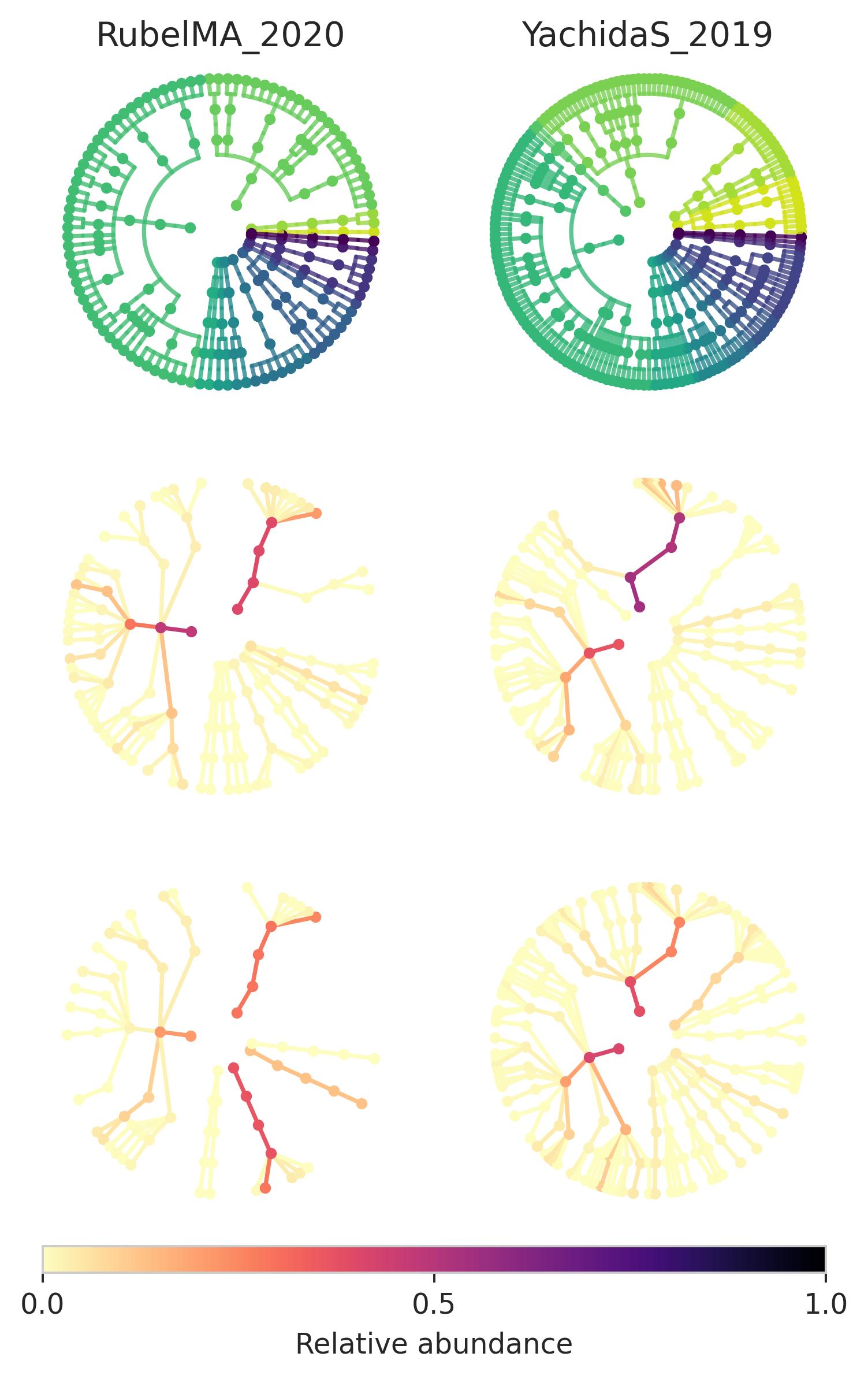}
    \caption{Taxonomy and taxa-abundance sample illustrations from datasets of Table~\ref{tab:curated_dataset_desc}. First row displays the taxonomic tree used in our experiments for the considered dataset, spanning from the \textit{class} to the \textit{species} level. Second and third rows show taxa-abundance profile for a control and a disease-associated sample respectively. In the taxonomy, colors indicate taxa belonging to the same \textit{class}, coloring in abundance profiles represent relative abundances.}
    \label{fig:curated_samples}
\end{figure}

Using this package, we conduct a comprehensive evaluation of TaxaPLN augmentation on 9 studies spanning several conditions and displaying varying amount of samples and imbalance. To mitigate overfitting and reduce noise from spurious taxa which are critical concerns in low-sample-size microbial studies, we perform prevalence filtering at the \textit{species} level, retaining only those taxa present in at least $15\%$ of the samples \citep{karwowska2025effects,asnicar2024machine}. The datasets composition and properties are provided in Table \ref{tab:curated_dataset_desc}.
\begin{table*}
    \centering
    \begin{tabular}{lccccccccc}
    \toprule
    {\textbf{Study ID}} & $\mathbf{n}$ & $\mathbf{p}$ & \textbf{Class 1 / Class 2} & \textbf{\# in 1} & \textbf{\# in 2} & \textbf{Reference} \\
    \midrule
    WirbelJ\_2018 & 125 & 189 & CRC / control & 60 & 65 & \citep{wirbel2019meta} \\
    KosticAD\_2015 & 120 & 104 & T1D / control & 31 & 89 & \citep{kostic2015dynamics} \\
    RubelMA\_2020 & 175 & 103 & STH / control & 89 & 86 & \citep{rubel2020lifestyle} \\
    ZhuF\_2020 & 171 & 155 & schizophrenia / control & 90 & 81 & \citep{zhu2020metagenome} \\
    ZellerG\_2014 & 103 & 216 & adenoma / control & 42 & 61 & \citep{zeller2014potential} \\
    YachidaS\_2019 & 318 & 177 & adenoma / control & 67 & 251 & \citep{yachida2019metagenomic} \\
    YuJ\_2015 & 128 & 198 & CRC / control & 74 & 54 & \citep{yu2017metagenomic} \\
    NielsenHB\_2014 & 396 & 174 & IBD / control & 148 & 248 & \citep{nielsen2014identification} \\
    HMP\_2019\_ibdmdb & 1627 & 103 & IBD / control & 1201 & 426 & \citep{lloyd2019multi} \\
    \bottomrule
    \end{tabular}
    \caption{Curated metagenomics datasets considered in our experiments, extracted from \cite{curated} with prevalence filtering at threshold $15\%$. CRC: colorectal cancer, T1D: type 1 diabetes, STH: soil-transmitted helminth infection, 
    IBD: inflammatory bowel disease.}
    \label{tab:curated_dataset_desc}
\end{table*}
The resulting set of taxa annotations enable to determine the datasets' associated taxonomy which hierarchically organize the taxa-abundance samples. An illustration on two example datasets is provided in Figure \ref{fig:curated_samples} with related taxa-abundance samples.
Additionally, to extend the applicability of count-based models to microbial data presented in Section~\ref{sec:plntree}, we convert relative abundances into count data by sampling from a multinomial distribution with a total count of $100,000$ and probabilities defined by the relative abundances of each sample. The large total count ensures minimal information loss during this transformation.
Finally, when covariates are available for a given study, we collect the age, sex, body mass index (BMI) and country location of the patients for the conditional augmentation in Section~\ref{sec:conditional_augmentation}.

\subsection{Evaluation tasks}
\label{sec:tasks}
Before evaluating TaxaPLN on data augmentation tasks, we first assess the realism of the synthetic microbiome data it generates following the diversity benchmark methodology of \cite{chaussard2025tree}. The goal is to assert whether samples generated by TaxaPLN preserve the biological signature of microbial communities, as quantified by the $\alpha$-diversity and $\beta$-diversity metrics.
Resemblance in $\alpha$-diversity is evaluated using the Shannon and Simpson indices \citep{gotelli2001quantifying,nagendra2002opposite}, comparing synthetic and real cohorts via Mann-Whitney U tests and Kolmogorov–Smirnov (KS) divergences. To assess $\beta$-diversity similarity, we employ Principal Coordinates Analysis (PCoA) on Aitchison distance \citep{aitchison1982statistical} and Bray–Curtis dissimilarity \citep{beals1984bray}, two metrics commonly used in shotgun metagenomics studies.

Following the biological validation, we assess the utility of TaxaPLN augmentation in two supervised learning scenarios: vanilla and conditional. The vanilla setting evaluates TaxaPLN's ability to enhance classification performance through the generation of synthetic microbiome profiles without any exogenous input, relying solely on microbial composition for host trait prediction. In the conditional setting, a PLN-Tree model is trained using both microbial data and covariates via the FiLM architecture defined in Section~\ref{sec:plntree}, enabling TaxaPLN to generate synthetic microbiome samples conditioned on external variables. This setting aims to improve classification performance when both microbiome composition and covariates are used as features.
In both scenarios, the impact of data augmentation is quantified using the Area Under the Precision–Recall Curve (AUPRC), a metric well-suited for imbalanced datasets \citep{davis2006relationship}, as is commonly found in Table~\ref{tab:curated_dataset_desc}.

\subsection{Augmentation settings}
\label{sec:training_settings}
TaxaPLN augmentation is evaluated on four machine learning models commonly applied in microbial-trait association: logistic regression (LR), multi-layer perceptron (MLP), gradient boosting classifier (XGBoost), and random forests (RF). For all classifiers, we employ the default implementations from Scikit-learn \citep{pedregosa2011scikit} with parameters detailed in Appendix~\ref{tab:classifiers_parameterization}.  
For each study in Table \ref{tab:curated_dataset_desc}, we adopt a 5-fold cross-validation scheme repeated 25 times with different random seeds to ensure robust performance estimates. 
Within each five splits of a cross-validation iteration, we fit a label-specific TaxaPLN model on the training data using the hyperparameters detailed in Section~\ref{sec:practical_analysis}. For the taxonomy ablation study, we additionally fit two label-specific PLN models, implemented via the \texttt{pyPLNmodels} package \citep{batardiere2024pyplnmodels}. The learned label-specific models are then used to augment the training data with a fixed augmentation ratio $\beta = 2$.
For all augmentation strategies, microbiome data are first augmented in raw count form, then preprocessed using the centered log-ratio (CLR) transformation \citep{aitchison1982statistical} prior to classifier training.
Predictions are made on each test split of the cross-validation procedure, allowing the computation of one AUPRC value per split. For each iteration of the cross-validation procedure, we average the AUPRC scores across the five test splits, yielding one AUPRC value per iteration. Repeating the cross-validation while changing seed yields 25 AUPRC evaluations per dataset, which serve as the benchmark metric for comparing augmentation strategies.
Besides, Appendix~\ref{app:relative_preprocessing} presents alternative results using proportion data instead of CLR-transformed, assessing the robustness of TaxaPLN to different preprocessing choices.

\subsection{Baseline methods}
\label{sec:baselines}
TaxaPLN is benchmarked against several augmentation methods, each generating an equivalent number of synthetic microbiome samples for evaluation, both for biological metrics analysis and downstream classification performances. To assess the contribution of the taxonomy-informed prior, we first compare TaxaPLN to a PLN baseline using the VAMP generator described in Section~\ref{sec:generation_microbiomes}, with the parameterization introduced by \citet{chiquet_pln}. In addition, we evaluate TaxaPLN against state-of-the-art microbiome augmentation techniques, including Vanilla Mixup \citep{zhang2017mixup}, which generates convex combinations of microbiome compositions, Compositional Cutmix \citep{gordon2022data} which is an extension of Cutmix \citep{yun2019cutmix} adapted to compositional data via renormalization of mixed samples, and PhyloMix \citep{jiang2025phylomix}, a biologically informed variant of Compositional Cutmix in which partitions are guided by structured biological graphs, such as the taxonomy or the phylogeny. As phylogenies are not consistently defined for all sequenced taxa, our experiments with PhyloMix rely exclusively on taxonomic information provided in \texttt{curatedMetagenomicData} \citep{curated}. Moreover, since none of these mixup strategy support the integration of covariates, the conditional augmentation experiments include comparisons only between TaxaPLN and PLN using their respective conditional VAMP sampler.
All experiments are conducted using the original implementations provided by the respective authors \citep{batardiere2024pyplnmodels, gordon2022data, jiang2025phylomix}.

\section{Results}
\label{sec:results}

\subsection{TaxaPLN generates realistic microbiomes}
\label{sec:exp_realistic_microbiome}
TaxaPLN models, along with PLN baselines, are trained on the metagenomics datasets listed in Table~\ref{tab:curated_dataset_desc} using the 5-fold cross-validation strategy described in Section~\ref{sec:training_settings}, enabling the generation of synthetic microbiome samples conditioned on class labels. To evaluate generation quality, we select the models trained during the first cross-validation iteration and sample synthetic microbiome profiles using TaxaPLN and the other baseline augmentation methods considered in our benchmark.

\begin{figure*}[htbp]
    \centering
    \begin{subfigure}[b]{0.9\linewidth}
        \centering
        \includegraphics[width=\linewidth,trim=0 0 0 0,clip]{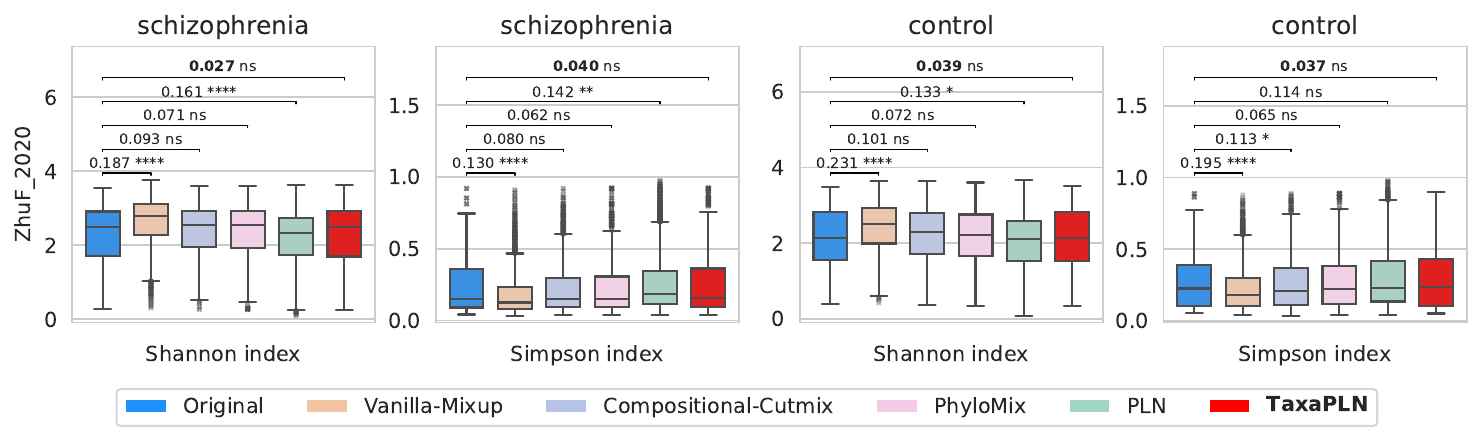}
    \end{subfigure}
    \caption{Shannon and Simpson $\alpha$-diversity distributions. Synthetic microbiome samples generated by TaxaPLN and baseline augmentation methods are evaluated on their $\alpha$-diversity consistency with the original microbiomes from two studies of Table~\ref{tab:curated_dataset_desc} based on Shannon index and Simpson index. Each method generates $500$ samples. Mann-Whitney U tests are performed to assess the statistical significance of distribution differences between generated and original samples for each method. Significance $P$-values thresholds are denoted by: ****$P \leq 0.0001$; ***$P \leq 0.001$; **$P \leq 0.01$; *$P \leq 0.05$; ns: $P > 0.05$. Kolmogorov-Smirnov divergence between original samples and generated microbiomes is provided with bold value indicating the minimal distance. Results on all datasets are reported in Appendix \ref{fig:vanilla_augmentation_alpha_diversity_full}.}
    \label{fig:vanilla_augmentation_alpha_diversity}
\end{figure*}

First, we assess the similarity between the generated and original microbiomes using $\alpha$-diversity metrics, specifically the Shannon and Simpson indices \citep{gotelli2001quantifying}. To evaluate whether the distributions of diversity indices differ significantly, we conduct Mann-Whitney U tests and provide empirical Kolmogorov-Smirnov divergence evaluations. Figure~\ref{fig:vanilla_augmentation_alpha_diversity} presents results for two representative studies, while complete results are provided in Appendix~\ref{fig:vanilla_augmentation_alpha_diversity_full}.
Our findings indicate that synthetic microbiomes generated by TaxaPLN always preserve the Shannon and Simpson $\alpha$-diversity signatures of the considered cohorts. Similarly, PhyloMix and Compositional Cutmix are almost always accepted, suggesting these methods also preserve key diversity characteristics. Conversely, PLN-generated data are always rejected on both metrics for all datasets, highlighting the interest of taxonomic relationships in microbiome modeling. Besides, Vanilla Mixup generations yield a significant $\alpha$-diversity shifts across all studies, being also consistently rejected. These results indicates that both PLN and Vanilla Mixup distort biological signatures of microbiome data, potentially impairing downstream analysis.

\begin{figure*}[htbp]
    \centering
    \begin{subfigure}[b]{0.98\linewidth}
        \centering
        \includegraphics[width=\linewidth,trim=0 40 0 0,clip]{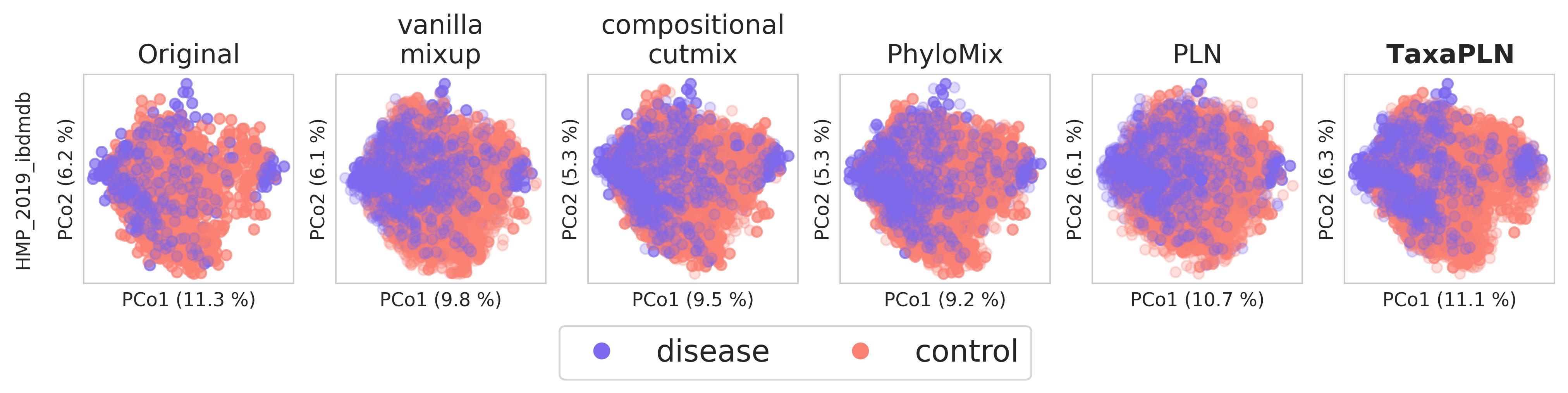}
        \caption{PCoA with Aitchison distance}
    \end{subfigure}
    \begin{subfigure}[b]{0.98\linewidth}
        \centering
        \includegraphics[width=\linewidth,trim=0 40 0 38,clip]{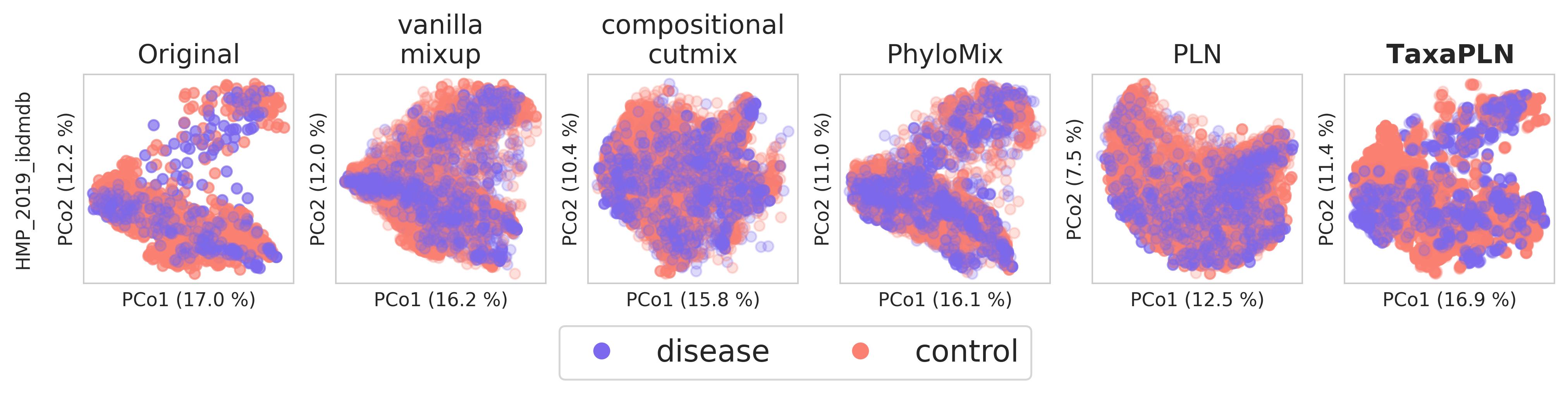}
        \caption{PCoA with Bray-Curtis dissimilarity}
    \end{subfigure}
    \begin{subfigure}[b]{0.98\linewidth}
        \centering
        \includegraphics[width=\linewidth,trim=0 0 0 170,clip]{braycurtis-PCoA-HMP_2019_ibdmdb}
    \end{subfigure}
    \caption{Bray–Curtis and Aitchison $\beta$-diversity PCoA plots of augmented datasets compared to the original cohort. Synthetic microbiome samples generated by TaxaPLN and baseline augmentation methods are evaluated on their $\beta$-diversity consistency with the original microbiomes. Each method augment the training fold with $\beta=2$. Prior to Bray–Curtis dissimilarity computation, all data are normalized into proportions, while Aitchison distance are computed using the Euclidean distance on CLR-transformed counts. Principal Coordinates Analysis (PCoA) is used to visualize the dissimilarity structure. Results on all datasets are reported in Appendix \ref{fig:aitchison_vanilla_augmentation_beta_diversity_full} and \ref{fig:braycurtis_vanilla_augmentation_beta_diversity_full}.}
    \label{fig:vanilla_augmentation_beta_diversity}
\end{figure*}

Additionally, we assess $\beta$-diversity alignments between generated and original microbiome samples using both the Aitchison distance \citep{aitchison1982statistical} and Bray–Curtis dissimilarity \citep{beals1984bray} through Principal Coordinates Analysis (PCoA).
As shown in Figure~\ref{fig:vanilla_augmentation_beta_diversity}, TaxaPLN preserve the $\beta$-diversity patterns of the original cohort, whereas other methods often introduce noticeable shifts. In particular, TaxaPLN-generated samples closely align with the real samples in the Bray–Curtis projected space, indicating more accurate preservation of the original cohort's ecological structure. In the Aitchison geometry, TaxaPLN further demonstrates the ability to explore the space of microbiome compositions while generalizing the geometric structures observed in the original data, contrary to other methods which tend to distort patterns. Together, these results suggest that TaxaPLN achieves a balance between exploration and ecological fidelity, enabling meaningful augmentation of microbiome cohorts.

\subsection{Data augmentation}

\subsubsection{Vanilla augmentation}
\label{sec:vanilla_augmentation}
\begin{figure*}[h]
    \centering
    \begin{subfigure}[b]{0.98\linewidth}
        \centering
        \includegraphics[width=\linewidth,trim=0 35 0 0,clip]{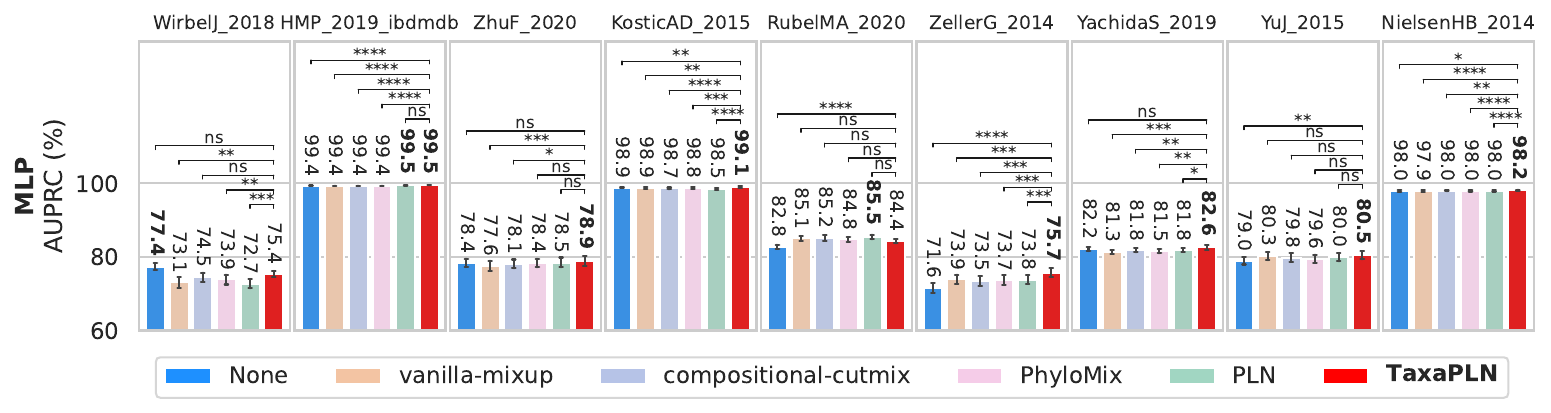}
    \end{subfigure}
    \begin{subfigure}[b]{0.98\linewidth}
        \centering
        \includegraphics[width=\linewidth,trim=0 35 0 20,clip]{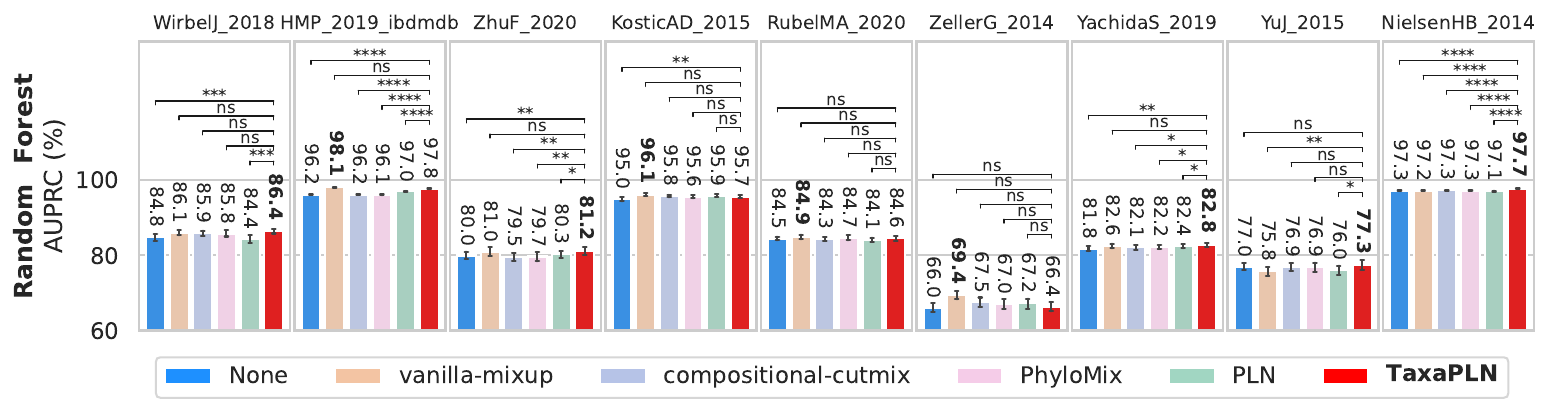}
    \end{subfigure}
    \begin{subfigure}[b]{0.98\linewidth}
        \centering
        \includegraphics[width=\linewidth,trim=0 35 0 20,clip]{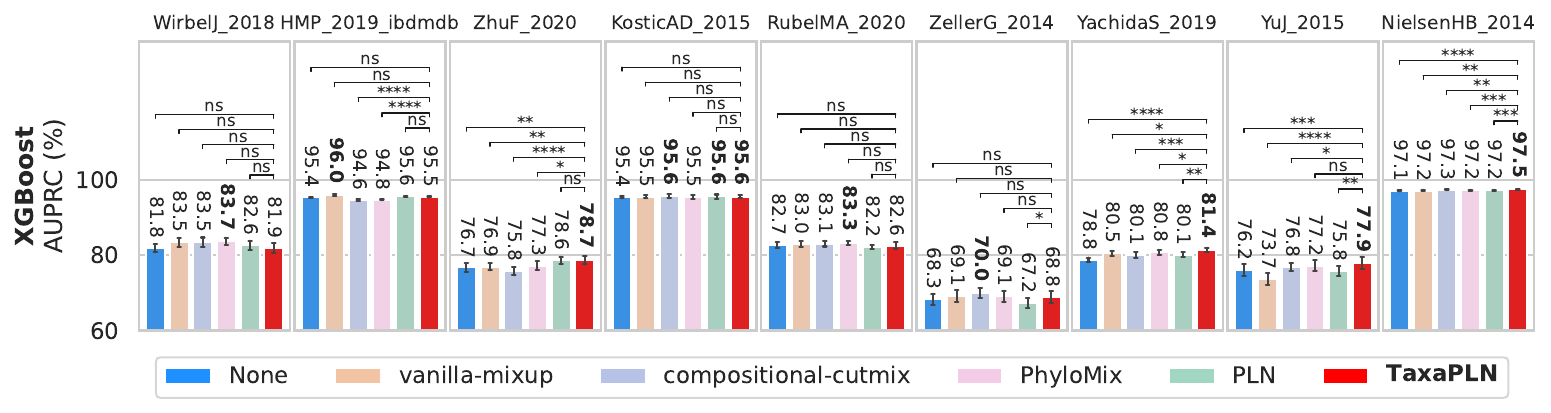}
    \end{subfigure}
    \begin{subfigure}[b]{0.98\linewidth}
        \centering
        \includegraphics[width=\linewidth, trim=0 0 0 20,clip]{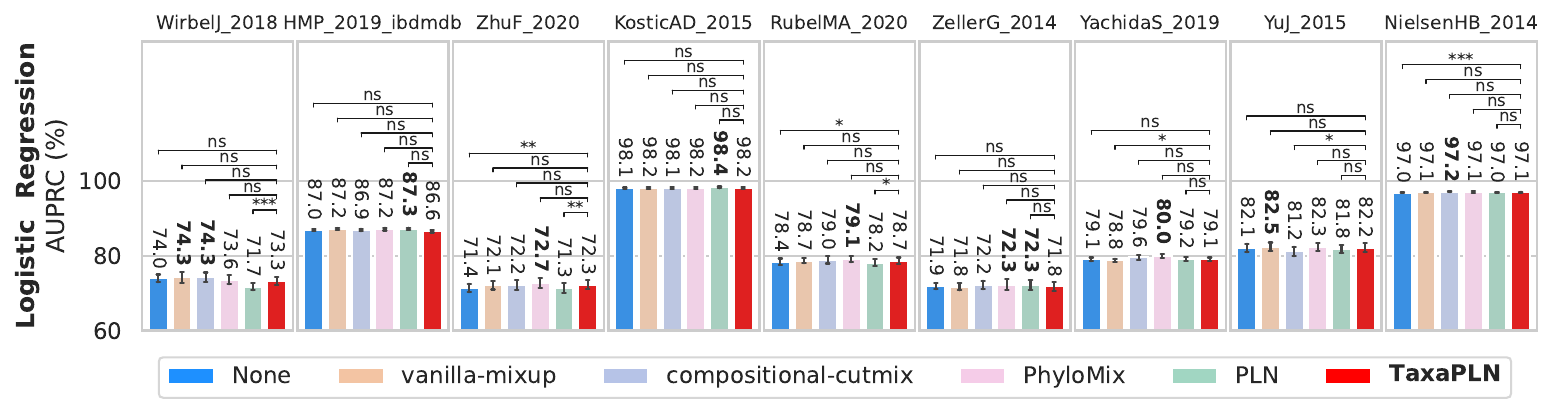}
    \end{subfigure}
    \caption{Vanilla augmentation performances on the supervised learning tasks. The evaluation was performed on real microbiome datasets from Table \ref{tab:curated_dataset_desc} sequenced at the \textit{species} level. TaxaPLN augmentation is evaluated with five classifiers and compared against four baseline methods according to Section~\ref{sec:vanilla_augmentation}. TaxaPLN augmentation generally preserves or enhance predictive performances in varying dimensionality and data availability contexts. Performance comparison between TaxaPLN and other methods is assessed using one-tailed paired $t$-tests with significance: ****$P$-value $\leq 0.0001$; ***$P$-value $\leq 0.001$; **$P$-value $\leq 0.01$; *$P$-value $\leq 0.05$; ns: $P$-value $>$ 0.05.}
    \label{fig:vanilla_augmentation_performances}
\end{figure*}
TaxaPLN augmentation is evaluated on $9$ microbiome-trait classification tasks drawn from studies summarized in Table~\ref{tab:curated_dataset_desc}. These datasets span a diverse range of microbiome-associated conditions, including inflammatory bowel disease, schizophrenia, type 1 diabetes, adenoma, parasitic infections, and colorectal cancer, thus providing a comprehensive benchmark for assessing our method across varied biomedical contexts.

Performance is measured using a 5-fold cross-validation scheme repeated 25 times, as detailed in Section~\ref{sec:training_settings}. The resulting AUPRC scores are reported in Figure~\ref{fig:vanilla_augmentation_performances}, along with their $95\%$ confidence intervals. To assess statistical significance, we conduct one-tailed paired $t$-tests comparing the mean AUPRC of TaxaPLN with that of each baseline augmentation strategy.

The augmentation results indicate that TaxaPLN generally enhances or preserves the predictive performance of various classifiers. Specifically, TaxaPLN achieves maximum test AUPRC improvements of $4.1\%$, $2.6\%$, $1.6\%$, and $0.9\%$ when used with MLP, XGBoost, Random Forest, and Logistic Regression classifiers respectively. In comparison, the second-best augmentation method, PhyloMix, achieves maximum gains of $2.1\%$, $2.0\%$, $1.0\%$, and $1.3\%$ with the same classifiers.
For non-linear models, TaxaPLN leads to performance degradation in only $2$ out of $30$ experiments, with a maximum AUPRC drop of $-2.0\%$. In contrast, PhyloMix, Compositional CutMix, and Vanilla Mixup lead to performance drops in $8/30$, $10/30$, and $5/30$ experiments, with respective maximum declines of $-3.5\%$, $-2.8\%$, and $-4.3\%$.
On average across studies, TaxaPLN ranks first among all augmentation methods when used with the MLP and XGBoost classifiers, yielding respective mean test AUPRC gains of $0.74\%$ and $0.83\%$. It also ranks second with Random Forest, just behind Vanilla Mixup, with an average gain of $0.82\%$, but ranks last with Logistic Regression, showing an average gain $0.01\%$. As observed for all methods, relative gains are generally modest, since most studies already achieve AUPRC scores around $80\%$ or higher likely due to the high quality and curation of the \texttt{curatedMetagenomicData} microbiome database and relative simplicity of the diagnoses tasks \citep{curated}.

Assessing the significance of TaxaPLN gains over PhyloMix, one-tailed paired $t$-tests show that TaxaPLN achieves significantly better performance on $6$ studies with MLP, $4$ with Random Forest, $4$ with XGBoost, and $0$ with Logistic Regression. These results suggest that TaxaPLN performs at least as well as PhyloMix with non-linear classifiers, and offers significant improvements in roughly half of the tested scenarios, while it is less efficient with linear classifiers.
Additionally, we observe that TaxaPLN generally performs better than PLN, with significantly better results in $18$ out of the $36$ experiments. Taken together with the biological conservation demonstrated by TaxaPLN relative to PLN, these findings highlight the benefit of incorporating taxonomic structure to generate synthetic microbiome data that are both biologically relevant and effective for data augmentation, as previously observed with PhyloMix \citep{jiang2025phylomix}.

\subsubsection{Conditional augmentation}
\label{sec:conditional_augmentation}
Incorporating host-related external covariates into microbiome-trait classification allows to investigate how environmental and physiological factors influence microbial communities \citep{wu2025machine,julien2022impact,wu2025microbiome}. Unlike non-parametric augmentation methods in the baseline, TaxaPLN augmentation is built on a probabilistic generative model that can naturally accommodates exogenous information via conditional sampling, as detailed in Section~\ref{sec:plntree}.

\begin{figure}[htbp]
    \centering
    \begin{subfigure}[b]{0.98\linewidth}
        \centering
        \includegraphics[width=\linewidth,trim=0 50 0 0,clip]{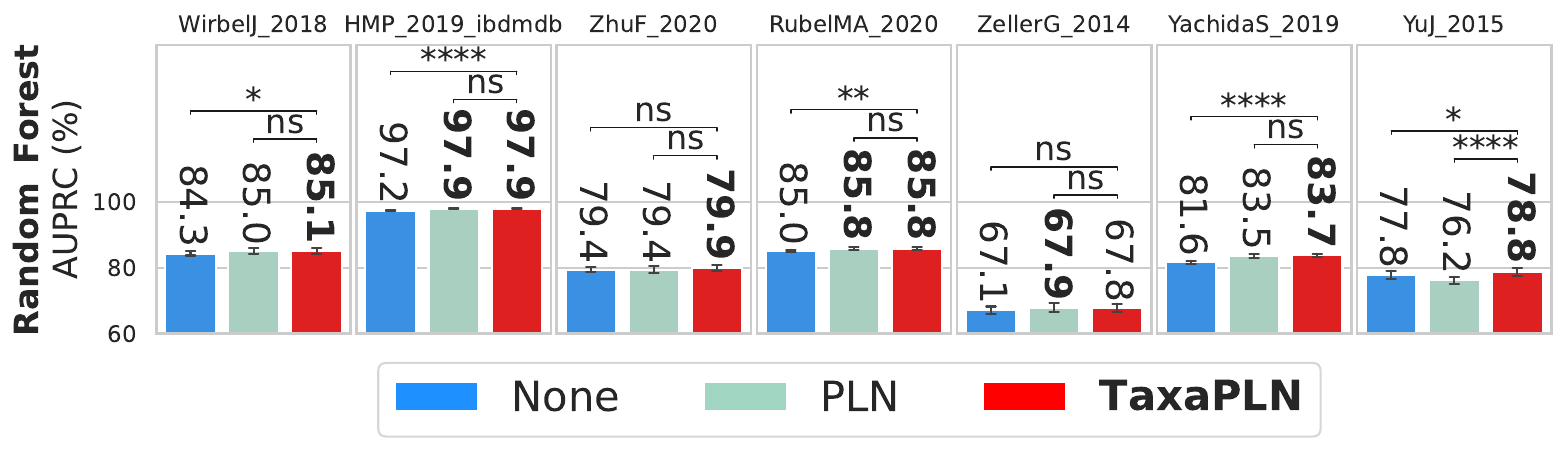}
    \end{subfigure}
    \begin{subfigure}[b]{0.98\linewidth}
        \centering
        \includegraphics[width=\linewidth,trim=0 0 0 20,clip]{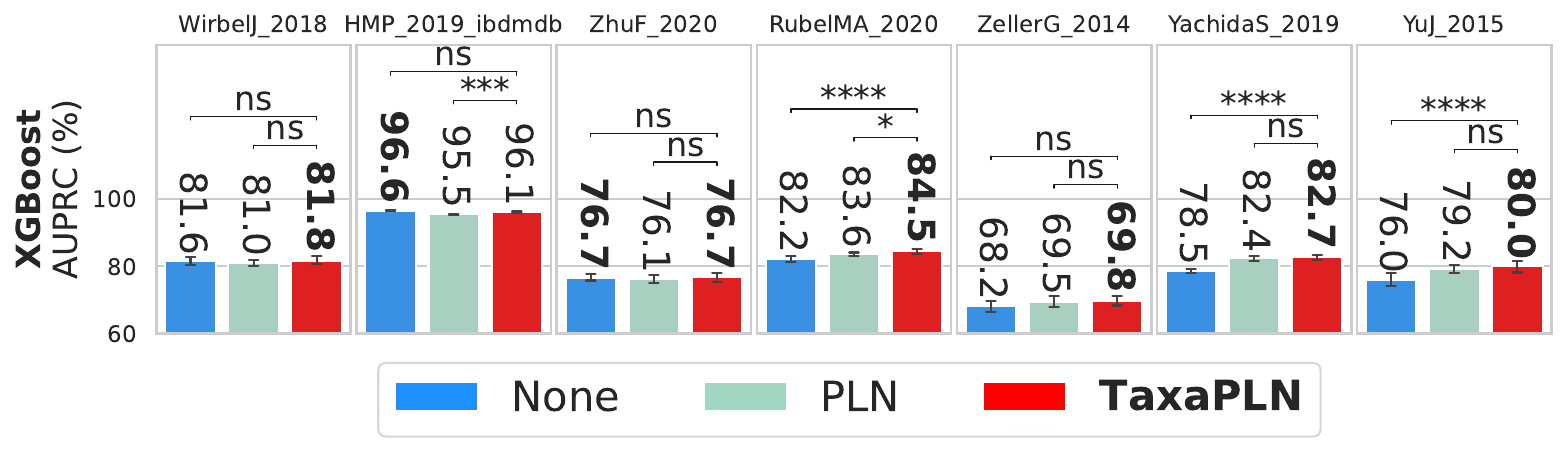}
    \end{subfigure}
    \caption{Conditional augmentation performances on the supervised learning tasks. The evaluation was performed on real microbiome datasets from Table \ref{tab:curated_dataset_desc} sequenced at the \textit{species} level with covariates described in Appendix \ref{tab:covariates}. TaxaPLN augmentation is evaluated with two classifiers and compared against a conditional PLN model using VAMP sampler. TaxaPLN augmentation generally preserves or enhance predictive performances in varying dimensionality and data availability contexts. Performance comparison between TaxaPLN and other methods is assessed using one-tailed paired $t$-tests with significance: ****$P$-value $\leq 0.0001$; ***$P$-value $\leq 0.001$; **$P$-value $\leq 0.01$; *$P$-value $\leq 0.05$; ns: $P$-value $>$ 0.05.}
    \label{fig:conditional_augmentation_performances}
\end{figure}

We evaluate TaxaPLN's conditional augmentation on $7$ out of the $9$ datasets from Table~\ref{tab:curated_dataset_desc} that all include information on patients' age, sex, body mass index (BMI), and country of residence. Prior to model training, these covariates are preprocessed according to Appendix~\ref{tab:covariates}. Performance evaluation follows the 5-fold cross-validation protocol described in Section~\ref{sec:training_settings}. AUPRC scores are reported in Figure~\ref{fig:conditional_augmentation_performances}, alongside 95\% confidence intervals and one-tailed paired $t$-tests assessing the statistical significance of TaxaPLN’s mean AUPRC improvement over baselines.

The comparative baseline is limited to PLN, as no other augmentation strategy explicitly supports the incorporation of covariates into microbiome generation to the best of our knowledge. Moreover, the set of benchmarked classifiers is restricted to Random Forest and XGBoost, as these models naturally accommodate the heterogeneous data types resulting from the concatenation of covariates and microbiome compositions.

The results show that conditional TaxaPLN augmentation consistently enhances or preserves the predictive performance of both Random Forest and XGBoost classifiers. Specifically, TaxaPLN achieves maximum test AUPRC improvements of $2.1\%$ and $4.0\%$ on Random Forest and XGBoost respectively, with corresponding average gains of $1.6\%$ and $0.9\%$. In comparison, the PLN baseline yields average AUPRC improvements of $1.2\%$ on Random Forest and $0.4\%$ on XGBoost, with more frequent negative impact than TaxaPLN.
Statistical analysis using one-tailed paired $t$-tests indicates that TaxaPLN significantly outperforms the baseline in over half of the experiments, and in two cases for PLN. A generative benchmark comparing synthetic microbiomes produced by TaxaPLN and PLN in Appendix~\ref{app:conditional_diversity_benchmark} further compliments these findings by showing that TaxaPLN preserves biological statistics of the data while PLN exhibits significant biological shifts.
These results demonstrate that TaxaPLN provides an effective and biologically grounded strategy for conditional data augmentation, with the potential to improve downstream classification and to enhance our understanding of the relationships between exogenous covariates and microbiome composition.

\subsection{Practical analysis of TaxaPLN}
\label{sec:practical_analysis}
\bmhead{Model parametrization} Choosing a PLN-Tree architecture involves specifying the parameterization of both the latent dynamics and the variational approximation. In our experiments, we adopt a similar deep architecture to that of \cite{chaussard2025tree} by using a linear latent dynamics and a residual backward variational approximation, as detailed in Appendix \ref{app:architecture_selection}. Despite not being tailored to individual datasets, this architecture displays consistent performance gains across varied studies, thus highlighting the flexibility of the TaxaPLN framework.
Besides, while a comprehensive hyperparameter tuning study is beyond the scope of this work, we note that appropriate architecture selection could further optimize augmentation performances. In particular, smaller datasets would benefit from lower-capacity models with regularization to prevent overfitting, whereas larger datasets could leverage more expressive architectures to capture fine-grained patterns in microbiome composition. Additionally, the number of taxa relative to the cohort size should also inform the choice of PLN-Tree architecture, in particular in high-dimensional low-sample-size contexts where overparameterization in deep learning modules can hinder generalization \citep{liu2017deep,chadebec2022data}. Ultimately, model parameterization plays a central role in the time efficiency of the augmentation pipeline. In our experiments, training a PLN-Tree model takes between $7$ and $15$ minutes, depending on dataset size, performing $10{,}000$ epochs on a NVIDIA TITAN X 12GB GPU, with similar performances on i5-1335U$\times$12 CPU. This introduces a non-negligible computational overhead compared to non-parametric methods, but remains comparable to that of other deep-based models such as MB-GAN \citep{rong2021mb}.

\begin{figure}[htbp]
    \centering
    \includegraphics[width=0.98\linewidth]{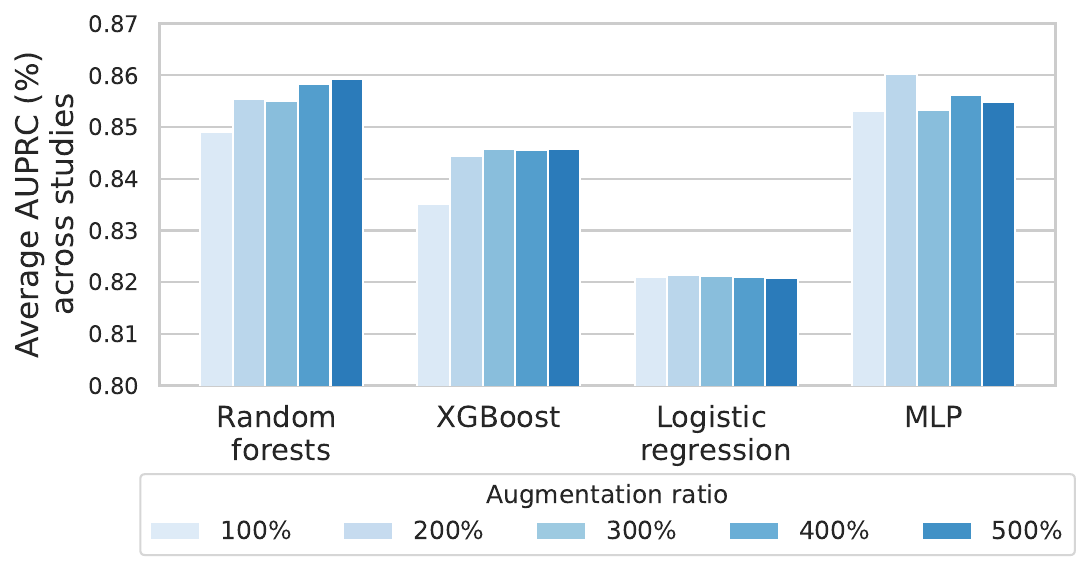}
    \caption{Impact of augmentation ratio on average classification performances with TaxaPLN. Averages are obtained from the performances on all datasets in Table~\ref{tab:curated_dataset_desc}. The augmentation ratio $\beta$ takes value spanning between $1$ (no augmentation) and $5$. Random Forests and XGBoost show stabilizing performance increase with $\beta$, while MLP has an optimum in $\beta = 2$, and linear models only show performance decrease with $\beta$.}
    \label{fig:impact_beta}
\end{figure}

\bmhead{Augmentation ratio} As observed with mixup-based augmentation methods \citep{zhang2017mixup,gordon2022data,jiang2025phylomix}, the number of generated microbiome samples has a measurable impact on classifiers performances that depends on the augmentation strategy and classifier. To investigate this effect with TaxaPLN, we conduct an analysis by varying the augmentation ratio $\beta$ and evaluating average classification performance across all studies.
Results shown in Figure~\ref{fig:impact_beta} reveal distinct behaviors across classifiers. Tree-based models such as Random Forest and XGBoost exhibit consistently improved performances as $\beta$ increases before plateauing. In contrast, linear models tend to degrade in performances with higher $\beta$ values, while the MLP classifier displays a critical peak in performances at $\beta=2$, followed by a drop back to baseline performance when an excessive number of synthetic samples are generated.
Overall, we find that $\beta = 2$ offers a reasonable compromise across the various studies and classifier types. Even so, specific applications would benefit of cross-validation studies to determine an optimal value of $\beta$ to maximize downstream performances.

\begin{figure}[htbp]
    \centering
    \includegraphics[width=0.9\linewidth]{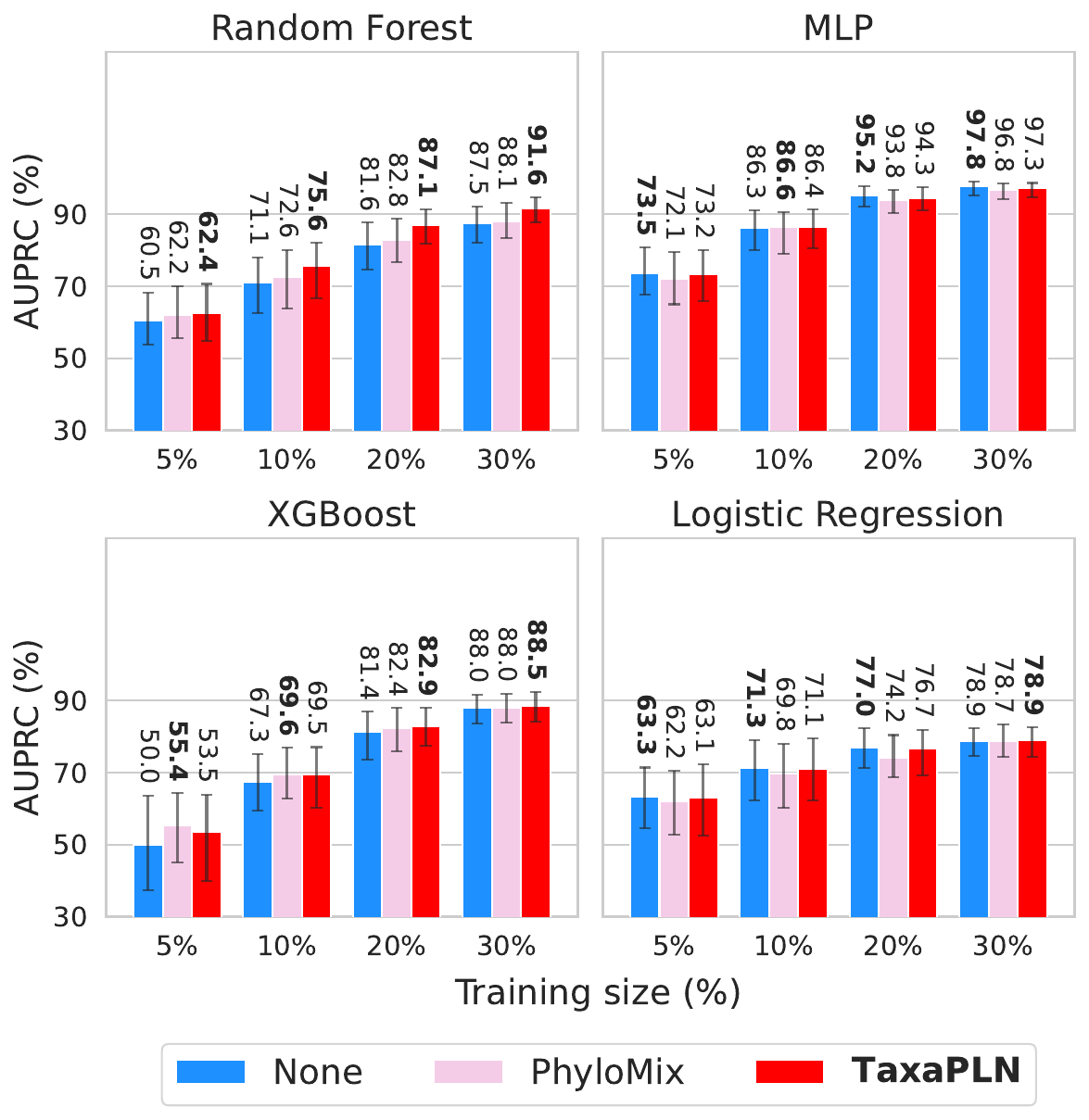}
    \caption{Effect of augmentation using TaxaPLN and PhyloMix with varying numbers of training samples from the \texttt{HMP\_2019\_ibdmdb} cohort, with augmentation ratio $\beta = 2$.
    Average performance is computed over $100$ hold-out splits at each specified training size, with performance measured through average AUPRC $\pm 95\%$ confidence intervals. At smaller training sizes, Random Forest and XGBoost classifiers exhibit notable relative gains from augmentation, which diminish as the amount of training data increases. On average, TaxaPLN augmentation achieves comparable or superior performances to PhyloMix across training sizes.}
    \label{fig:impact_samples}
\end{figure}

\bmhead{Cohort size effect} We investigate the impact of training set size on the effectiveness of TaxaPLN augmentation. Specifically, we conduct a control experiment using the \texttt{HMP\_2019\_ibdmbd} dataset, where subsets of the training data are randomly selected to represent $5\%$ to $30\%$ of the full cohort. This procedure is repeated 100 times with different random seeds to assess variability in performance. For each subset, a TaxaPLN model is trained and used for vanilla data augmentation with $\beta = 2$, followed by classification with different models.
Performance results are reported in Figure~\ref{fig:impact_samples}. We observe that although data augmentation on larger training sets generally yields better performances, the relative improvements brought by TaxaPLN are more pronounced when training data are limited, highlighting the practical relevance of TaxaPLN in low-resource settings. Besides, the performance gains are generally comparable to those obtained with PhyloMix, with TaxaPLN achieving superior average performance in 13 out of 16 scenarios, thus underpinning the interest of TaxaPLN for data augmentation across varying sample size settings.

\section{Discussion}
\label{sec:discussion}
In this work, we propose TaxaPLN, a novel data augmentation method designed for microbiome data to enhance predictive performance while enabling the incorporation of exogenous information. At its core, TaxaPLN leverages taxonomic relationships among microbial taxa as an informative structural prior, modeling their distribution using a PLN-Tree generative model \citep{chaussard2025tree}. Building on the variational approximation obtained during model fitting, we introduce an improved sampling strategy inspired by the VAMP prior \citep{tomczak2018vae}, showcasing biologically faithful generations. Furthermore, we extend the PLN-Tree model with a FiLM-based conditional architecture \citep{perez2018film}, enabling the incorporation of covariates and conditional sampling. To facilitate applications and reproducibility, we also release a Python package on PyPI implementing PLN-Tree and TaxaPLN.

We evaluated TaxaPLN augmentation on a diverse set of high-quality real-world microbiome datasets drawn from \texttt{curatedMetagenomicData} \citep{curated}, displaying varying sample sizes and feature dimensionality applications. Using $\alpha$-diversity and $\beta$-diversity metrics, we demonstrated that TaxaPLN augmentation preserves the biological integrity of the cohorts, ensuring compatibility with downstream analysis, contrary to PLN that lacks the taxonomic information. 
Subsequently, we performed a comprehensive data augmentation benchmark involving several machine learning models commonly used in microbiome research, including logistic regression, random forests, XGBoost and deep neural networks. Overall, TaxaPLN augmentation enhances or preserves predictive performance, particularly with non-linear models where it consistently outperforms baseline non-parametric methods such as PhyloMix and Compositional Cutmix.
Moreover, TaxaPLN improves classification performance in conditional data augmentation tasks involving exogenous covariates such as patient age, BMI, sex, and country of residence, establishing a novel baseline for covariate-aware microbiome modeling.

While TaxaPLN shows consistent empirical performance, several aspects offer room for improvement and future development. In our study, we adopted a general-purpose architecture and fixed augmentation ratio across datasets to ensure consistency and simplicity. Nonetheless, performance may benefit from dataset-specific hyperparameter tuning, particularly in low-data settings. Additionally, as in most variational models, the quality of generated samples depends heavily on the fit of the variational approximation, which can lead to more conservative sampling and limited diversity when model capacity is not tailored to data availability constraints. Finally, although the conditional architecture enables covariate-aware generation, the interest of TaxaPLN for analyzing microbiome–covariate relationships remains unexplored.

As such, this study opens several promising directions for future work. The generative framework of TaxaPLN offers a broad design space with numerous variants yet to be explored. For instance, implementing $\beta$-VAE regularization \citep{kumar2020implicit} or early stopping strategies could improve augmentation quality by regularizing the variational approximation, namely in low-data regimes. Besides, leveraging the latent space could open opportunities for advanced sampling strategies, such as latent mixup \citep{mangla2021benefits} or Riemannian geodesic interpolation \citep{chadebec2022data}, which could enhance exploration while improving augmentation performances.
As a model-based approach, TaxaPLN also offers tools for investigating the biology of the gut microbiome. Exploring structured parameterizations in PLN-Tree, namely sparse or block-diagonal precision matrices, could yield interpretable insights into microbial community structures, revealing interactions that guide synthetic data generation. Finally, automatically tuning the augmentation ratio to maximize downstream performance could further enhance the practical impact of augmentation strategies. Thus, exploring data-driven methods to identify the optimal number of synthetic samples would yield substantial interest for practical applications. 

\backmatter

\section*{Declarations}
\bmhead{Funding} A. C. PhD thesis public funds are provided by Institute of Computing and Data Sciences (ISCD) from Sorbonne Universit\'e.
\bmhead{Author contributions} A. C. proposed the methodological development under the supervision of S. L. C. and A. B. while H. S. supervised the metagenomics aspects. A. C. developed the code, carried out the experiments, and wrote the first draft of the manuscript. Reviewing and editing has been performed by all authors.
\bmhead{Code availability} We develop a PyPI package \texttt{plntree} allowing for PLN-Tree inference and TaxaPLN augmentation. Installation procedure as well as tutorial and documentations are available on the GitHub repository.

\bmhead{Data availability} All microbiome datasets are pulled from the \texttt{curatedMetagenomicData} R package \citep{curated} on open access. Specific data used in our experiments are available on our GitHub.
\bmhead{Conflict of interest} The authors declare no conflict of interest.


\bibliography{sn-bibliography}


\begin{thebibliography}{64}
\ifx \bisbn   \undefined \def \bisbn  #1{ISBN #1}\fi
\ifx \binits  \undefined \def \binits#1{#1}\fi
\ifx \bauthor  \undefined \def \bauthor#1{#1}\fi
\ifx \batitle  \undefined \def \batitle#1{#1}\fi
\ifx \bjtitle  \undefined \def \bjtitle#1{#1}\fi
\ifx \bvolume  \undefined \def \bvolume#1{\textbf{#1}}\fi
\ifx \byear  \undefined \def \byear#1{#1}\fi
\ifx \bissue  \undefined \def \bissue#1{#1}\fi
\ifx \bfpage  \undefined \def \bfpage#1{#1}\fi
\ifx \blpage  \undefined \def \blpage #1{#1}\fi
\ifx \burl  \undefined \def \burl#1{\textsf{#1}}\fi
\ifx \doiurl  \undefined \def \doiurl#1{\url{https://doi.org/#1}}\fi
\ifx \betal  \undefined \def \betal{\textit{et al.}}\fi
\ifx \binstitute  \undefined \def \binstitute#1{#1}\fi
\ifx \binstitutionaled  \undefined \def \binstitutionaled#1{#1}\fi
\ifx \bctitle  \undefined \def \bctitle#1{#1}\fi
\ifx \beditor  \undefined \def \beditor#1{#1}\fi
\ifx \bpublisher  \undefined \def \bpublisher#1{#1}\fi
\ifx \bbtitle  \undefined \def \bbtitle#1{#1}\fi
\ifx \bedition  \undefined \def \bedition#1{#1}\fi
\ifx \bseriesno  \undefined \def \bseriesno#1{#1}\fi
\ifx \blocation  \undefined \def \blocation#1{#1}\fi
\ifx \bsertitle  \undefined \def \bsertitle#1{#1}\fi
\ifx \bsnm \undefined \def \bsnm#1{#1}\fi
\ifx \bsuffix \undefined \def \bsuffix#1{#1}\fi
\ifx \bparticle \undefined \def \bparticle#1{#1}\fi
\ifx \barticle \undefined \def \barticle#1{#1}\fi
\bibcommenthead
\ifx \bconfdate \undefined \def \bconfdate #1{#1}\fi
\ifx \botherref \undefined \def \botherref #1{#1}\fi
\ifx \url \undefined \def \url#1{\textsf{#1}}\fi
\ifx \bchapter \undefined \def \bchapter#1{#1}\fi
\ifx \bbook \undefined \def \bbook#1{#1}\fi
\ifx \bcomment \undefined \def \bcomment#1{#1}\fi
\ifx \oauthor \undefined \def \oauthor#1{#1}\fi
\ifx \citeauthoryear \undefined \def \citeauthoryear#1{#1}\fi
\ifx \endbibitem  \undefined \def \endbibitem {}\fi
\ifx \bconflocation  \undefined \def \bconflocation#1{#1}\fi
\ifx \arxivurl  \undefined \def \arxivurl#1{\textsf{#1}}\fi
\csname PreBibitemsHook\endcsname

\bibitem[\protect\citeauthoryear{Aitchison and
  Ho}{1989}]{aitchison1989multivariate}
\begin{barticle}
\bauthor{\bsnm{Aitchison}, \binits{J.}},
\bauthor{\bsnm{Ho}, \binits{C.H.}}:
\batitle{The multivariate {Poisson-Log Normal} distribution}.
\bjtitle{Biometrika}
\bvolume{76}(\bissue{4}),
\bfpage{643}--\blpage{653}
(\byear{1989})
\end{barticle}
\endbibitem

\bibitem[\protect\citeauthoryear{Ai et~al.}{2021}]{ai2021bype}
\begin{bchapter}
\bauthor{\bsnm{Ai}, \binits{Q.}},
\bauthor{\bsnm{He}, \binits{L.}},
\bauthor{\bsnm{Liu}, \binits{S.}},
\bauthor{\bsnm{Xu}, \binits{Z.}}:
\bctitle{{ByPE-VAE}: Bayesian pseudocoresets exemplar {VAE}}.
In: \beditor{\bsnm{Ranzato}, \binits{M.}},
\beditor{\bsnm{Beygelzimer}, \binits{A.}},
\beditor{\bsnm{Dauphin}, \binits{Y.N.}},
\beditor{\bsnm{Liang}, \binits{P.}},
\beditor{\bsnm{Vaughan}, \binits{J.W.}} (eds.)
\bbtitle{Advances in Neural Information Processing Systems 34: Annual
  Conference on Neural Information Processing Systems 2021, NeurIPS 2021,
  December 6-14, 2021, Virtual},
pp. \bfpage{5910}--\blpage{5920}
(\byear{2021})
\end{bchapter}
\endbibitem

\bibitem[\protect\citeauthoryear{Aitchison}{1982}]{aitchison1982statistical}
\begin{barticle}
\bauthor{\bsnm{Aitchison}, \binits{J.}}:
\batitle{The statistical analysis of compositional data}.
\bjtitle{Journal of the Royal Statistical Society: Series B (Methodological)}
\bvolume{44}(\bissue{2}),
\bfpage{139}--\blpage{160}
(\byear{1982})
\end{barticle}
\endbibitem

\bibitem[\protect\citeauthoryear{Asnicar et~al.}{2024}]{asnicar2024machine}
\begin{barticle}
\bauthor{\bsnm{Asnicar}, \binits{F.}},
\bauthor{\bsnm{Thomas}, \binits{A.M.}},
\bauthor{\bsnm{Passerini}, \binits{A.}},
\bauthor{\bsnm{Waldron}, \binits{L.}},
\bauthor{\bsnm{Segata}, \binits{N.}}:
\batitle{Machine learning for microbiologists}.
\bjtitle{Nature Reviews Microbiology}
\bvolume{22}(\bissue{4}),
\bfpage{191}--\blpage{205}
(\byear{2024})
\end{barticle}
\endbibitem

\bibitem[\protect\citeauthoryear{Beals}{1984}]{beals1984bray}
\begin{bchapter}
\bauthor{\bsnm{Beals}, \binits{E.W.}}:
\bctitle{{Bray-Curtis} ordination: an effective strategy for analysis of
  multivariate ecological data}.
In: \bbtitle{Advances in Ecological Research},
vol. \bseriesno{14},
pp. \bfpage{1}--\blpage{55}.
\bpublisher{Elsevier},
\blocation{Amsterdam}
(\byear{1984})
\end{bchapter}
\endbibitem

\bibitem[\protect\citeauthoryear{Batardiere
  et~al.}{2024}]{batardiere2024pyplnmodels}
\begin{barticle}
\bauthor{\bsnm{Batardiere}, \binits{B.}},
\bauthor{\bsnm{Kwon}, \binits{J.}},
\bauthor{\bsnm{Chiquet}, \binits{J.}}:
\batitle{{pyPLNmodels}: A python package to analyze multivariate
  high-dimensional count data}.
\bjtitle{Journal of Open Source Software}
\bvolume{9}(\bissue{104}),
\bfpage{6969}
(\byear{2024})
\end{barticle}
\endbibitem

\bibitem[\protect\citeauthoryear{Beghini et~al.}{2021}]{metaphlan}
\begin{barticle}
\bauthor{\bsnm{Beghini}, \binits{F.}},
\bauthor{\bsnm{McIver}, \binits{L.J.}},
\bauthor{\bsnm{Blanco-Míguez}, \binits{A.}},
\bauthor{\bsnm{Dubois}, \binits{L.}},
\bauthor{\bsnm{Asnicar}, \binits{F.}},
\bauthor{\bsnm{Maharjan}, \binits{S.}},
\bauthor{\bsnm{Mailyan}, \binits{A.}},
\bauthor{\bsnm{Manghi}, \binits{P.}},
\bauthor{\bsnm{Scholz}, \binits{M.}},
\bauthor{\bsnm{Thomas}, \binits{A.M.}},
\bauthor{\bsnm{Valles-Colomer}, \binits{M.}},
\bauthor{\bsnm{Weingart}, \binits{G.}},
\bauthor{\bsnm{Zhang}, \binits{Y.}},
\bauthor{\bsnm{Zolfo}, \binits{M.}},
\bauthor{\bsnm{Huttenhower}, \binits{C.}},
\bauthor{\bsnm{Franzosa}, \binits{E.A.}},
\bauthor{\bsnm{Segata}, \binits{N.}}:
\batitle{Integrating taxonomic, functional, and strain-level profiling of
  diverse microbial communities with {bioBakery} 3}.
\bjtitle{eLife}
\bvolume{10},
\bfpage{65088}
(\byear{2021})
\end{barticle}
\endbibitem

\bibitem[\protect\citeauthoryear{Chaussard et~al.}{2025}]{chaussard2025tree}
\begin{barticle}
\bauthor{\bsnm{Chaussard}, \binits{A.}},
\bauthor{\bsnm{Bonnet}, \binits{A.}},
\bauthor{\bsnm{Gassiat}, \binits{E.}},
\bauthor{\bsnm{Le~Corff}, \binits{S.}}:
\batitle{Tree-based variational inference for poisson log-normal models}.
\bjtitle{Statistics and Computing}
\bvolume{35}(\bissue{5}),
\bfpage{1}--\blpage{35}
(\byear{2025})
\end{barticle}
\endbibitem

\bibitem[\protect\citeauthoryear{Chawla et~al.}{2002}]{chawla2002smote}
\begin{barticle}
\bauthor{\bsnm{Chawla}, \binits{N.V.}},
\bauthor{\bsnm{Bowyer}, \binits{K.W.}},
\bauthor{\bsnm{Hall}, \binits{L.O.}},
\bauthor{\bsnm{Kegelmeyer}, \binits{W.P.}}:
\batitle{{SMOTE}: synthetic minority over-sampling technique}.
\bjtitle{Journal of artificial intelligence research}
\bvolume{16},
\bfpage{321}--\blpage{357}
(\byear{2002})
\end{barticle}
\endbibitem

\bibitem[\protect\citeauthoryear{Chiquet et~al.}{2018}]{chiquet_plnpca}
\begin{barticle}
\bauthor{\bsnm{Chiquet}, \binits{J.}},
\bauthor{\bsnm{Mariadassou}, \binits{M.}},
\bauthor{\bsnm{Robin}, \binits{S.}}:
\batitle{{Variational inference for probabilistic {Poisson} PCA}}.
\bjtitle{The Annals of Applied Statistics}
\bvolume{12}(\bissue{4}),
\bfpage{2674}--\blpage{2698}
(\byear{2018})
\end{barticle}
\endbibitem

\bibitem[\protect\citeauthoryear{Chiquet et~al.}{2021}]{chiquet_pln}
\begin{botherref}
\oauthor{\bsnm{Chiquet}, \binits{J.}},
\oauthor{\bsnm{Mariadassou}, \binits{M.}},
\oauthor{\bsnm{Robin}, \binits{S.}}:
The {Poisson-Lognormal} model as a versatile framework for the joint analysis
  of species abundances.
Frontiers in Ecology and Evolution
\textbf{9}
(2021)
\end{botherref}
\endbibitem

\bibitem[\protect\citeauthoryear{Chiquet et~al.}{2019}]{chiquet2019variational}
\begin{bchapter}
\bauthor{\bsnm{Chiquet}, \binits{J.}},
\bauthor{\bsnm{Robin}, \binits{S.}},
\bauthor{\bsnm{Mariadassou}, \binits{M.}}:
\bctitle{Variational inference for sparse network reconstruction from count
  data}.
In: \beditor{\bsnm{Chaudhuri}, \binits{K.}},
\beditor{\bsnm{Salakhutdinov}, \binits{R.}} (eds.)
\bbtitle{Proceedings of the 36th International Conference on Machine Learning,
  {ICML} 2019, 9-15 June 2019}.
\bsertitle{Proceedings of Machine Learning Research},
vol. \bseriesno{97},
pp. \bfpage{1162}--\blpage{1171}.
\bpublisher{{PMLR}},
\blocation{Long Beach, California, {USA}}
(\byear{2019})
\end{bchapter}
\endbibitem

\bibitem[\protect\citeauthoryear{Chadebec et~al.}{2022}]{chadebec2022data}
\begin{barticle}
\bauthor{\bsnm{Chadebec}, \binits{C.}},
\bauthor{\bsnm{Thibeau-Sutre}, \binits{E.}},
\bauthor{\bsnm{Burgos}, \binits{N.}},
\bauthor{\bsnm{Allassonni{\`e}re}, \binits{S.}}:
\batitle{Data augmentation in high dimensional low sample size setting using a
  geometry-based variational autoencoder}.
\bjtitle{IEEE Transactions on Pattern Analysis and Machine Intelligence}
\bvolume{45}(\bissue{3}),
\bfpage{2879}--\blpage{2896}
(\byear{2022})
\end{barticle}
\endbibitem

\bibitem[\protect\citeauthoryear{Davis and
  Goadrich}{2006}]{davis2006relationship}
\begin{bchapter}
\bauthor{\bsnm{Davis}, \binits{J.}},
\bauthor{\bsnm{Goadrich}, \binits{M.H.}}:
\bctitle{The relationship between {Precision-Recall} and {ROC} curves}.
In: \beditor{\bsnm{Cohen}, \binits{W.W.}},
\beditor{\bsnm{Moore}, \binits{A.W.}} (eds.)
\bbtitle{Machine Learning, Proceedings of the Twenty-Third International
  Conference {(ICML} 2006), June 25-29, 2006}.
\bsertitle{{ACM} International Conference Proceeding Series},
vol. \bseriesno{148},
pp. \bfpage{233}--\blpage{240}.
\bpublisher{{ACM}},
\blocation{, Pittsburgh, Pennsylvania, USA}
(\byear{2006})
\end{bchapter}
\endbibitem

\bibitem[\protect\citeauthoryear{D'Amour
  et~al.}{2022}]{d2022underspecification}
\begin{barticle}
\bauthor{\bsnm{D'Amour}, \binits{A.}},
\bauthor{\bsnm{Heller}, \binits{K.A.}},
\bauthor{\bsnm{Moldovan}, \binits{D.}},
\bauthor{\bsnm{Adlam}, \binits{B.}},
\bauthor{\bsnm{Alipanahi}, \binits{B.}},
\bauthor{\bsnm{Beutel}, \binits{A.}},
\bauthor{\bsnm{Chen}, \binits{C.}},
\bauthor{\bsnm{Deaton}, \binits{J.}},
\bauthor{\bsnm{Eisenstein}, \binits{J.}},
\bauthor{\bsnm{Hoffman}, \binits{M.D.}},
\bauthor{\bsnm{Hormozdiari}, \binits{F.}},
\bauthor{\bsnm{Houlsby}, \binits{N.}},
\bauthor{\bsnm{Hou}, \binits{S.}},
\bauthor{\bsnm{Jerfel}, \binits{G.}},
\bauthor{\bsnm{Karthikesalingam}, \binits{A.}},
\bauthor{\bsnm{Lucic}, \binits{M.}},
\bauthor{\bsnm{Ma}, \binits{Y.}},
\bauthor{\bsnm{McLean}, \binits{C.Y.}},
\bauthor{\bsnm{Mincu}, \binits{D.}},
\bauthor{\bsnm{Mitani}, \binits{A.}},
\bauthor{\bsnm{Montanari}, \binits{A.}},
\bauthor{\bsnm{Nado}, \binits{Z.}},
\bauthor{\bsnm{Natarajan}, \binits{V.}},
\bauthor{\bsnm{Nielson}, \binits{C.}},
\bauthor{\bsnm{Osborne}, \binits{T.F.}},
\bauthor{\bsnm{Raman}, \binits{R.}},
\bauthor{\bsnm{Ramasamy}, \binits{K.}},
\bauthor{\bsnm{Sayres}, \binits{R.}},
\bauthor{\bsnm{Schrouff}, \binits{J.}},
\bauthor{\bsnm{Seneviratne}, \binits{M.}},
\bauthor{\bsnm{Sequeira}, \binits{S.}},
\bauthor{\bsnm{Suresh}, \binits{H.}},
\bauthor{\bsnm{Veitch}, \binits{V.}},
\bauthor{\bsnm{Vladymyrov}, \binits{M.}},
\bauthor{\bsnm{Wang}, \binits{X.}},
\bauthor{\bsnm{Webster}, \binits{K.}},
\bauthor{\bsnm{Yadlowsky}, \binits{S.}},
\bauthor{\bsnm{Yun}, \binits{T.}},
\bauthor{\bsnm{Zhai}, \binits{X.}},
\bauthor{\bsnm{Sculley}, \binits{D.}}:
\batitle{Underspecification presents challenges for credibility in modern
  machine learning}.
\bjtitle{Journal of Machine Learning Research}
\bvolume{23},
\bfpage{226}--\blpage{122661}
(\byear{2022})
\end{barticle}
\endbibitem

\bibitem[\protect\citeauthoryear{Dumoulin et~al.}{2018}]{dumoulin2018feature}
\begin{barticle}
\bauthor{\bsnm{Dumoulin}, \binits{V.}},
\bauthor{\bsnm{Perez}, \binits{E.}},
\bauthor{\bsnm{Schucher}, \binits{N.}},
\bauthor{\bsnm{Strub}, \binits{F.}},
\bauthor{\bsnm{Vries}, \binits{H.d.}},
\bauthor{\bsnm{Courville}, \binits{A.}},
\bauthor{\bsnm{Bengio}, \binits{Y.}}:
\batitle{Feature-wise transformations}.
\bjtitle{Distill}
\bvolume{3}(\bissue{7}),
\bfpage{11}
(\byear{2018})
\end{barticle}
\endbibitem

\bibitem[\protect\citeauthoryear{Gilbert et~al.}{2018}]{gilbert2018current}
\begin{barticle}
\bauthor{\bsnm{Gilbert}, \binits{J.A.}},
\bauthor{\bsnm{Blaser}, \binits{M.J.}},
\bauthor{\bsnm{Caporaso}, \binits{J.G.}},
\bauthor{\bsnm{Jansson}, \binits{J.K.}},
\bauthor{\bsnm{Lynch}, \binits{S.V.}},
\bauthor{\bsnm{Knight}, \binits{R.}}:
\batitle{Current understanding of the human microbiome}.
\bjtitle{Nature medicine}
\bvolume{24}(\bissue{4}),
\bfpage{392}--\blpage{400}
(\byear{2018})
\end{barticle}
\endbibitem

\bibitem[\protect\citeauthoryear{Gotelli and
  Colwell}{2001}]{gotelli2001quantifying}
\begin{barticle}
\bauthor{\bsnm{Gotelli}, \binits{N.J.}},
\bauthor{\bsnm{Colwell}, \binits{R.K.}}:
\batitle{Quantifying biodiversity: procedures and pitfalls in the measurement
  and comparison of species richness}.
\bjtitle{Ecology Letters}
\bvolume{4}(\bissue{4}),
\bfpage{379}--\blpage{391}
(\byear{2001})
\end{barticle}
\endbibitem

\bibitem[\protect\citeauthoryear{Gassiat
  et~al.}{2020}]{gassiat2020identifiability}
\begin{barticle}
\bauthor{\bsnm{Gassiat}, \binits{E.}},
\bauthor{\bsnm{Le\;Corff}, \binits{S.}},
\bauthor{\bsnm{Leh{\'{e}}ricy}, \binits{L.}}:
\batitle{Identifiability and consistent estimation of nonparametric translation
  hidden {Markov} models with general state space}.
\bjtitle{Journal of Machine Learning Research}
\bvolume{21},
\bfpage{115}--\blpage{111540}
(\byear{2020})
\end{barticle}
\endbibitem

\bibitem[\protect\citeauthoryear{Gloor et~al.}{2017}]{gloor2017microbiome}
\begin{barticle}
\bauthor{\bsnm{Gloor}, \binits{G.B.}},
\bauthor{\bsnm{Macklaim}, \binits{J.M.}},
\bauthor{\bsnm{Pawlowsky-Glahn}, \binits{V.}},
\bauthor{\bsnm{Egozcue}, \binits{J.J.}}:
\batitle{Microbiome datasets are compositional: and this is not optional}.
\bjtitle{Frontiers in microbiology}
\bvolume{8},
\bfpage{2224}
(\byear{2017})
\end{barticle}
\endbibitem

\bibitem[\protect\citeauthoryear{Goodfellow
  et~al.}{2014}]{goodfellow2014generative}
\begin{bchapter}
\bauthor{\bsnm{Goodfellow}, \binits{I.J.}},
\bauthor{\bsnm{Pouget{-}Abadie}, \binits{J.}},
\bauthor{\bsnm{Mirza}, \binits{M.}},
\bauthor{\bsnm{Xu}, \binits{B.}},
\bauthor{\bsnm{Warde{-}Farley}, \binits{D.}},
\bauthor{\bsnm{Ozair}, \binits{S.}},
\bauthor{\bsnm{Courville}, \binits{A.C.}},
\bauthor{\bsnm{Bengio}, \binits{Y.}}:
\bctitle{Generative adversarial nets}.
In: \beditor{\bsnm{Ghahramani}, \binits{Z.}},
\beditor{\bsnm{Welling}, \binits{M.}},
\beditor{\bsnm{Cortes}, \binits{C.}},
\beditor{\bsnm{Lawrence}, \binits{N.D.}},
\beditor{\bsnm{Weinberger}, \binits{K.Q.}} (eds.)
\bbtitle{Advances in Neural Information Processing Systems 27: Annual
  Conference on Neural Information Processing Systems 2014, December 8-13 2014,
  Montreal},
\bconflocation{Quebec, Canada},
pp. \bfpage{2672}--\blpage{2680}
(\byear{2014})
\end{bchapter}
\endbibitem

\bibitem[\protect\citeauthoryear{Gordon{-}Rodr{\'{\i}}guez
  et~al.}{2022}]{gordon2022data}
\begin{bchapter}
\bauthor{\bsnm{Gordon{-}Rodr{\'{\i}}guez}, \binits{E.}},
\bauthor{\bsnm{Quinn}, \binits{T.P.}},
\bauthor{\bsnm{Cunningham}, \binits{J.P.}}:
\bctitle{Data augmentation for compositional data: Advancing predictive models
  of the microbiome}.
In: \beditor{\bsnm{Koyejo}, \binits{S.}},
\beditor{\bsnm{Mohamed}, \binits{S.}},
\beditor{\bsnm{Agarwal}, \binits{A.}},
\beditor{\bsnm{Belgrave}, \binits{D.}},
\beditor{\bsnm{Cho}, \binits{K.}},
\beditor{\bsnm{Oh}, \binits{A.}} (eds.)
\bbtitle{Advances in Neural Information Processing Systems 35: Annual
  Conference on Neural Information Processing Systems 2022, NeurIPS 2022,
  November 28 - December 9, 2022},
\bconflocation{New Orleans, LA, USA}
(\byear{2022})
\end{bchapter}
\endbibitem

\bibitem[\protect\citeauthoryear{Holmes et~al.}{2012}]{holmes2012dirichlet}
\begin{barticle}
\bauthor{\bsnm{Holmes}, \binits{I.}},
\bauthor{\bsnm{Harris}, \binits{K.}},
\bauthor{\bsnm{Quince}, \binits{C.}}:
\batitle{{Dirichlet} multinomial mixtures: generative models for microbial
  metagenomics}.
\bjtitle{PloS ONE}
\bvolume{7}(\bissue{2}),
\bfpage{30126}
(\byear{2012})
\end{barticle}
\endbibitem

\bibitem[\protect\citeauthoryear{H{\"{a}}lv{\"{a}}
  et~al.}{2021}]{halva2021disentangling}
\begin{bchapter}
\bauthor{\bsnm{H{\"{a}}lv{\"{a}}}, \binits{H.}},
\bauthor{\bsnm{Le\;Corff}, \binits{S.}},
\bauthor{\bsnm{Leh{\'{e}}ricy}, \binits{L.}},
\bauthor{\bsnm{So}, \binits{J.}},
\bauthor{\bsnm{Zhu}, \binits{Y.}},
\bauthor{\bsnm{Gassiat}, \binits{E.}},
\bauthor{\bsnm{Hyv{\"{a}}rinen}, \binits{A.}}:
\bctitle{Disentangling identifiable features from noisy data with structured
  nonlinear {ICA}}.
In: \beditor{\bsnm{Ranzato}, \binits{M.}},
\beditor{\bsnm{Beygelzimer}, \binits{A.}},
\beditor{\bsnm{Dauphin}, \binits{Y.N.}},
\beditor{\bsnm{Liang}, \binits{P.}},
\beditor{\bsnm{Vaughan}, \binits{J.W.}} (eds.)
\bbtitle{Advances in Neural Information Processing Systems 34: Annual
  Conference on Neural Information Processing Systems 2021, NeurIPS 2021,
  December 6-14, 2021, Virtual},
pp. \bfpage{1624}--\blpage{1633}
(\byear{2021})
\end{bchapter}
\endbibitem

\bibitem[\protect\citeauthoryear{Julien et~al.}{2022}]{julien2022impact}
\begin{barticle}
\bauthor{\bsnm{Julien}, \binits{C.}},
\bauthor{\bsnm{Anakok}, \binits{E.}},
\bauthor{\bsnm{Treton}, \binits{X.}},
\bauthor{\bsnm{Nachury}, \binits{M.}},
\bauthor{\bsnm{Nancey}, \binits{S.}},
\bauthor{\bsnm{Buisson}, \binits{A.}},
\bauthor{\bsnm{Fumery}, \binits{M.}},
\bauthor{\bsnm{Filippi}, \binits{J.}},
\bauthor{\bsnm{Maggiori}, \binits{L.}},
\bauthor{\bsnm{Panis}, \binits{Y.}}, \betal:
\batitle{Impact of the ileal microbiota on surgical site infections in
  {Crohn}’s disease: a nationwide prospective cohort}.
\bjtitle{Journal of Crohn's and Colitis}
\bvolume{16}(\bissue{8}),
\bfpage{1211}--\blpage{1221}
(\byear{2022})
\end{barticle}
\endbibitem

\bibitem[\protect\citeauthoryear{Jiang et~al.}{2025}]{jiang2025modeling}
\begin{barticle}
\bauthor{\bsnm{Jiang}, \binits{Y.}},
\bauthor{\bsnm{Aton}, \binits{M.}},
\bauthor{\bsnm{Zhu}, \binits{Q.}},
\bauthor{\bsnm{Lu}, \binits{Y.Y.}}:
\batitle{Modeling microbiome-trait associations with taxonomy-adaptive neural
  networks}.
\bjtitle{Microbiome}
\bvolume{13}(\bissue{1}),
\bfpage{87}
(\byear{2025})
\end{barticle}
\endbibitem

\bibitem[\protect\citeauthoryear{Jiang et~al.}{2025}]{jiang2025phylomix}
\begin{botherref}
\oauthor{\bsnm{Jiang}, \binits{Y.}},
\oauthor{\bsnm{Liao}, \binits{D.}},
\oauthor{\bsnm{Zhu}, \binits{Q.}},
\oauthor{\bsnm{Lu}, \binits{Y.Y.}}:
{PhyloMix}: Enhancing microbiome-trait association prediction through
  phylogeny-mixing augmentation.
Bioinformatics,
014
(2025)
\end{botherref}
\endbibitem

\bibitem[\protect\citeauthoryear{Karwowska et~al.}{2025}]{karwowska2025effects}
\begin{barticle}
\bauthor{\bsnm{Karwowska}, \binits{Z.}},
\bauthor{\bsnm{Aasmets}, \binits{O.}},
\bauthor{\bsnm{Metspalu Mait Metspalu Andres Milani Lili Esko~T{\~o}nu},
  \binits{E.B.}},
\bauthor{\bsnm{Kosciolek}, \binits{T.}},
\bauthor{\bsnm{Org}, \binits{E.}}:
\batitle{Effects of data transformation and model selection on feature
  importance in microbiome classification data}.
\bjtitle{Microbiome}
\bvolume{13}(\bissue{1}),
\bfpage{2}
(\byear{2025})
\end{barticle}
\endbibitem

\bibitem[\protect\citeauthoryear{Kingma and Ba}{2015}]{kingma2015adam}
\begin{bchapter}
\bauthor{\bsnm{Kingma}, \binits{D.P.}},
\bauthor{\bsnm{Ba}, \binits{J.}}:
\bctitle{Adam: {A} method for stochastic optimization}.
In: \beditor{\bsnm{Bengio}, \binits{Y.}},
\beditor{\bsnm{LeCun}, \binits{Y.}} (eds.)
\bbtitle{3rd International Conference on Learning Representations, {ICLR} 2015,
  May 7-9, 2015, Conference Track Proceedings},
\bconflocation{San Diego, CA, USA}
(\byear{2015})
\end{bchapter}
\endbibitem

\bibitem[\protect\citeauthoryear{Kostic et~al.}{2015}]{kostic2015dynamics}
\begin{barticle}
\bauthor{\bsnm{Kostic}, \binits{A.D.}},
\bauthor{\bsnm{Gevers}, \binits{D.}},
\bauthor{\bsnm{Siljander}, \binits{H.}},
\bauthor{\bsnm{Vatanen}, \binits{T.}},
\bauthor{\bsnm{Hy{\"o}tyl{\"a}inen}, \binits{T.}},
\bauthor{\bsnm{H{\"a}m{\"a}l{\"a}inen}, \binits{A.-M.}},
\bauthor{\bsnm{Peet}, \binits{A.}},
\bauthor{\bsnm{Tillmann}, \binits{V.}},
\bauthor{\bsnm{P{\"o}h{\"o}}, \binits{P.}},
\bauthor{\bsnm{Mattila}, \binits{I.}}, \betal:
\batitle{The dynamics of the human infant gut microbiome in development and in
  progression toward type 1 diabetes}.
\bjtitle{Cell host \& microbe}
\bvolume{17}(\bissue{2}),
\bfpage{260}--\blpage{273}
(\byear{2015})
\end{barticle}
\endbibitem

\bibitem[\protect\citeauthoryear{Kumar and Poole}{2020}]{kumar2020implicit}
\begin{bchapter}
\bauthor{\bsnm{Kumar}, \binits{A.}},
\bauthor{\bsnm{Poole}, \binits{B.}}:
\bctitle{On implicit regularization in {\(\beta\)}-{VAE}s}.
In: \bbtitle{Proceedings of the 37th International Conference on Machine
  Learning, {ICML} 2020, 13-18 July 2020}.
\bsertitle{Proceedings of Machine Learning Research},
vol. \bseriesno{119},
pp. \bfpage{5480}--\blpage{5490}.
\bpublisher{{PMLR}},
\blocation{Virtual Event}
(\byear{2020})
\end{bchapter}
\endbibitem

\bibitem[\protect\citeauthoryear{Knight et~al.}{2018}]{knight2018best}
\begin{barticle}
\bauthor{\bsnm{Knight}, \binits{R.}},
\bauthor{\bsnm{Vrbanac}, \binits{A.}},
\bauthor{\bsnm{Taylor}, \binits{B.C.}},
\bauthor{\bsnm{Aksenov}, \binits{A.}},
\bauthor{\bsnm{Callewaert}, \binits{C.}},
\bauthor{\bsnm{Debelius}, \binits{J.}},
\bauthor{\bsnm{Gonzalez}, \binits{A.}},
\bauthor{\bsnm{Kosciolek}, \binits{T.}},
\bauthor{\bsnm{McCall}, \binits{L.-I.}},
\bauthor{\bsnm{McDonald}, \binits{D.}}, \betal:
\batitle{Best practices for analysing microbiomes}.
\bjtitle{Nature Reviews Microbiology}
\bvolume{16}(\bissue{7}),
\bfpage{410}--\blpage{422}
(\byear{2018})
\end{barticle}
\endbibitem

\bibitem[\protect\citeauthoryear{Lloyd-Price et~al.}{2019}]{lloyd2019multi}
\begin{barticle}
\bauthor{\bsnm{Lloyd-Price}, \binits{J.}},
\bauthor{\bsnm{Arze}, \binits{C.}},
\bauthor{\bsnm{Ananthakrishnan}, \binits{A.N.}},
\bauthor{\bsnm{Schirmer}, \binits{M.}},
\bauthor{\bsnm{Avila-Pacheco}, \binits{J.}},
\bauthor{\bsnm{Poon}, \binits{T.W.}},
\bauthor{\bsnm{Andrews}, \binits{E.}},
\bauthor{\bsnm{Ajami}, \binits{N.J.}},
\bauthor{\bsnm{Bonham}, \binits{K.S.}},
\bauthor{\bsnm{Brislawn}, \binits{C.J.}}, \betal:
\batitle{Multi-omics of the gut microbial ecosystem in inflammatory bowel
  diseases}.
\bjtitle{Nature}
\bvolume{569}(\bissue{7758}),
\bfpage{655}--\blpage{662}
(\byear{2019})
\end{barticle}
\endbibitem

\bibitem[\protect\citeauthoryear{Liu et~al.}{2017}]{liu2017deep}
\begin{bchapter}
\bauthor{\bsnm{Liu}, \binits{B.}},
\bauthor{\bsnm{Wei}, \binits{Y.}},
\bauthor{\bsnm{Zhang}, \binits{Y.}},
\bauthor{\bsnm{Yang}, \binits{Q.}}:
\bctitle{Deep neural networks for high dimension, low sample size data}.
In: \beditor{\bsnm{Sierra}, \binits{C.}} (ed.)
\bbtitle{Proceedings of the Twenty-Sixth International Joint Conference on
  Artificial Intelligence, {IJCAI} 2017, August 19-25, 2017},
pp. \bfpage{2287}--\blpage{2293}.
\bpublisher{ijcai.org},
\blocation{Melbourne, Australia}
(\byear{2017})
\end{bchapter}
\endbibitem

\bibitem[\protect\citeauthoryear{Mumuni and Mumuni}{2022}]{mumuni2022data}
\begin{barticle}
\bauthor{\bsnm{Mumuni}, \binits{A.}},
\bauthor{\bsnm{Mumuni}, \binits{F.}}:
\batitle{Data augmentation: a comprehensive survey of modern approaches}.
\bjtitle{Array}
\bvolume{16},
\bfpage{100258}
(\byear{2022})
\end{barticle}
\endbibitem

\bibitem[\protect\citeauthoryear{Mangla et~al.}{2021}]{mangla2021benefits}
\begin{barticle}
\bauthor{\bsnm{Mangla}, \binits{P.}},
\bauthor{\bsnm{Singh}, \binits{V.}},
\bauthor{\bsnm{Havaldar}, \binits{S.}},
\bauthor{\bsnm{Balasubramanian}, \binits{V.}}:
\batitle{On the benefits of defining vicinal distributions in latent space}.
\bjtitle{Pattern Recognition Letters}
\bvolume{152},
\bfpage{382}--\blpage{390}
(\byear{2021})
\end{barticle}
\endbibitem

\bibitem[\protect\citeauthoryear{Marcos-Zambrano
  et~al.}{2021}]{marcos2021applications}
\begin{barticle}
\bauthor{\bsnm{Marcos-Zambrano}, \binits{L.J.}},
\bauthor{\bsnm{Karaduzovic-Hadziabdic}, \binits{K.}},
\bauthor{\bsnm{Loncar~Turukalo}, \binits{T.}},
\bauthor{\bsnm{Przymus}, \binits{P.}},
\bauthor{\bsnm{Trajkovik}, \binits{V.}},
\bauthor{\bsnm{Aasmets}, \binits{O.}},
\bauthor{\bsnm{Berland}, \binits{M.}},
\bauthor{\bsnm{Gruca}, \binits{A.}},
\bauthor{\bsnm{Hasic}, \binits{J.}},
\bauthor{\bsnm{Hron}, \binits{K.}}, \betal:
\batitle{Applications of machine learning in human microbiome studies: a review
  on feature selection, biomarker identification, disease prediction and
  treatment}.
\bjtitle{Frontiers in microbiology}
\bvolume{12},
\bfpage{634511}
(\byear{2021})
\end{barticle}
\endbibitem

\bibitem[\protect\citeauthoryear{Nagendra}{2002}]{nagendra2002opposite}
\begin{barticle}
\bauthor{\bsnm{Nagendra}, \binits{H.}}:
\batitle{Opposite trends in response for the {Shannon and Simpson} indices of
  landscape diversity}.
\bjtitle{Applied Geography}
\bvolume{22}(\bissue{2}),
\bfpage{175}--\blpage{186}
(\byear{2002})
\end{barticle}
\endbibitem

\bibitem[\protect\citeauthoryear{Nielsen
  et~al.}{2014}]{nielsen2014identification}
\begin{barticle}
\bauthor{\bsnm{Nielsen}, \binits{H.B.}},
\bauthor{\bsnm{Almeida}, \binits{M.}},
\bauthor{\bsnm{Juncker}, \binits{A.S.}},
\bauthor{\bsnm{Rasmussen}, \binits{S.}},
\bauthor{\bsnm{Li}, \binits{J.}},
\bauthor{\bsnm{Sunagawa}, \binits{S.}},
\bauthor{\bsnm{Plichta}, \binits{D.R.}},
\bauthor{\bsnm{Gautier}, \binits{L.}},
\bauthor{\bsnm{Pedersen}, \binits{A.G.}},
\bauthor{\bsnm{Le~Chatelier}, \binits{E.}}, \betal:
\batitle{Identification and assembly of genomes and genetic elements in complex
  metagenomic samples without using reference genomes}.
\bjtitle{Nature biotechnology}
\bvolume{32}(\bissue{8}),
\bfpage{822}--\blpage{828}
(\byear{2014})
\end{barticle}
\endbibitem

\bibitem[\protect\citeauthoryear{Norouzi et~al.}{2020}]{norouzi2020examplar}
\begin{bchapter}
\bauthor{\bsnm{Norouzi}, \binits{S.}},
\bauthor{\bsnm{Fleet}, \binits{D.J.}},
\bauthor{\bsnm{Norouzi}, \binits{M.}}:
\bctitle{Exemplar {VAE:} linking generative models, nearest neighbor retrieval,
  and data augmentation}.
In: \beditor{\bsnm{Larochelle}, \binits{H.}},
\beditor{\bsnm{Ranzato}, \binits{M.}},
\beditor{\bsnm{Hadsell}, \binits{R.}},
\beditor{\bsnm{Balcan}, \binits{M.}},
\beditor{\bsnm{Lin}, \binits{H.}} (eds.)
\bbtitle{Advances in Neural Information Processing Systems 33: Annual
  Conference on Neural Information Processing Systems 2020, NeurIPS 2020,
  December 6-12, 2020, Virtual}
(\byear{2020})
\end{bchapter}
\endbibitem

\bibitem[\protect\citeauthoryear{Paszke et~al.}{2019}]{paszke2019pytorch}
\begin{bchapter}
\bauthor{\bsnm{Paszke}, \binits{A.}},
\bauthor{\bsnm{Gross}, \binits{S.}},
\bauthor{\bsnm{Massa}, \binits{F.}},
\bauthor{\bsnm{Lerer}, \binits{A.}},
\bauthor{\bsnm{Bradbury}, \binits{J.}},
\bauthor{\bsnm{Chanan}, \binits{G.}},
\bauthor{\bsnm{Killeen}, \binits{T.}},
\bauthor{\bsnm{Lin}, \binits{Z.}},
\bauthor{\bsnm{Gimelshein}, \binits{N.}},
\bauthor{\bsnm{Antiga}, \binits{L.}},
\bauthor{\bsnm{Desmaison}, \binits{A.}},
\bauthor{\bsnm{K{\"{o}}pf}, \binits{A.}},
\bauthor{\bsnm{Yang}, \binits{E.Z.}},
\bauthor{\bsnm{DeVito}, \binits{Z.}},
\bauthor{\bsnm{Raison}, \binits{M.}},
\bauthor{\bsnm{Tejani}, \binits{A.}},
\bauthor{\bsnm{Chilamkurthy}, \binits{S.}},
\bauthor{\bsnm{Steiner}, \binits{B.}},
\bauthor{\bsnm{Fang}, \binits{L.}},
\bauthor{\bsnm{Bai}, \binits{J.}},
\bauthor{\bsnm{Chintala}, \binits{S.}}:
\bctitle{{PyTorch}: An imperative style, high-performance deep learning
  library}.
In: \beditor{\bsnm{Wallach}, \binits{H.M.}},
\beditor{\bsnm{Larochelle}, \binits{H.}},
\beditor{\bsnm{Beygelzimer}, \binits{A.}},
\beditor{\bsnm{d'Alch{\'{e}}{-}Buc}, \binits{F.}},
\beditor{\bsnm{Fox}, \binits{E.B.}},
\beditor{\bsnm{Garnett}, \binits{R.}} (eds.)
\bbtitle{Advances in Neural Information Processing Systems 32: Annual
  Conference on Neural Information Processing Systems 2019, NeurIPS 2019,
  December 8-14, 2019},
\bconflocation{Vancouver, BC, Canada},
pp. \bfpage{8024}--\blpage{8035}
(\byear{2019})
\end{bchapter}
\endbibitem

\bibitem[\protect\citeauthoryear{Perez et~al.}{2018}]{perez2018film}
\begin{bchapter}
\bauthor{\bsnm{Perez}, \binits{E.}},
\bauthor{\bsnm{Strub}, \binits{F.}},
\bauthor{\bsnm{Vries}, \binits{H.}},
\bauthor{\bsnm{Dumoulin}, \binits{V.}},
\bauthor{\bsnm{Courville}, \binits{A.C.}}:
\bctitle{{FiLM}: Visual reasoning with a general conditioning layer}.
In: \beditor{\bsnm{McIlraith}, \binits{S.A.}},
\beditor{\bsnm{Weinberger}, \binits{K.Q.}} (eds.)
\bbtitle{Proceedings of the Thirty-Second {AAAI} Conference on Artificial
  Intelligence, (AAAI-18), the 30th Innovative Applications of Artificial
  Intelligence (IAAI-18), and the 8th {AAAI} Symposium on Educational Advances
  in Artificial Intelligence (EAAI-18), February 2-7, 2018},
pp. \bfpage{3942}--\blpage{3951}.
\bpublisher{{AAAI} Press},
\blocation{New Orleans, Louisiana, USA}
(\byear{2018})
\end{bchapter}
\endbibitem

\bibitem[\protect\citeauthoryear{Pasolli et~al.}{2017}]{curated}
\begin{barticle}
\bauthor{\bsnm{Pasolli}, \binits{E.}},
\bauthor{\bsnm{Schiffer}, \binits{L.}},
\bauthor{\bsnm{Manghi}, \binits{P.}},
\bauthor{\bsnm{Renson}, \binits{A.}},
\bauthor{\bsnm{Obenchain}, \binits{V.}},
\bauthor{\bsnm{Truong}, \binits{D.T.}},
\bauthor{\bsnm{Beghini}, \binits{F.}},
\bauthor{\bsnm{Malik}, \binits{F.}},
\bauthor{\bsnm{Ramos}, \binits{M.}},
\bauthor{\bsnm{Dowd}, \binits{J.B.}},
\bauthor{\bsnm{Huttenhower}, \binits{M.}},
\bauthor{\bsnm{Segata}, \binits{W.}}:
\batitle{Accessible, curated metagenomic data through experimenthub}.
\bjtitle{Nature methods}
\bvolume{14}(\bissue{11}),
\bfpage{1023}--\blpage{1024}
(\byear{2017})
\end{barticle}
\endbibitem

\bibitem[\protect\citeauthoryear{Pedregosa et~al.}{2011}]{pedregosa2011scikit}
\begin{barticle}
\bauthor{\bsnm{Pedregosa}, \binits{F.}},
\bauthor{\bsnm{Varoquaux}, \binits{G.}},
\bauthor{\bsnm{Gramfort}, \binits{A.}},
\bauthor{\bsnm{Michel}, \binits{V.}},
\bauthor{\bsnm{Thirion}, \binits{B.}},
\bauthor{\bsnm{Grisel}, \binits{O.}},
\bauthor{\bsnm{Blondel}, \binits{M.}},
\bauthor{\bsnm{Prettenhofer}, \binits{P.}},
\bauthor{\bsnm{Weiss}, \binits{R.}},
\bauthor{\bsnm{Dubourg}, \binits{V.}},
\bauthor{\bsnm{VanderPlas}, \binits{J.}},
\bauthor{\bsnm{Passos}, \binits{A.}},
\bauthor{\bsnm{Cournapeau}, \binits{D.}},
\bauthor{\bsnm{Brucher}, \binits{M.}},
\bauthor{\bsnm{Perrot}, \binits{M.}},
\bauthor{\bsnm{Duchesnay}, \binits{E.}}:
\batitle{Scikit-learn: Machine learning in python}.
\bjtitle{Journal of Machine Learning Research}
\bvolume{12},
\bfpage{2825}--\blpage{2830}
(\byear{2011})
\end{barticle}
\endbibitem

\bibitem[\protect\citeauthoryear{Quinn et~al.}{2018}]{quinn2018understanding}
\begin{barticle}
\bauthor{\bsnm{Quinn}, \binits{T.P.}},
\bauthor{\bsnm{Erb}, \binits{I.}},
\bauthor{\bsnm{Richardson}, \binits{M.F.}},
\bauthor{\bsnm{Crowley}, \binits{T.M.}}:
\batitle{Understanding sequencing data as compositions: an outlook and review}.
\bjtitle{Bioinformatics}
\bvolume{34}(\bissue{16}),
\bfpage{2870}--\blpage{2878}
(\byear{2018})
\end{barticle}
\endbibitem

\bibitem[\protect\citeauthoryear{Qian et~al.}{2024}]{qian2024splang}
\begin{barticle}
\bauthor{\bsnm{Qian}, \binits{W.}},
\bauthor{\bsnm{Stanley}, \binits{K.G.}},
\bauthor{\bsnm{Aziz}, \binits{Z.}},
\bauthor{\bsnm{Aziz}, \binits{U.}},
\bauthor{\bsnm{Siciliano}, \binits{S.D.}}:
\batitle{{SPLANG}—a synthetic {Poisson-Lognormal}-based abundance and network
  generative model for microbial interaction inference algorithms}.
\bjtitle{Scientific Reports}
\bvolume{14}(\bissue{1}),
\bfpage{25099}
(\byear{2024})
\end{barticle}
\endbibitem

\bibitem[\protect\citeauthoryear{Rubel et~al.}{2020}]{rubel2020lifestyle}
\begin{barticle}
\bauthor{\bsnm{Rubel}, \binits{M.A.}},
\bauthor{\bsnm{Abbas}, \binits{A.}},
\bauthor{\bsnm{Taylor}, \binits{L.J.}},
\bauthor{\bsnm{Connell}, \binits{A.}},
\bauthor{\bsnm{Tanes}, \binits{C.}},
\bauthor{\bsnm{Bittinger}, \binits{K.}},
\bauthor{\bsnm{Ndze}, \binits{V.N.}},
\bauthor{\bsnm{Fonsah}, \binits{J.Y.}},
\bauthor{\bsnm{Ngwang}, \binits{E.}},
\bauthor{\bsnm{Essiane}, \binits{A.}}, \betal:
\batitle{Lifestyle and the presence of helminths is associated with gut
  microbiome composition in {Cameroonians}}.
\bjtitle{Genome biology}
\bvolume{21},
\bfpage{1}--\blpage{32}
(\byear{2020})
\end{barticle}
\endbibitem

\bibitem[\protect\citeauthoryear{Rong et~al.}{2021}]{rong2021mb}
\begin{barticle}
\bauthor{\bsnm{Rong}, \binits{R.}},
\bauthor{\bsnm{Jiang}, \binits{S.}},
\bauthor{\bsnm{Xu}, \binits{L.}},
\bauthor{\bsnm{Xiao}, \binits{G.}},
\bauthor{\bsnm{Xie}, \binits{Y.}},
\bauthor{\bsnm{Liu}, \binits{D.J.}},
\bauthor{\bsnm{Li}, \binits{Q.}},
\bauthor{\bsnm{Zhan}, \binits{X.}}:
\batitle{{MB-GAN}: microbiome simulation via generative adversarial network}.
\bjtitle{GigaScience}
\bvolume{10}(\bissue{2}),
\bfpage{005}
(\byear{2021})
\end{barticle}
\endbibitem

\bibitem[\protect\citeauthoryear{Rosenberg}{2024}]{rosenberg2024diversity}
\begin{barticle}
\bauthor{\bsnm{Rosenberg}, \binits{E.}}:
\batitle{Diversity of bacteria within the human gut and its contribution to the
  functional unity of holobionts}.
\bjtitle{npj Biofilms and Microbiomes}
\bvolume{10}(\bissue{1}),
\bfpage{134}
(\byear{2024})
\end{barticle}
\endbibitem

\bibitem[\protect\citeauthoryear{Sayyari et~al.}{2019}]{sayyari2019tada}
\begin{barticle}
\bauthor{\bsnm{Sayyari}, \binits{E.}},
\bauthor{\bsnm{Kawas}, \binits{B.}},
\bauthor{\bsnm{Mirarab}, \binits{S.}}:
\batitle{{TADA}: phylogenetic augmentation of microbiome samples enhances
  phenotype classification}.
\bjtitle{Bioinformatics}
\bvolume{35}(\bissue{14}),
\bfpage{31}--\blpage{40}
(\byear{2019})
\end{barticle}
\endbibitem

\bibitem[\protect\citeauthoryear{Sharma et~al.}{2024}]{sharma2024phylagan}
\begin{barticle}
\bauthor{\bsnm{Sharma}, \binits{D.}},
\bauthor{\bsnm{Lou}, \binits{W.}},
\bauthor{\bsnm{Xu}, \binits{W.}}:
\batitle{{phylaGAN}: data augmentation through conditional {GAN}s and
  autoencoders for improving disease prediction accuracy using microbiome
  data}.
\bjtitle{Bioinformatics}
\bvolume{40}(\bissue{4}),
\bfpage{161}
(\byear{2024})
\end{barticle}
\endbibitem

\bibitem[\protect\citeauthoryear{Shi et~al.}{2016}]{shi2016regression}
\begin{barticle}
\bauthor{\bsnm{Shi}, \binits{P.}},
\bauthor{\bsnm{Zhang}, \binits{A.}},
\bauthor{\bsnm{Li}, \binits{H.}}:
\batitle{{Regression analysis for microbiome compositional data}}.
\bjtitle{The Annals of Applied Statistics}
\bvolume{10}(\bissue{2}),
\bfpage{1019}--\blpage{1040}
(\byear{2016})
\end{barticle}
\endbibitem

\bibitem[\protect\citeauthoryear{Tomczak and Welling}{2018}]{tomczak2018vae}
\begin{bchapter}
\bauthor{\bsnm{Tomczak}, \binits{J.M.}},
\bauthor{\bsnm{Welling}, \binits{M.}}:
\bctitle{{VAE} with a {VampPrior}}.
In: \beditor{\bsnm{Storkey}, \binits{A.J.}},
\beditor{\bsnm{P{\'{e}}rez{-}Cruz}, \binits{F.}} (eds.)
\bbtitle{International Conference on Artificial Intelligence and Statistics,
  {AISTATS} 2018, 9-11 April 2018}.
\bsertitle{Proceedings of Machine Learning Research},
vol. \bseriesno{84},
pp. \bfpage{1214}--\blpage{1223}.
\bpublisher{{PMLR}},
\blocation{Playa Blanca, Lanzarote, Canary Islands, Spain}
(\byear{2018})
\end{bchapter}
\endbibitem

\bibitem[\protect\citeauthoryear{Vaswani et~al.}{2017}]{vaswani2017attention}
\begin{bchapter}
\bauthor{\bsnm{Vaswani}, \binits{A.}},
\bauthor{\bsnm{Shazeer}, \binits{N.}},
\bauthor{\bsnm{Parmar}, \binits{N.}},
\bauthor{\bsnm{Uszkoreit}, \binits{J.}},
\bauthor{\bsnm{Jones}, \binits{L.}},
\bauthor{\bsnm{Gomez}, \binits{A.N.}},
\bauthor{\bsnm{Kaiser}, \binits{L.}},
\bauthor{\bsnm{Polosukhin}, \binits{I.}}:
\bctitle{Attention is all you need}.
In: \beditor{\bsnm{Guyon}, \binits{I.}},
\beditor{\bsnm{Luxburg}, \binits{U.}},
\beditor{\bsnm{Bengio}, \binits{S.}},
\beditor{\bsnm{Wallach}, \binits{H.M.}},
\beditor{\bsnm{Fergus}, \binits{R.}},
\beditor{\bsnm{Vishwanathan}, \binits{S.V.N.}},
\beditor{\bsnm{Garnett}, \binits{R.}} (eds.)
\bbtitle{Advances in Neural Information Processing Systems 30: Annual
  Conference on Neural Information Processing Systems 2017, December 4-9,
  2017},
\bconflocation{Long Beach, CA, {USA}},
pp. \bfpage{5998}--\blpage{6008}
(\byear{2017})
\end{bchapter}
\endbibitem

\bibitem[\protect\citeauthoryear{Wu et~al.}{2025}]{wu2025machine}
\begin{barticle}
\bauthor{\bsnm{Wu}, \binits{H.}},
\bauthor{\bsnm{Li}, \binits{Y.}},
\bauthor{\bsnm{Jiang}, \binits{Y.}},
\bauthor{\bsnm{Li}, \binits{X.}},
\bauthor{\bsnm{Wang}, \binits{S.}},
\bauthor{\bsnm{Zhao}, \binits{C.}},
\bauthor{\bsnm{Yang}, \binits{X.}},
\bauthor{\bsnm{Chang}, \binits{B.}},
\bauthor{\bsnm{Yang}, \binits{J.}},
\bauthor{\bsnm{Qiao}, \binits{J.}}:
\batitle{Machine learning prediction of obesity-associated gut microbiota:
  identifying {Bifidobacterium} pseudocatenulatum as a potential therapeutic
  target}.
\bjtitle{Frontiers in Microbiology}
\bvolume{15},
\bfpage{1488656}
(\byear{2025})
\end{barticle}
\endbibitem

\bibitem[\protect\citeauthoryear{Wu et~al.}{2025}]{wu2025microbiome}
\begin{botherref}
\oauthor{\bsnm{Wu}, \binits{H.}},
\oauthor{\bsnm{Lv}, \binits{B.}},
\oauthor{\bsnm{Zhi}, \binits{L.}},
\oauthor{\bsnm{Shao}, \binits{Y.}},
\oauthor{\bsnm{Liu}, \binits{X.}},
\oauthor{\bsnm{Mitteregger}, \binits{M.}},
\oauthor{\bsnm{Chakaroun}, \binits{R.}},
\oauthor{\bsnm{Tremaroli}, \binits{V.}},
\oauthor{\bsnm{Hazen}, \binits{S.L.}},
\oauthor{\bsnm{Wang}, \binits{R.}}, et al.:
Microbiome--metabolome dynamics associated with impaired glucose control and
  responses to lifestyle changes.
Nature Medicine,
1--10
(2025)
\end{botherref}
\endbibitem

\bibitem[\protect\citeauthoryear{Wirbel et~al.}{2019}]{wirbel2019meta}
\begin{barticle}
\bauthor{\bsnm{Wirbel}, \binits{J.}},
\bauthor{\bsnm{Pyl}, \binits{P.T.}},
\bauthor{\bsnm{Kartal}, \binits{E.}},
\bauthor{\bsnm{Zych}, \binits{K.}},
\bauthor{\bsnm{Kashani}, \binits{A.}},
\bauthor{\bsnm{Milanese}, \binits{A.}},
\bauthor{\bsnm{Fleck}, \binits{J.S.}},
\bauthor{\bsnm{Voigt}, \binits{A.Y.}},
\bauthor{\bsnm{Palleja}, \binits{A.}},
\bauthor{\bsnm{Ponnudurai}, \binits{R.}}, \betal:
\batitle{Meta-analysis of fecal metagenomes reveals global microbial signatures
  that are specific for colorectal cancer}.
\bjtitle{Nature medicine}
\bvolume{25}(\bissue{4}),
\bfpage{679}--\blpage{689}
(\byear{2019})
\end{barticle}
\endbibitem

\bibitem[\protect\citeauthoryear{Yu et~al.}{2017}]{yu2017metagenomic}
\begin{barticle}
\bauthor{\bsnm{Yu}, \binits{J.}},
\bauthor{\bsnm{Feng}, \binits{Q.}},
\bauthor{\bsnm{Wong}, \binits{S.H.}},
\bauthor{\bsnm{Zhang}, \binits{D.}},
\bauthor{\bsnm{Liang}, \binits{Q.}},
\bauthor{\bsnm{Qin}, \binits{Y.}},
\bauthor{\bsnm{Tang}, \binits{L.}},
\bauthor{\bsnm{Zhao}, \binits{H.}},
\bauthor{\bsnm{Stenvang}, \binits{J.}},
\bauthor{\bsnm{Li}, \binits{Y.}}, \betal:
\batitle{Metagenomic analysis of faecal microbiome as a tool towards targeted
  non-invasive biomarkers for colorectal cancer}.
\bjtitle{Gut}
\bvolume{66}(\bissue{1}),
\bfpage{70}--\blpage{78}
(\byear{2017})
\end{barticle}
\endbibitem

\bibitem[\protect\citeauthoryear{Yun et~al.}{2019}]{yun2019cutmix}
\begin{bchapter}
\bauthor{\bsnm{Yun}, \binits{S.}},
\bauthor{\bsnm{Han}, \binits{D.}},
\bauthor{\bsnm{Chun}, \binits{S.}},
\bauthor{\bsnm{Oh}, \binits{S.J.}},
\bauthor{\bsnm{Yoo}, \binits{Y.}},
\bauthor{\bsnm{Choe}, \binits{J.}}:
\bctitle{Cutmix: Regularization strategy to train strong classifiers with
  localizable features}.
In: \bbtitle{2019 {IEEE/CVF} International Conference on Computer Vision,
  {ICCV} 2019, October 27 - November 2, 2019},
pp. \bfpage{6022}--\blpage{6031}.
\bpublisher{{IEEE}},
\blocation{Seoul, Korea (South)}
(\byear{2019})
\end{bchapter}
\endbibitem

\bibitem[\protect\citeauthoryear{Yachida et~al.}{2019}]{yachida2019metagenomic}
\begin{barticle}
\bauthor{\bsnm{Yachida}, \binits{S.}},
\bauthor{\bsnm{Mizutani}, \binits{S.}},
\bauthor{\bsnm{Shiroma}, \binits{H.}},
\bauthor{\bsnm{Shiba}, \binits{S.}},
\bauthor{\bsnm{Nakajima}, \binits{T.}},
\bauthor{\bsnm{Sakamoto}, \binits{T.}},
\bauthor{\bsnm{Watanabe}, \binits{H.}},
\bauthor{\bsnm{Masuda}, \binits{K.}},
\bauthor{\bsnm{Nishimoto}, \binits{Y.}},
\bauthor{\bsnm{Kubo}, \binits{M.}}, \betal:
\batitle{Metagenomic and metabolomic analyses reveal distinct stage-specific
  phenotypes of the gut microbiota in colorectal cancer}.
\bjtitle{Nature medicine}
\bvolume{25}(\bissue{6}),
\bfpage{968}--\blpage{976}
(\byear{2019})
\end{barticle}
\endbibitem

\bibitem[\protect\citeauthoryear{Zhang et~al.}{2018}]{zhang2017mixup}
\begin{bchapter}
\bauthor{\bsnm{Zhang}, \binits{H.}},
\bauthor{\bsnm{Ciss{\'{e}}}, \binits{M.}},
\bauthor{\bsnm{Dauphin}, \binits{Y.N.}},
\bauthor{\bsnm{Lopez{-}Paz}, \binits{D.}}:
\bctitle{mixup: Beyond empirical risk minimization}.
In: \bbtitle{6th International Conference on Learning Representations, {ICLR}
  2018, April 30 - May 3, 2018, Conference Track Proceedings}.
\bpublisher{OpenReview.net},
\blocation{Vancouver, BC, Canada}
(\byear{2018})
\end{bchapter}
\endbibitem

\bibitem[\protect\citeauthoryear{Zhu et~al.}{2020}]{zhu2020metagenome}
\begin{barticle}
\bauthor{\bsnm{Zhu}, \binits{F.}},
\bauthor{\bsnm{Ju}, \binits{Y.}},
\bauthor{\bsnm{Wang}, \binits{W.}},
\bauthor{\bsnm{Wang}, \binits{Q.}},
\bauthor{\bsnm{Guo}, \binits{R.}},
\bauthor{\bsnm{Ma}, \binits{Q.}},
\bauthor{\bsnm{Sun}, \binits{Q.}},
\bauthor{\bsnm{Fan}, \binits{Y.}},
\bauthor{\bsnm{Xie}, \binits{Y.}},
\bauthor{\bsnm{Yang}, \binits{Z.}}, \betal:
\batitle{Metagenome-wide association of gut microbiome features for
  schizophrenia}.
\bjtitle{Nature communications}
\bvolume{11}(\bissue{1}),
\bfpage{1612}
(\byear{2020})
\end{barticle}
\endbibitem

\bibitem[\protect\citeauthoryear{Zeller et~al.}{2014}]{zeller2014potential}
\begin{barticle}
\bauthor{\bsnm{Zeller}, \binits{G.}},
\bauthor{\bsnm{Tap}, \binits{J.}},
\bauthor{\bsnm{Voigt}, \binits{A.Y.}},
\bauthor{\bsnm{Sunagawa}, \binits{S.}},
\bauthor{\bsnm{Kultima}, \binits{J.R.}},
\bauthor{\bsnm{Costea}, \binits{P.I.}},
\bauthor{\bsnm{Amiot}, \binits{A.}},
\bauthor{\bsnm{B{\"o}hm}, \binits{J.}},
\bauthor{\bsnm{Brunetti}, \binits{F.}},
\bauthor{\bsnm{Habermann}, \binits{N.}}, \betal:
\batitle{Potential of fecal microbiota for early-stage detection of colorectal
  cancer}.
\bjtitle{Molecular systems biology}
\bvolume{10}(\bissue{11}),
\bfpage{766}
(\byear{2014})
\end{barticle}
\endbibitem

\bibitem[\protect\citeauthoryear{Zhou and Wei}{2020}]{zhou2020learning}
\begin{bchapter}
\bauthor{\bsnm{Zhou}, \binits{D.}},
\bauthor{\bsnm{Wei}, \binits{X.-X.}}:
\bctitle{Learning identifiable and interpretable latent models of
  high-dimensional neural activity using {pi-VAE}}.
In: \beditor{\bsnm{Larochelle}, \binits{H.}},
\beditor{\bsnm{Ranzato}, \binits{M.}},
\beditor{\bsnm{Hadsell}, \binits{R.}},
\beditor{\bsnm{Balcan}, \binits{M.F.}},
\beditor{\bsnm{Lin}, \binits{H.}} (eds.)
\bbtitle{Advances in Neural Information Processing Systems 33: Annual
  Conference on Neural Information Processing Systems 2020, NeurIPS 2020,
  December 6-12, 2020},
vol. \bseriesno{33},
pp. \bfpage{7234}--\blpage{7247}.
\bpublisher{Curran Associates, Inc.},
\blocation{virtual}
(\byear{2020})
\end{bchapter}
\endbibitem

\end{thebibliography}


\onecolumn
\begin{appendices}

\section{Model parameters and training}

\subsection{PLN-Tree architecture}
\label{app:architecture_selection}
TaxaPLN builds upon the PLN-Tree framework to model microbiome count data and generate synthetic samples using the VAMP-based sampler proposed in Section~\ref{sec:generation_microbiomes}, making it a parametric data augmentation method. In this work, we adopt a single general-purpose parameterization across all datasets detailed in Table~\ref{tab:model_parameterization}. Although model hyperparameters are typically tuned to dataset-specific characteristics, this unified configuration displays consistent gains across our experiments, illustrating the flexibility and generalizability of the approach by avoiding over-specification.

\begin{figure}[H]
    \centering
    \includegraphics[width=0.95\linewidth]{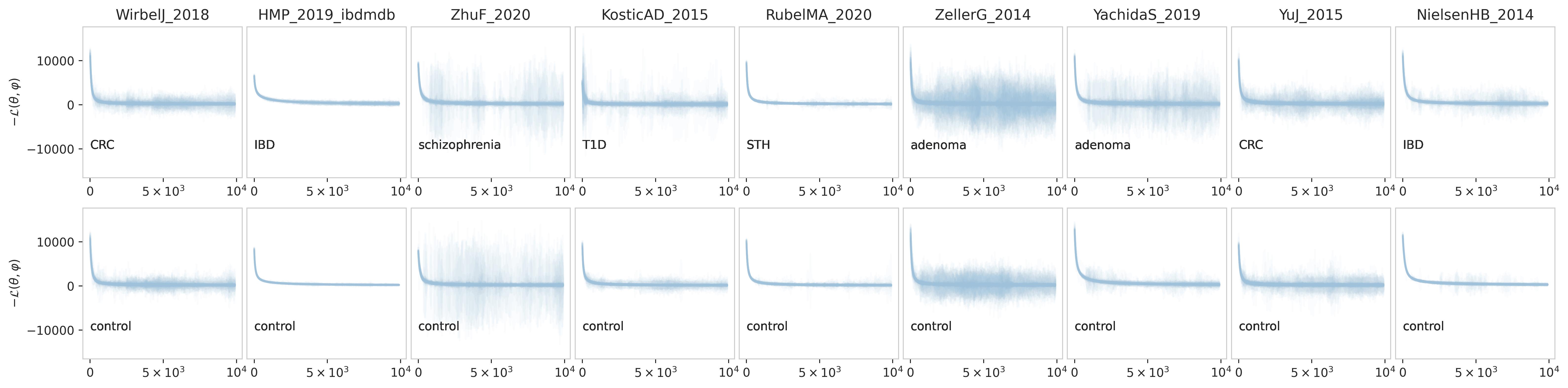}
    \caption{ELBO convergence over $10{,}000$ training epochs for PLN-Tree models across the first $5$ cross-validation (CV) iterations.
    Each CV iteration comprises $5$ folds, resulting in $5$ PLN-Tree models trained per iteration. The figure therefore displays $25$ learning curves per label for each study.}
    \label{fig:elbo_convergence}
\end{figure}

Model training follows the protocol of \cite{chaussard2025tree}, with parameters optimized by maximizing the ELBO using Adam optimizer \citep{kingma2015adam} with a learning rate of $10^{-3}$ and default parameters based on PyTorch framework \citep{paszke2019pytorch}, implemented with gradient clipping at $5.0$, and a batch size of $512$, over $10{,}000$ epochs for each cross-validation fold, as illustrated on Figure \ref{fig:elbo_convergence}. Training was performed on NVIDIA TITAN X 12GB GPU using CUDA.
\begin{table*}[htbp]
    \centering
    \begin{tabular}{lcccc}
    \toprule
    {\textbf{Distribution}} & \textbf{Architecture} &  \textbf{Parametrization}  \\
    \midrule 
    $p_{\btheta}(\bZ)$ & Per-layer Neural Network & $1$-layer NN \\
    $q_{\bphi}(\bZ\mid\bX)$ & Residual amortized backward & GRU $2\times32$ + $2$-layers NN \\
    $q_{\bphi}(\bZ\mid\bX,\bC)$ & Residual amortized backward & GRU $2\times32$ + FiLM $2$-layers NN + $2$-layers NN \\
    \bottomrule
    \end{tabular}
    \caption{PLN-Tree models parameterization in our experiments. The proposed architectures are derived from \citep{chaussard2025tree} benchmarks on microbiome data. NN: Neural Network, GRU: Gated-Recurrent Unit.}
    \label{tab:model_parameterization}
\end{table*}

\subsection{Data augmentation and classifiers parameters}
Following the learning phase of TaxaPLN, we perform data augmentation on the studies listed in Table~\ref{tab:curated_dataset_desc}. For each cross-validation fold, the corresponding TaxaPLN model is used to augment the training set by $100\%$ of its original size ($\beta = 2$). Classifiers from the \texttt{scikit-learn} library \citep{pedregosa2011scikit} are then trained on the augmented training sets, using the parameterizations detailed in Table~\ref{tab:classifiers_parameterization}.

We evaluate performance under two common preprocessing strategies for microbiome data: the centered log-ratio (CLR) transform and the relative abundance transform. The CLR transform normalizes counts based on their geometric mean, partially handling the compositional constraint and enabling the use of a wide range of classifiers. In contrast, the relative abundance transform projects the data onto the simplex, which introduces identifiability issues for parametric models. Consequently, we restrict evaluation under the relative abundance transform to classifiers that are well-suited for compositional data, such as Random Forests, XGBoost, and support vector machines with radial basis function (RBF) kernels.
\begin{table*}[htbp]
    \centering
    \begin{tabular}{lcccc}
    \toprule
    {\textbf{Classifier}} & \textbf{Preprocessing} &  \textbf{Parameters}  \\
    \midrule 
    Random Forest & CLR or relative & default: $100$ trees, Gini criterion, no max depth \\
    XGBoost & CLR or relative & learning rate: $0.05$, default: $100$ trees, max depth to $3$ \\
    MLP & CLR & hidden layers: $(256, 128)$, default: Adam, 200 max iterations \\
    Logistic Regression & CLR & default: penalty L2 with $C=1$ \\ 
    SVM (RBF) & relative & default: kernel RBF, penalty $C=1$ \\
    \bottomrule
    \end{tabular}
    \caption{Classifiers and parameterizations used in our experiments. All classifiers are implemented using the \texttt{scikit-learn} library \citep{pedregosa2011scikit}, with default parameters corresponding to their respective library implementations unless otherwise specified.}
    \label{tab:classifiers_parameterization}
\end{table*}

\clearpage
\section{Comparing TaxaPLN generator against PLN-Tree prior}
\label{app:compare_taxapln_prior}
\begin{figure}[H]
    \centering
    \begin{subfigure}[b]{0.48\linewidth}
        \centering
        \includegraphics[width=\linewidth,trim=0 30 0 0,clip]{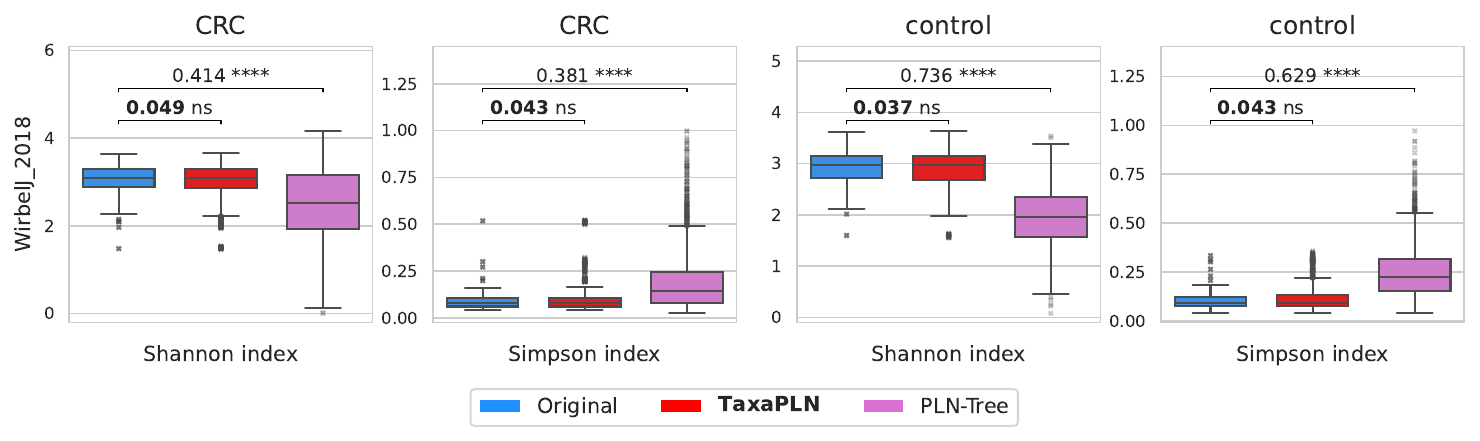}
    \end{subfigure}
    \begin{subfigure}[b]{0.48\linewidth}
        \centering
        \includegraphics[width=\linewidth,trim=0 30 0 0,clip]{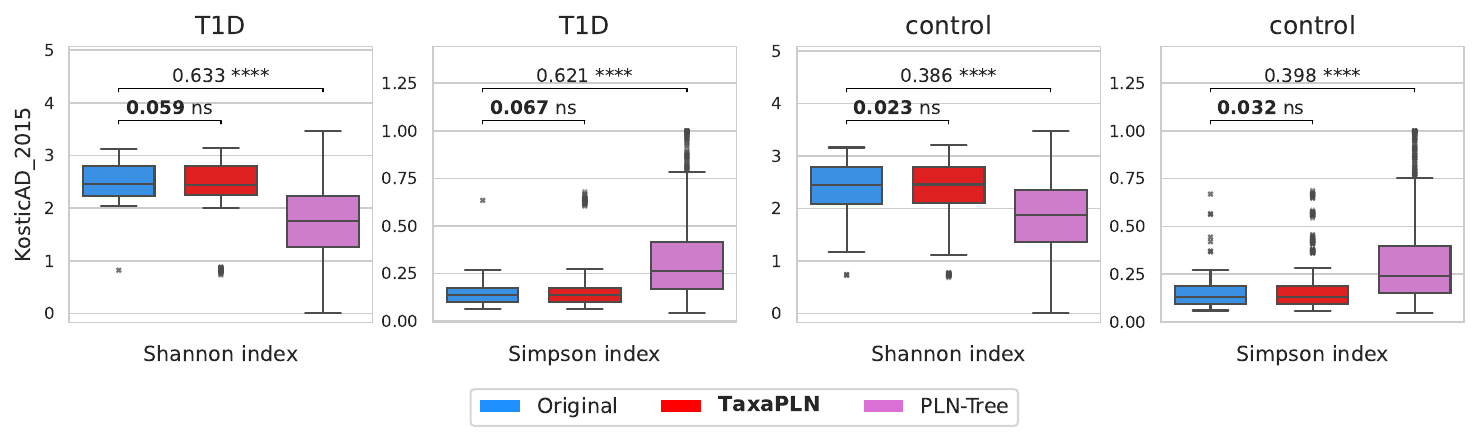}
    \end{subfigure}
    \begin{subfigure}[b]{0.48\linewidth}
        \centering
        \includegraphics[width=\linewidth,trim=0 30 0 0,clip]{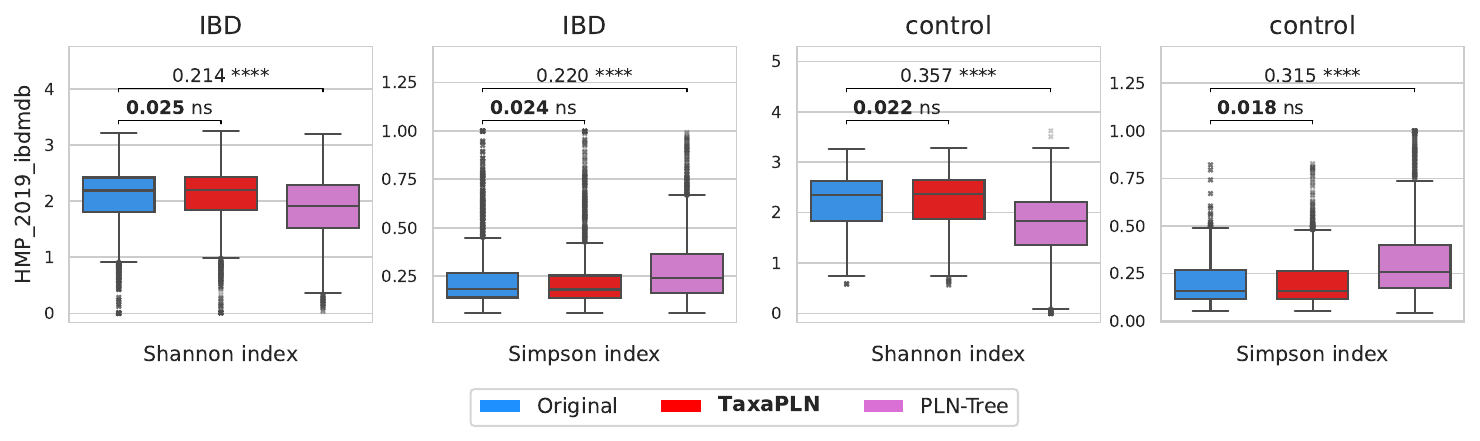}
    \end{subfigure}
    \begin{subfigure}[b]{0.48\linewidth}
        \centering
        \includegraphics[width=\linewidth,trim=0 30 0 0,clip]{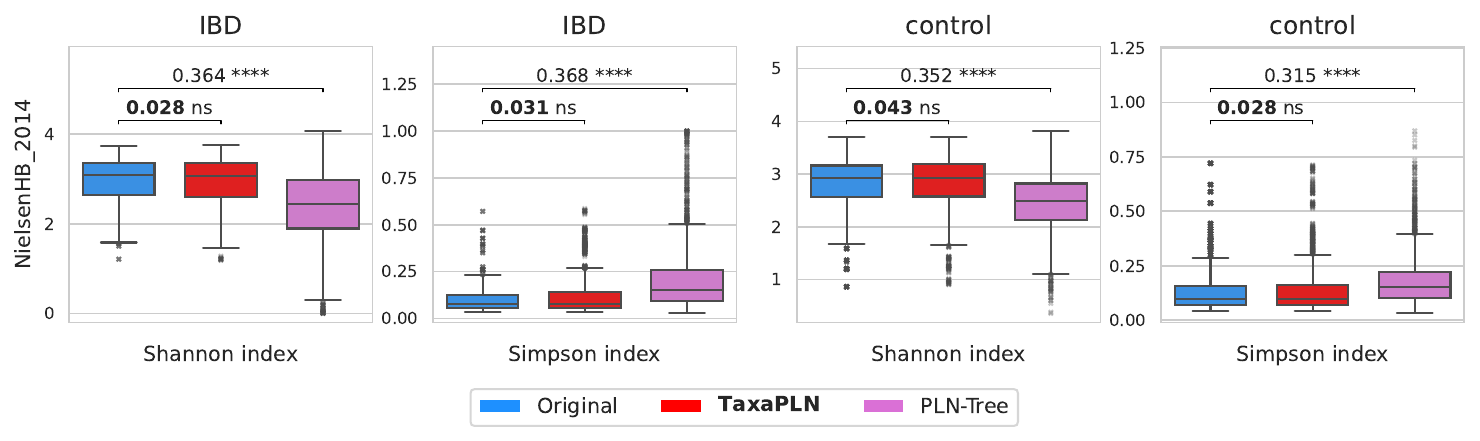}
    \end{subfigure}
    \begin{subfigure}[b]{0.48\linewidth}
        \centering
        \includegraphics[width=\linewidth,trim=0 30 0 0,clip]{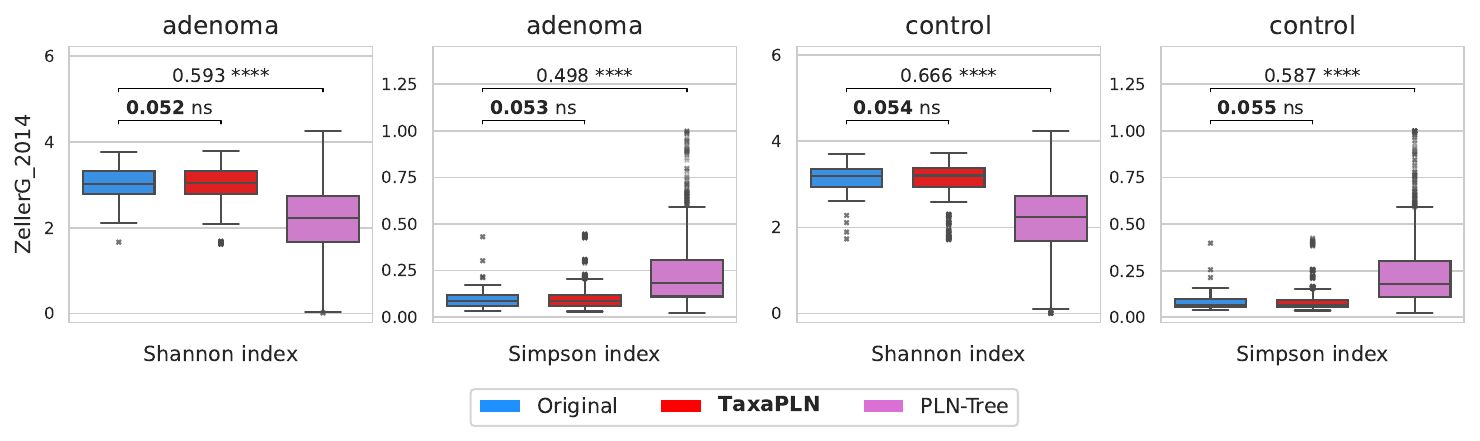}
    \end{subfigure}
    \begin{subfigure}[b]{0.48\linewidth}
        \centering
        \includegraphics[width=\linewidth,trim=0 30 0 0,clip]{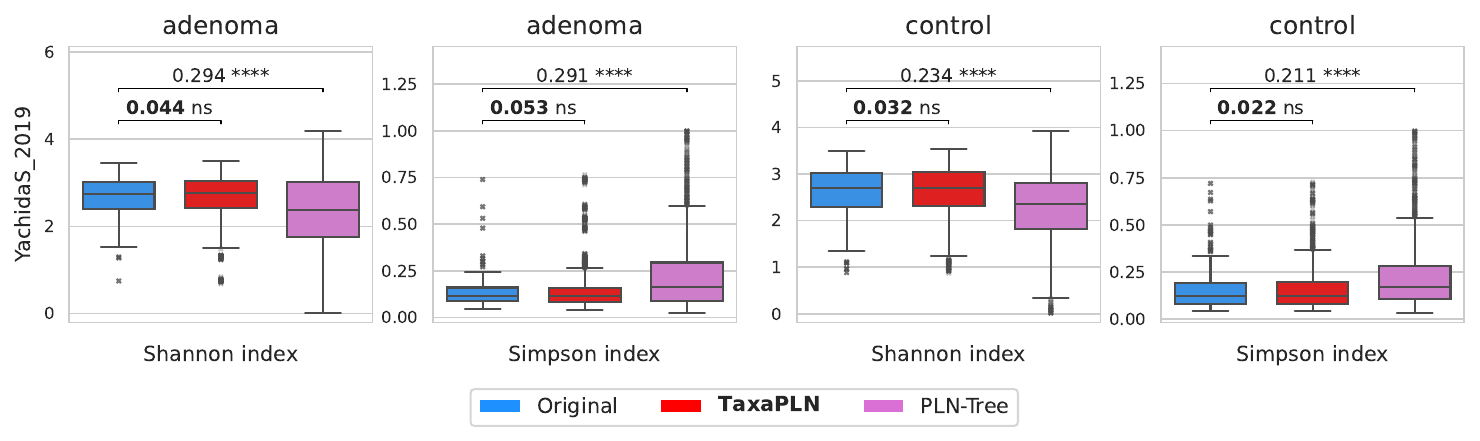}
    \end{subfigure}
    \begin{subfigure}[b]{0.48\linewidth}
        \centering
        \includegraphics[width=\linewidth,trim=0 30 0 0,clip]{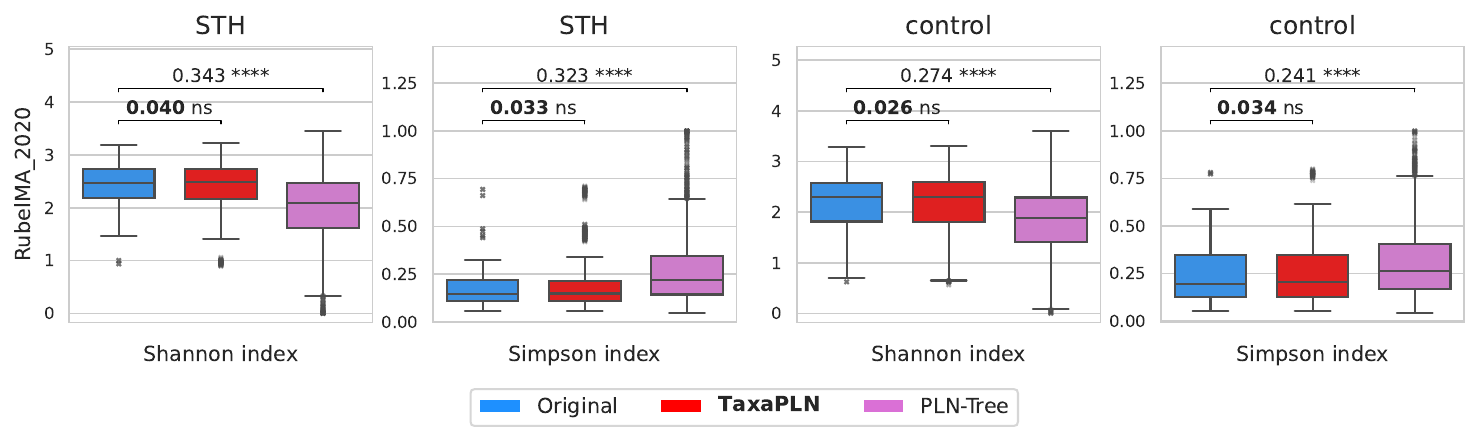}
    \end{subfigure}
    \begin{subfigure}[b]{0.48\linewidth}
        \centering
        \includegraphics[width=\linewidth,trim=0 30 0 0,clip]{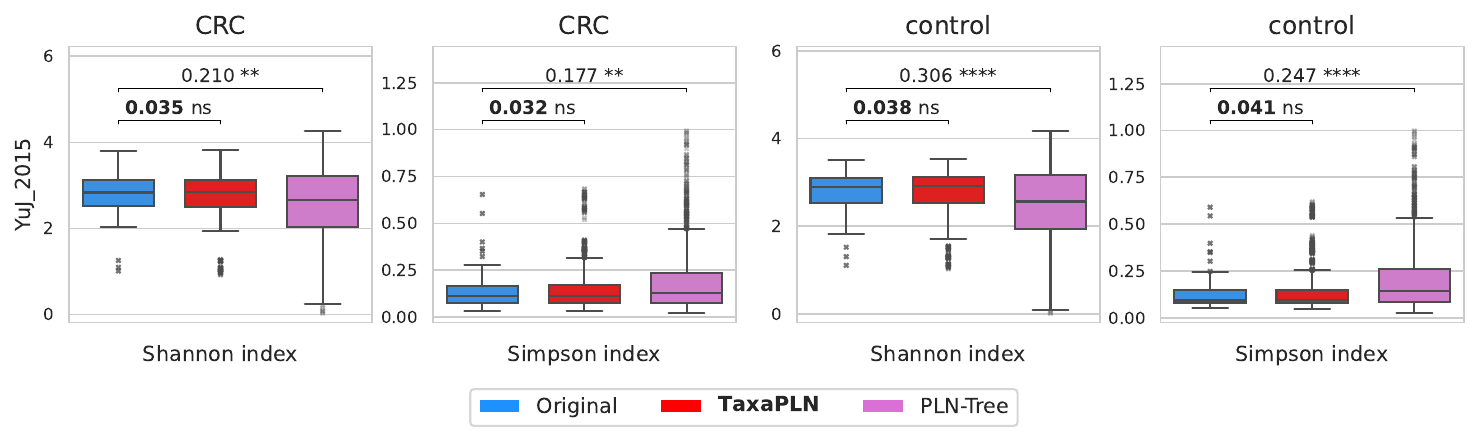}
    \end{subfigure}
    \begin{subfigure}[b]{0.8\linewidth}
        \centering
        \includegraphics[width=\linewidth,trim=0 30 0 0,clip]{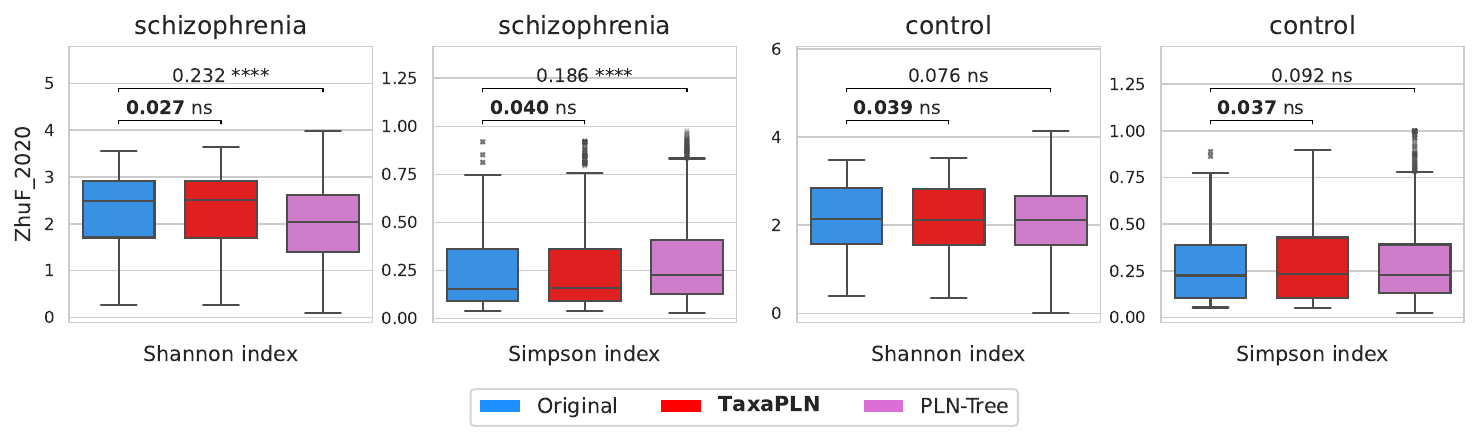}
    \end{subfigure}
    \begin{subfigure}[b]{0.9\linewidth}
        \centering
        \includegraphics[width=\linewidth,trim=0 0 0 180,clip]{alpha_diversity_benchmark_TaxaPLN_vs_prior_ZhuF_2020_Shannon_index_Simpson_index}
    \end{subfigure}
    \caption{Shannon and Simpson $\alpha$-diversity distributions. Synthetic microbiome samples generated by TaxaPLN and PLN-Tree prior \citep{chaussard2025tree} are evaluated on their $\alpha$-diversity consistency with the original microbiomes from Table~\ref{tab:curated_dataset_desc} based on Shannon index and Simpson index. Each method generates $500$ samples. Mann-Whitney U tests are performed to assess the statistical significance of distribution differences between generated and original samples for each method. Significance $P$-values thresholds are denoted by: ****$P \leq 0.0001$; ***$P \leq 0.001$; **$P \leq 0.01$; *$P \leq 0.05$; ns: $P > 0.05$. Kolmogorov-Smirnov divergence between original samples and generated microbiomes is provided with bold value indicating the minimal distance.}
    \label{fig:taxaplnvsprior_augmentation_alpha_diversity_full}
\end{figure}

\begin{figure}[H]
    \centering
    \begin{subfigure}[b]{0.48\linewidth}
        \centering
        \includegraphics[width=\linewidth,trim=0 40 0 0,clip]{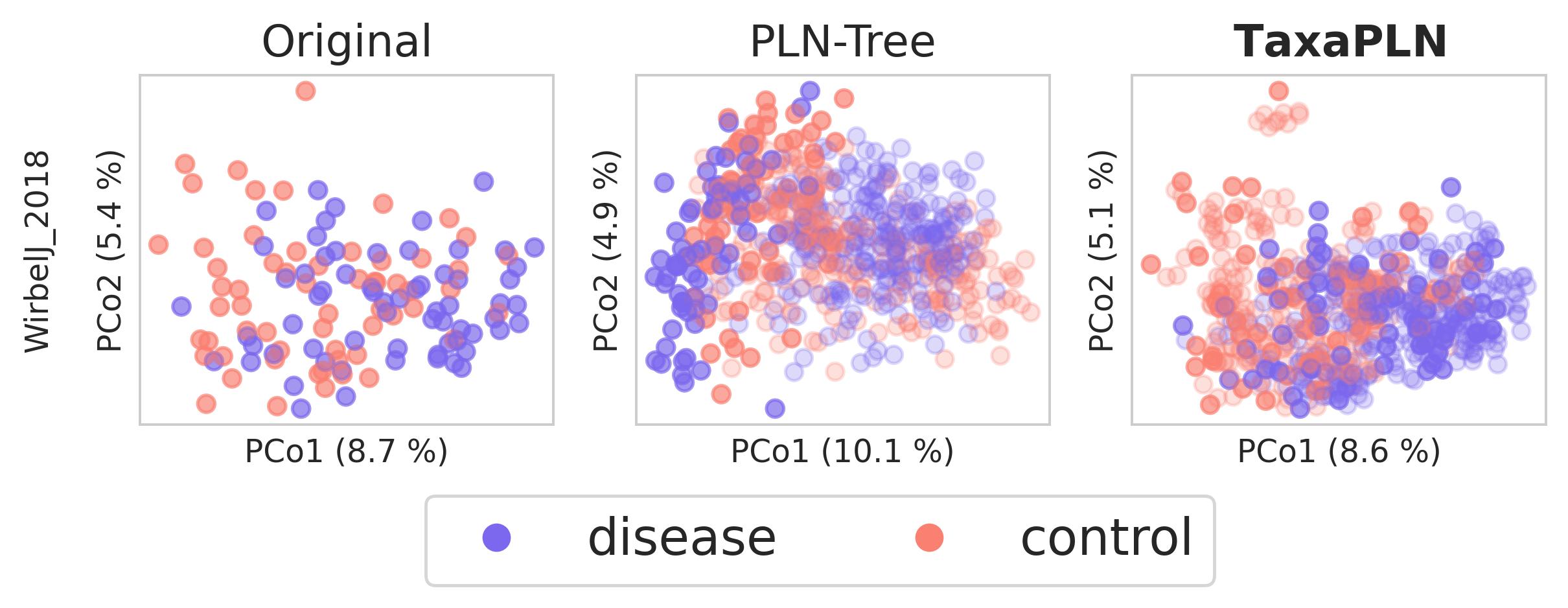}
    \end{subfigure}
    \begin{subfigure}[b]{0.48\linewidth}
        \centering
        \includegraphics[width=\linewidth,trim=0 40 0 0,clip]{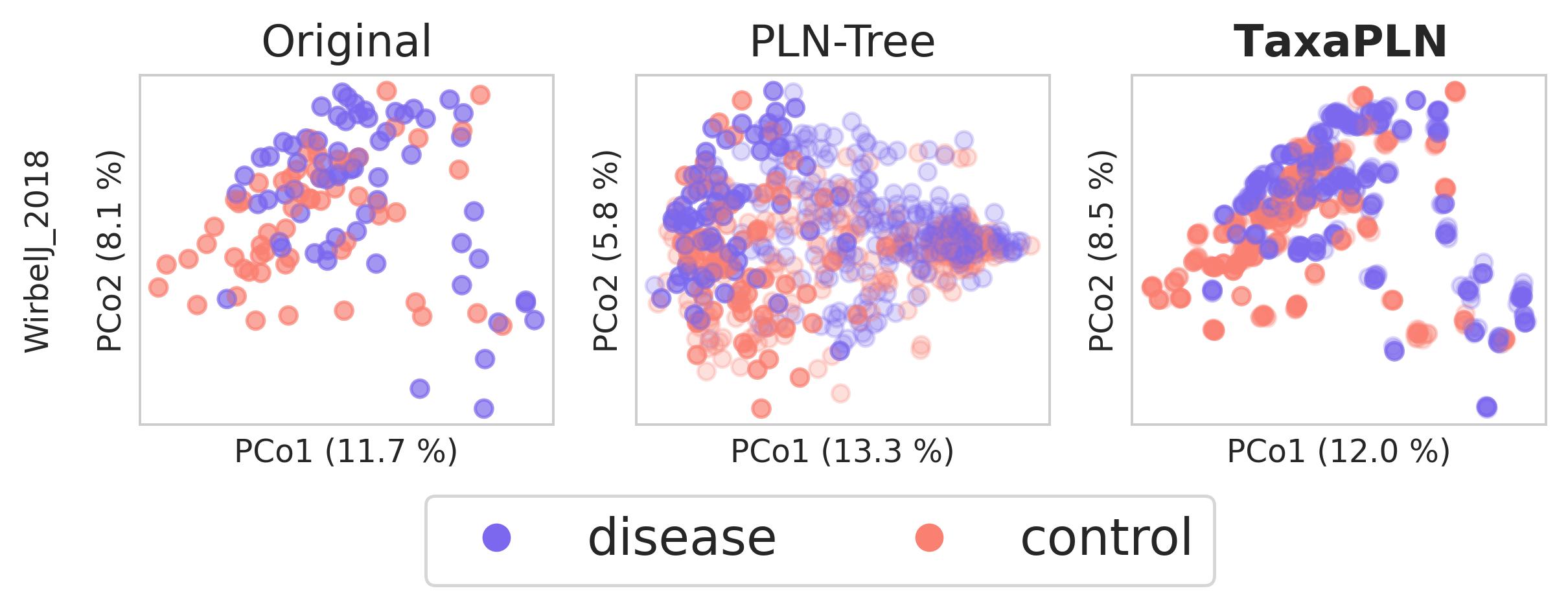}
    \end{subfigure}

    \begin{subfigure}[b]{0.48\linewidth}
        \centering
        \includegraphics[width=\linewidth,trim=0 40 0 22,clip]{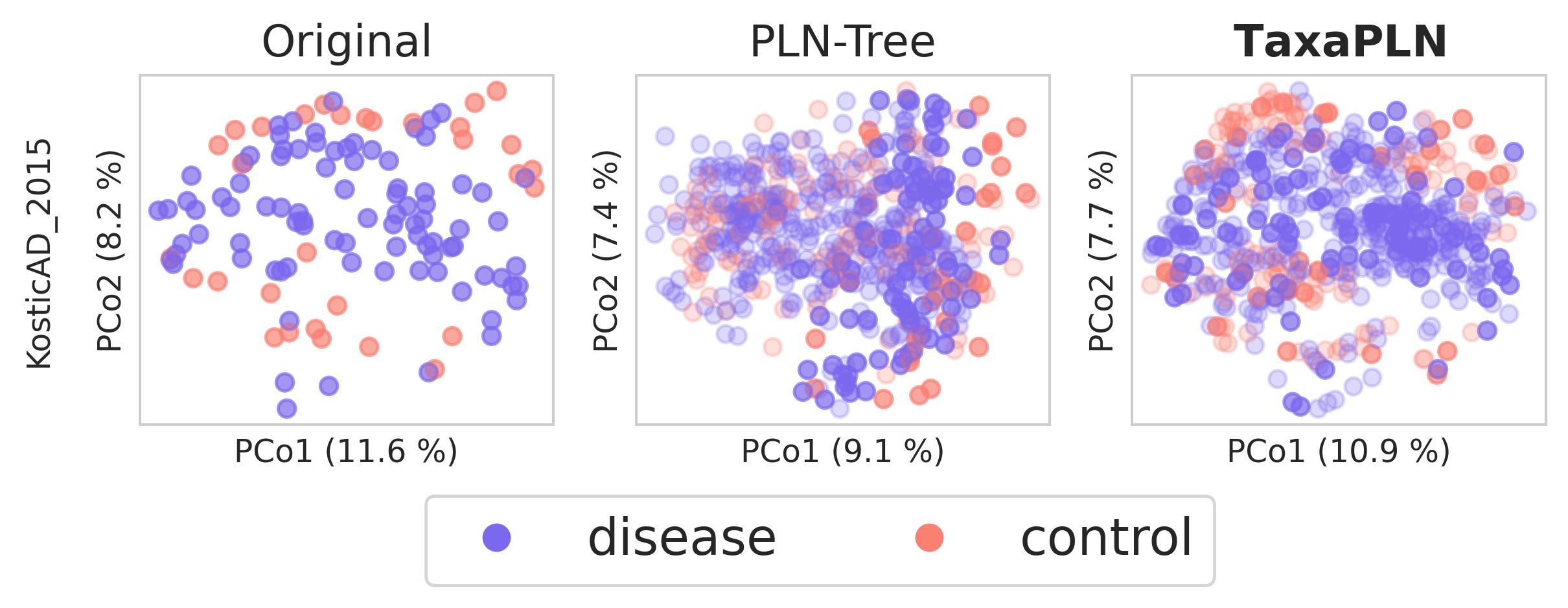}
    \end{subfigure}
    \begin{subfigure}[b]{0.48\linewidth}
        \centering
        \includegraphics[width=\linewidth,trim=0 40 0 22,clip]{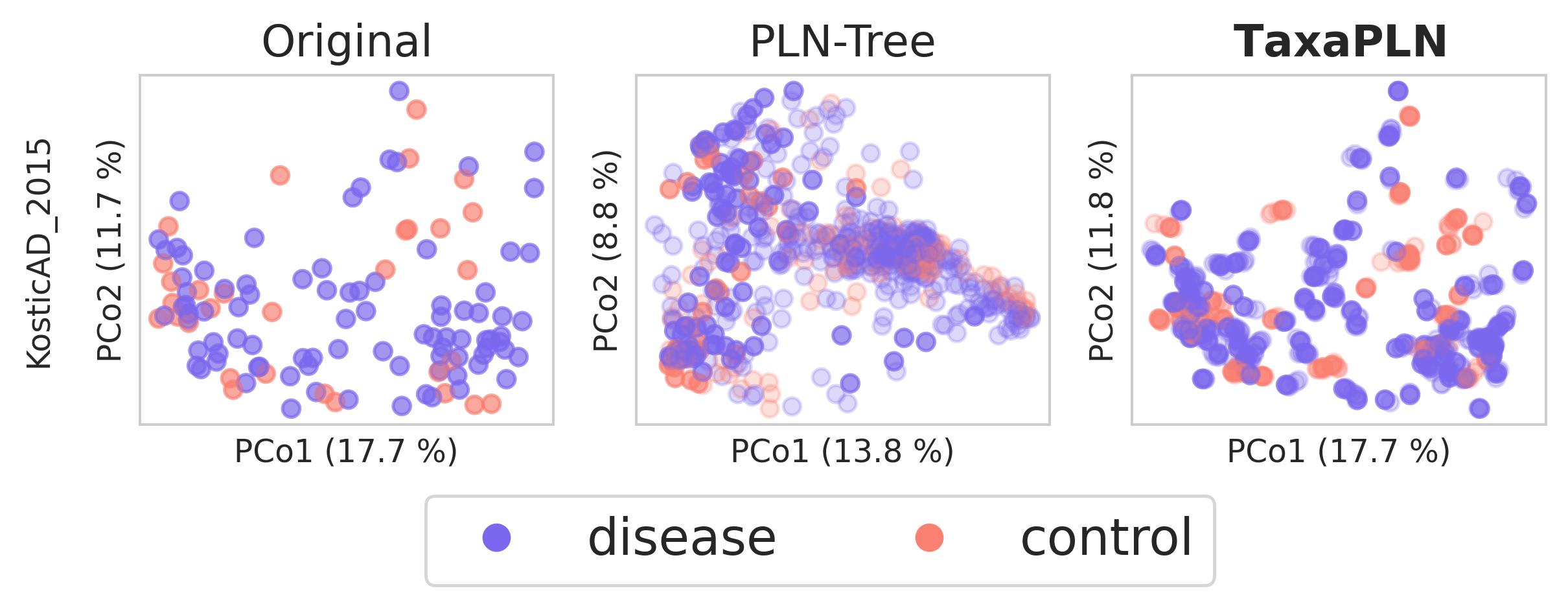}
    \end{subfigure}
    
    \begin{subfigure}[b]{0.48\linewidth}
        \centering
        \includegraphics[width=\linewidth,trim=0 40 0 22,clip]{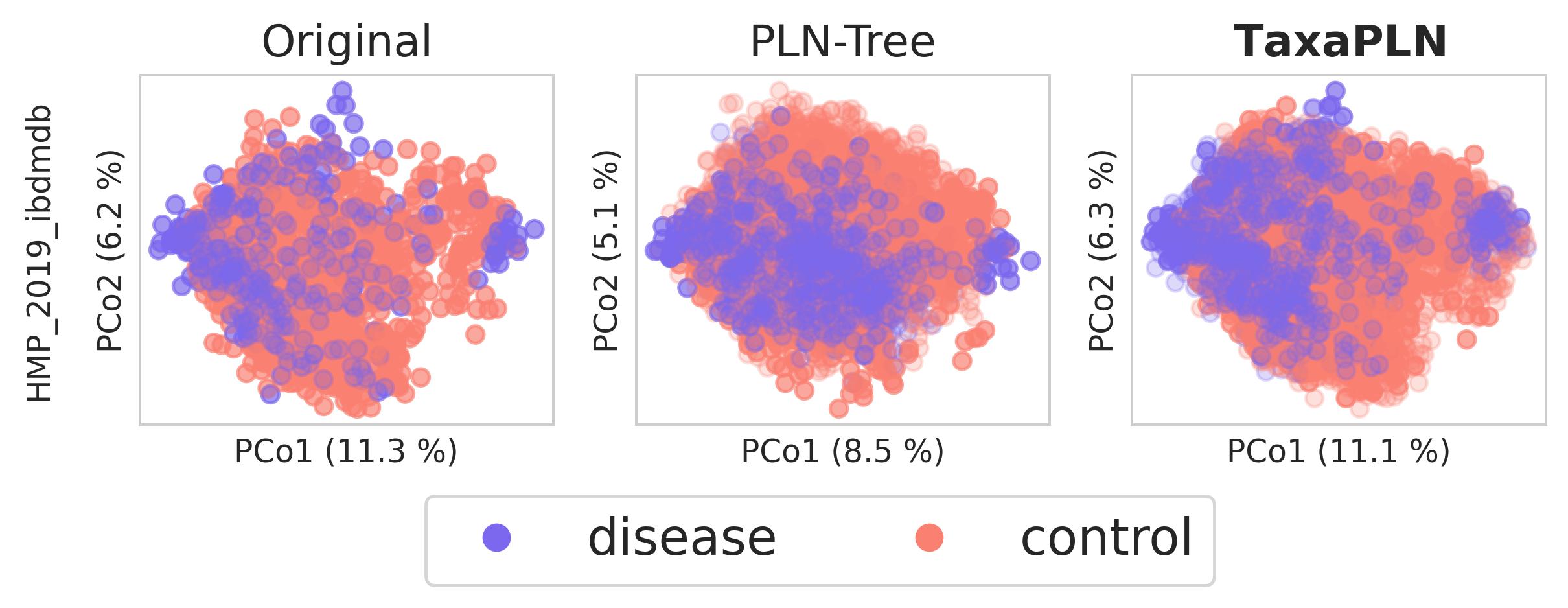}
    \end{subfigure}
    \begin{subfigure}[b]{0.48\linewidth}
        \centering
        \includegraphics[width=\linewidth,trim=0 40 0 22,clip]{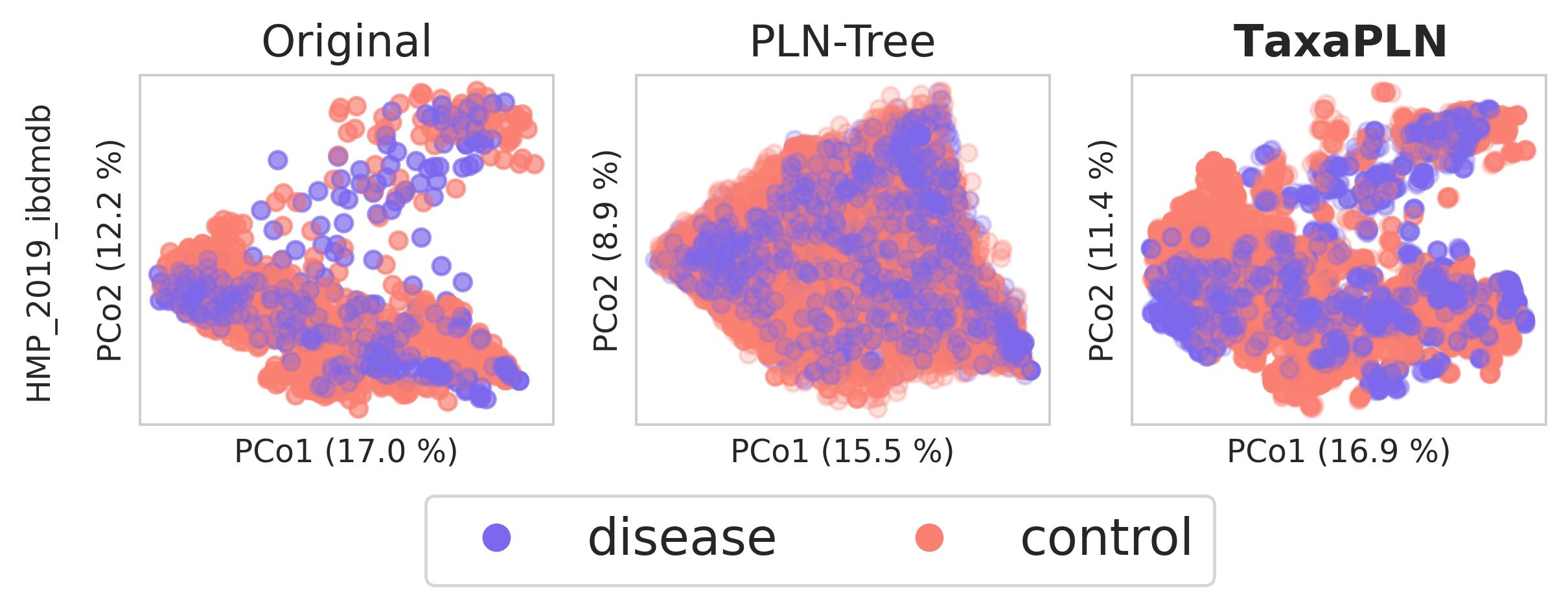}
    \end{subfigure}
    
    \begin{subfigure}[b]{0.48\linewidth}
        \centering
        \includegraphics[width=\linewidth,trim=0 40 0 22,clip]{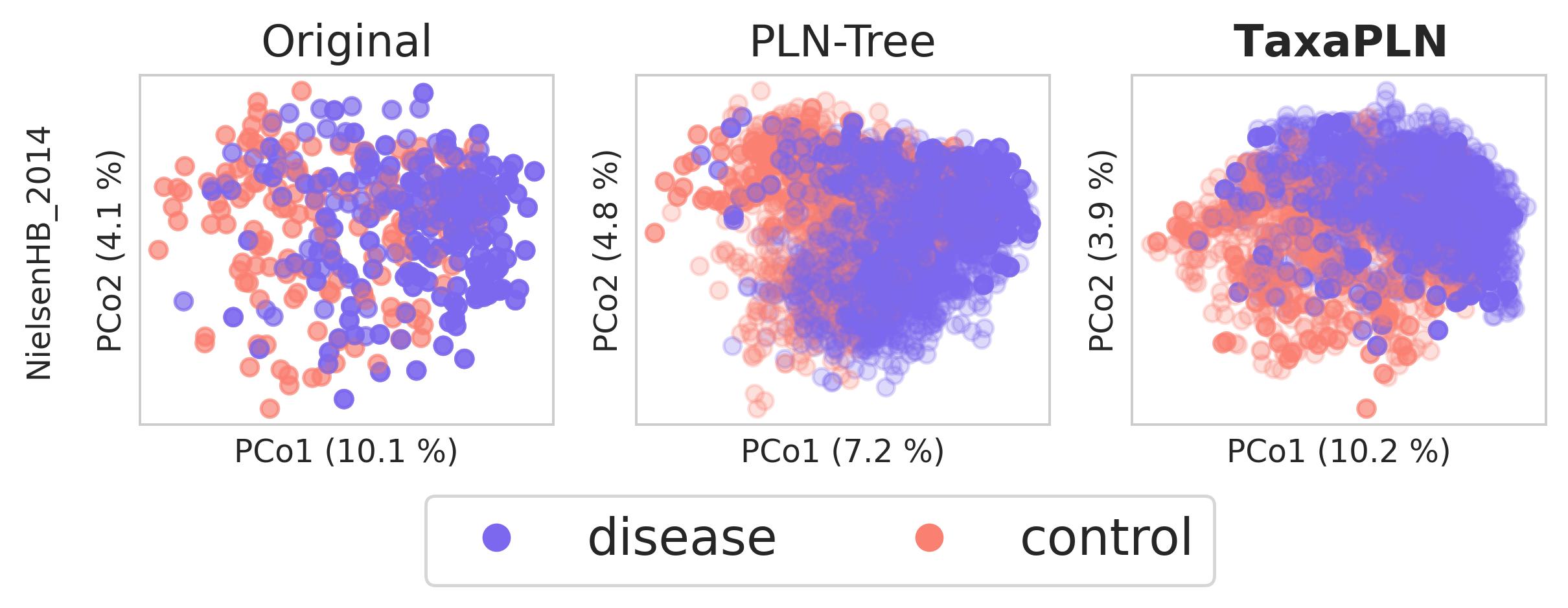}
    \end{subfigure}
    \begin{subfigure}[b]{0.48\linewidth}
        \centering
        \includegraphics[width=\linewidth,trim=0 40 0 22,clip]{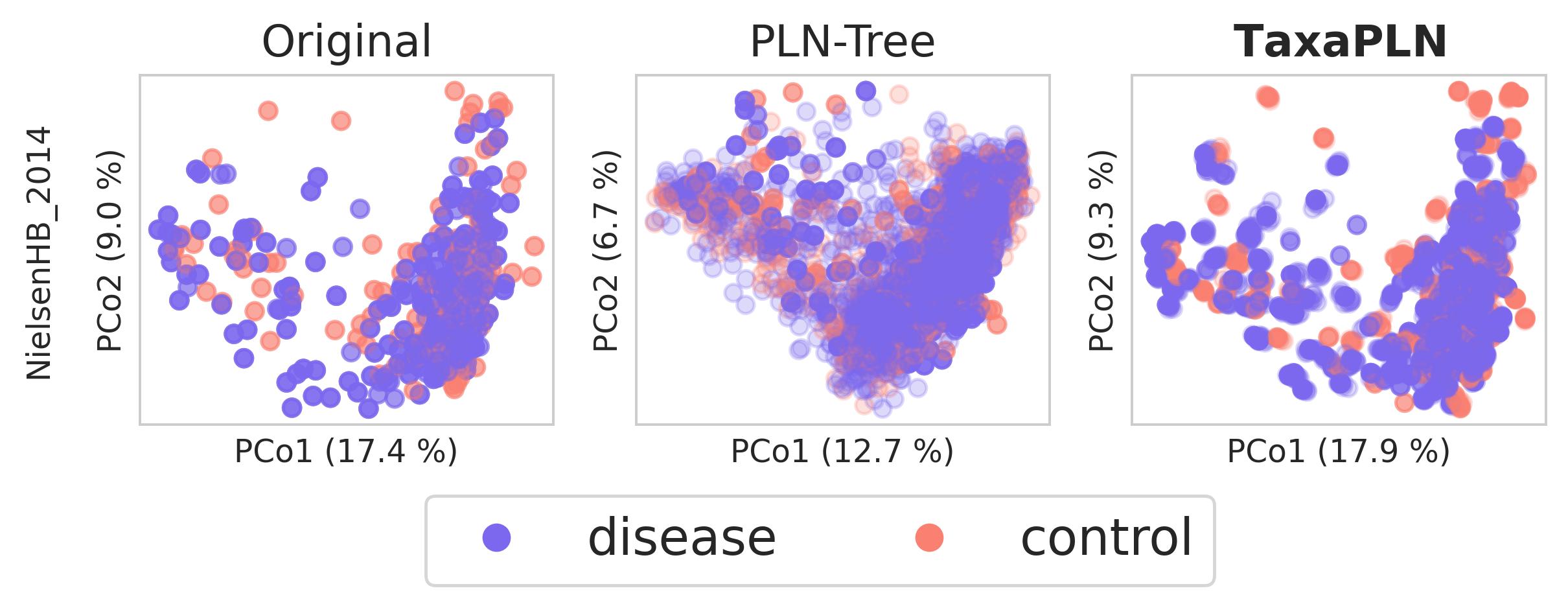}
    \end{subfigure}
    
    \begin{subfigure}[b]{0.48\linewidth}
        \centering
        \includegraphics[width=\linewidth,trim=0 40 0 22,clip]{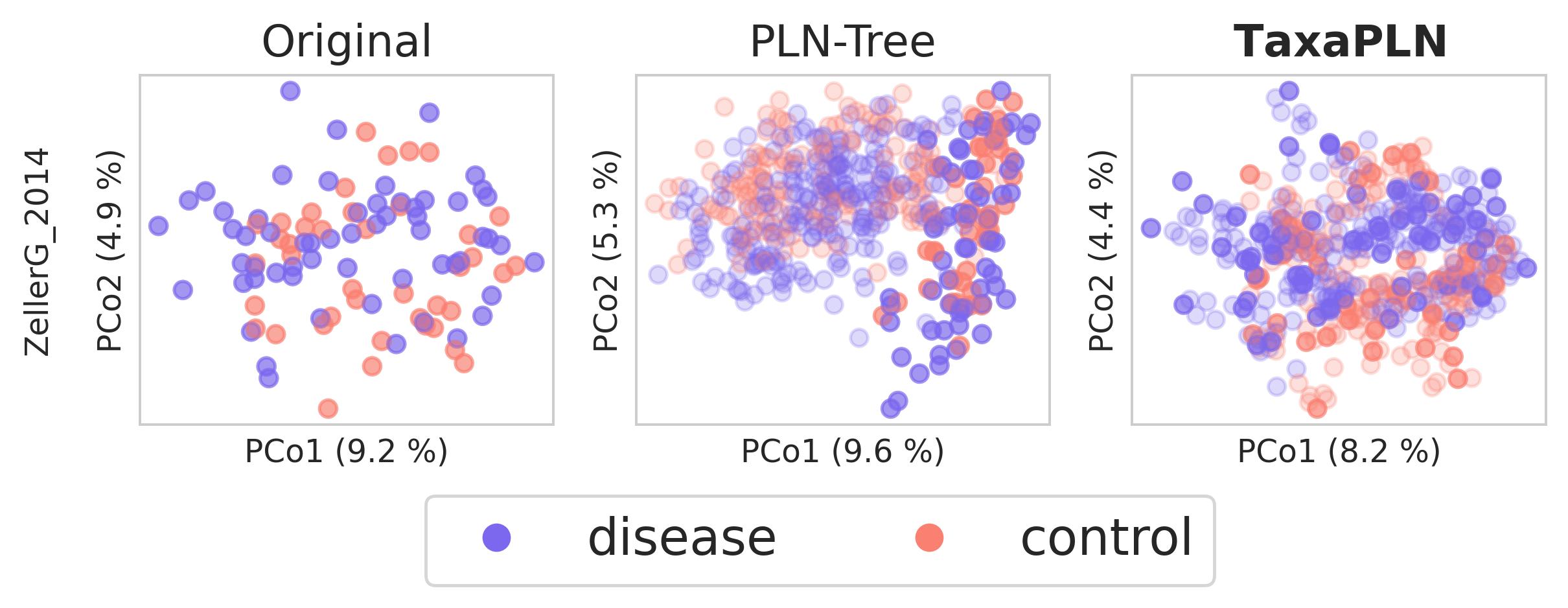}
    \end{subfigure}
    \begin{subfigure}[b]{0.48\linewidth}
        \centering
        \includegraphics[width=\linewidth,trim=0 40 0 22,clip]{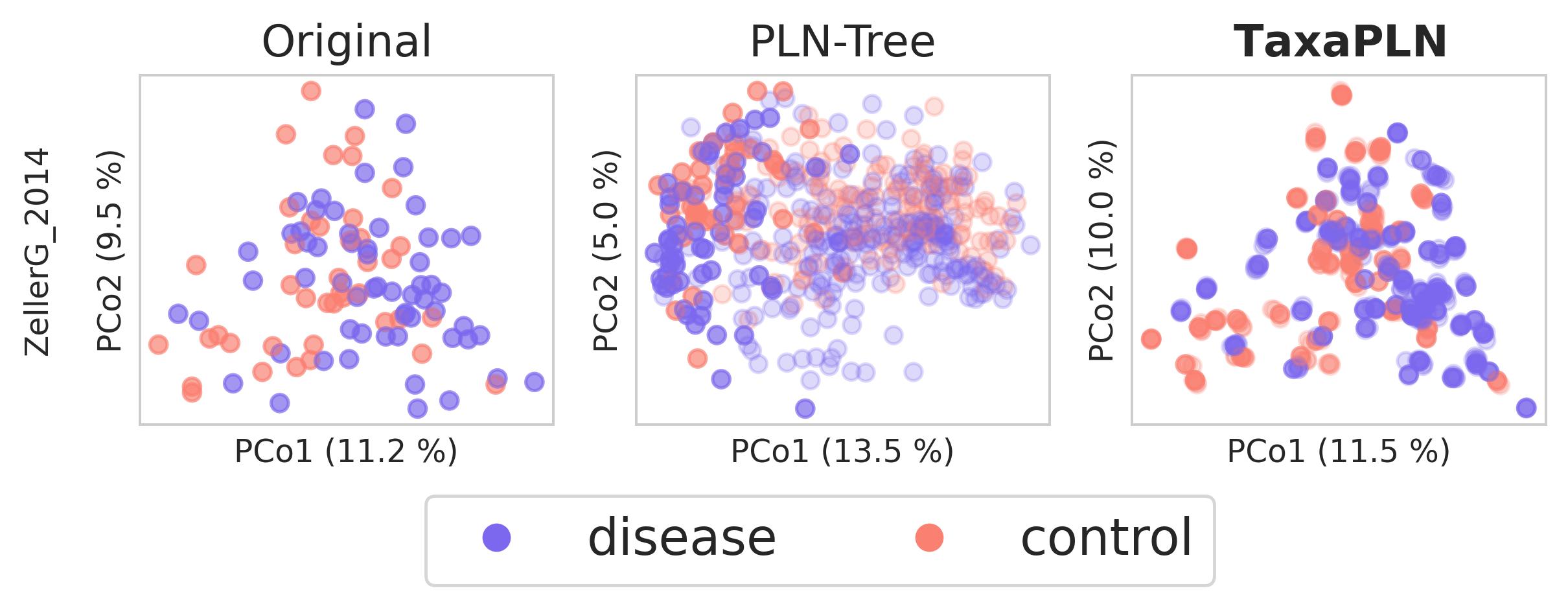}
    \end{subfigure}

    \begin{subfigure}[b]{0.48\linewidth}
        \centering
        \includegraphics[width=\linewidth,trim=0 40 0 22,clip]{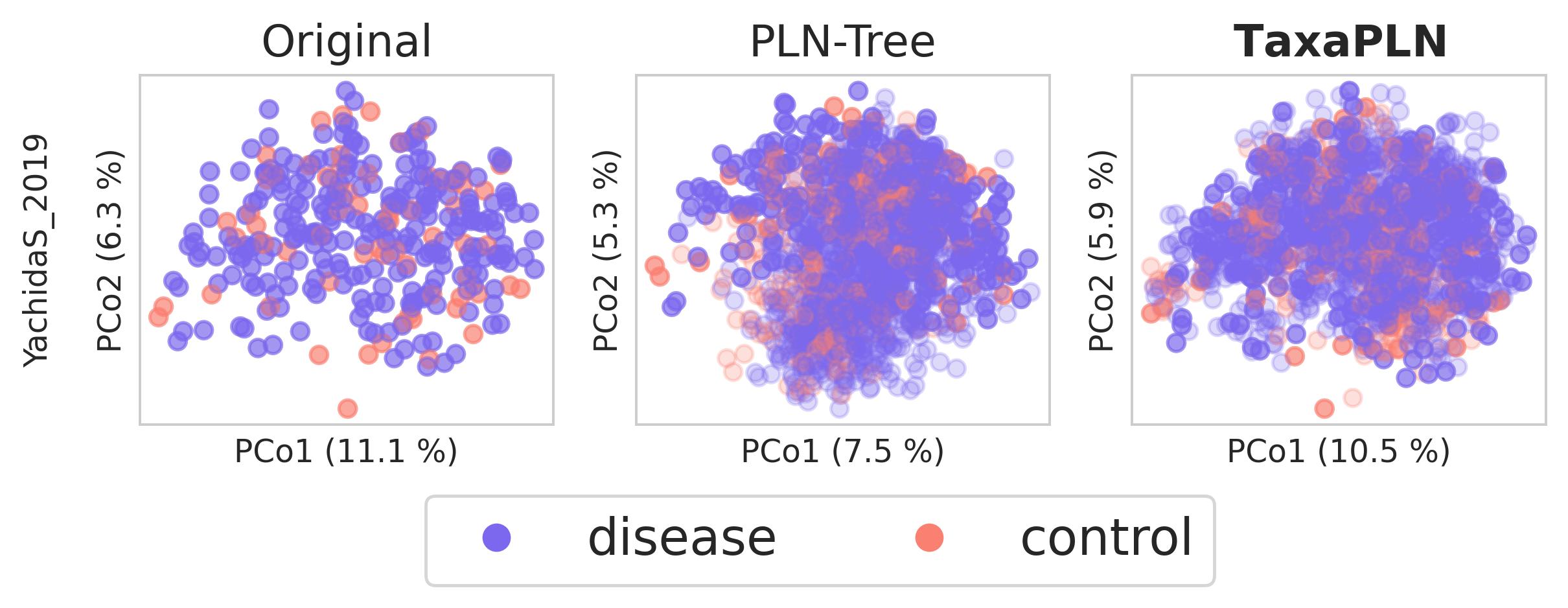}
    \end{subfigure}
    \begin{subfigure}[b]{0.48\linewidth}
        \centering
        \includegraphics[width=\linewidth,trim=0 40 0 22,clip]{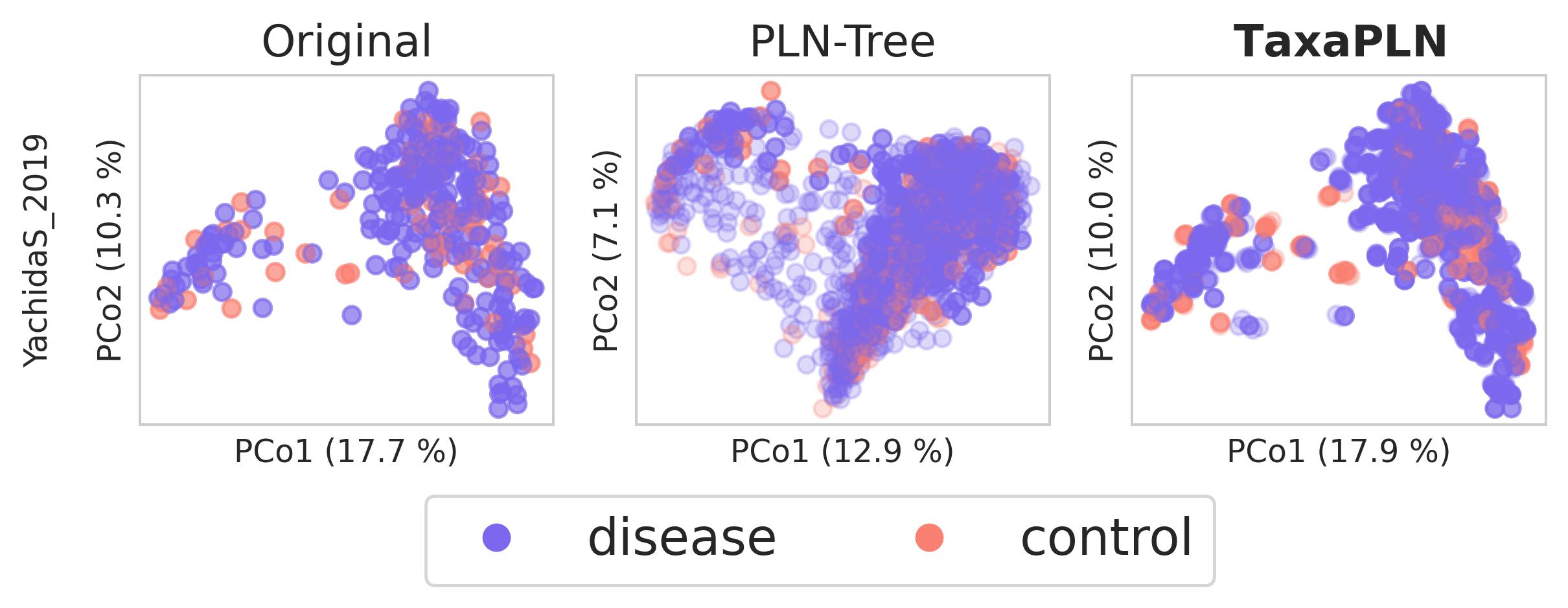}
    \end{subfigure}

    \begin{subfigure}[b]{0.48\linewidth}
        \centering
        \includegraphics[width=\linewidth,trim=0 40 0 22,clip]{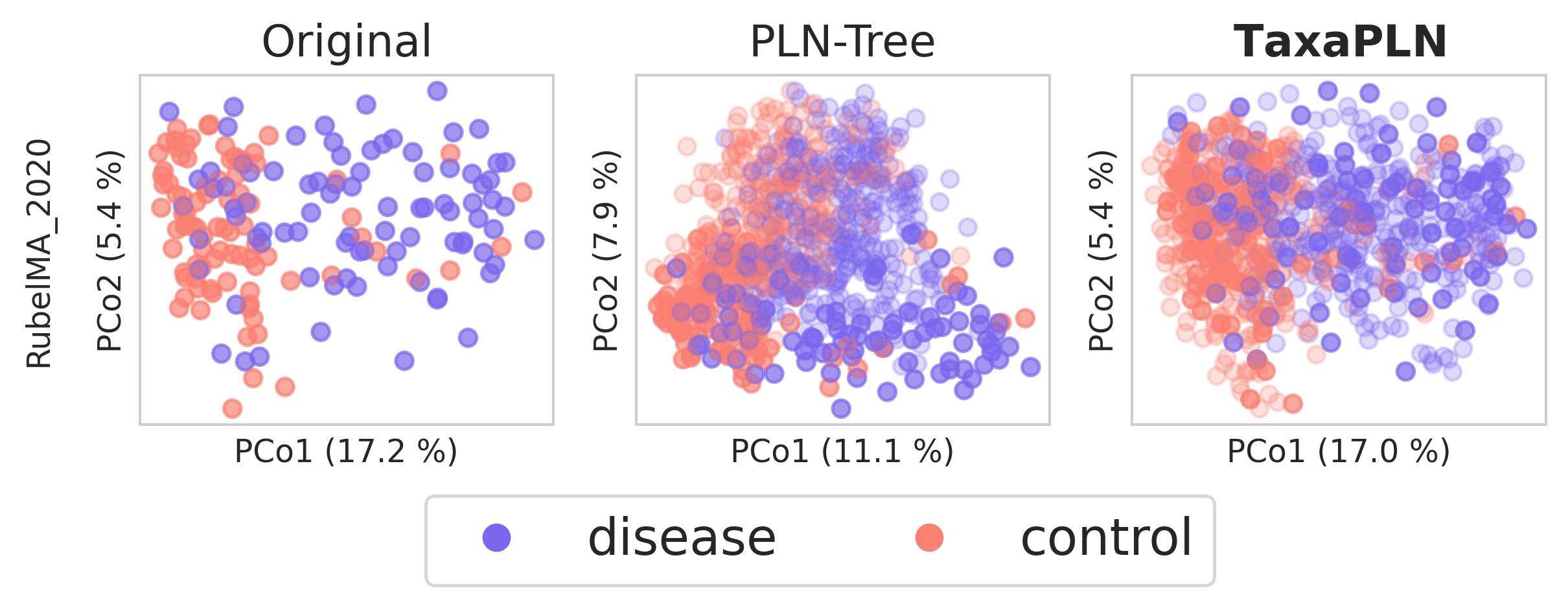}
    \end{subfigure}
    \begin{subfigure}[b]{0.48\linewidth}
        \centering
        \includegraphics[width=\linewidth,trim=0 40 0 22,clip]{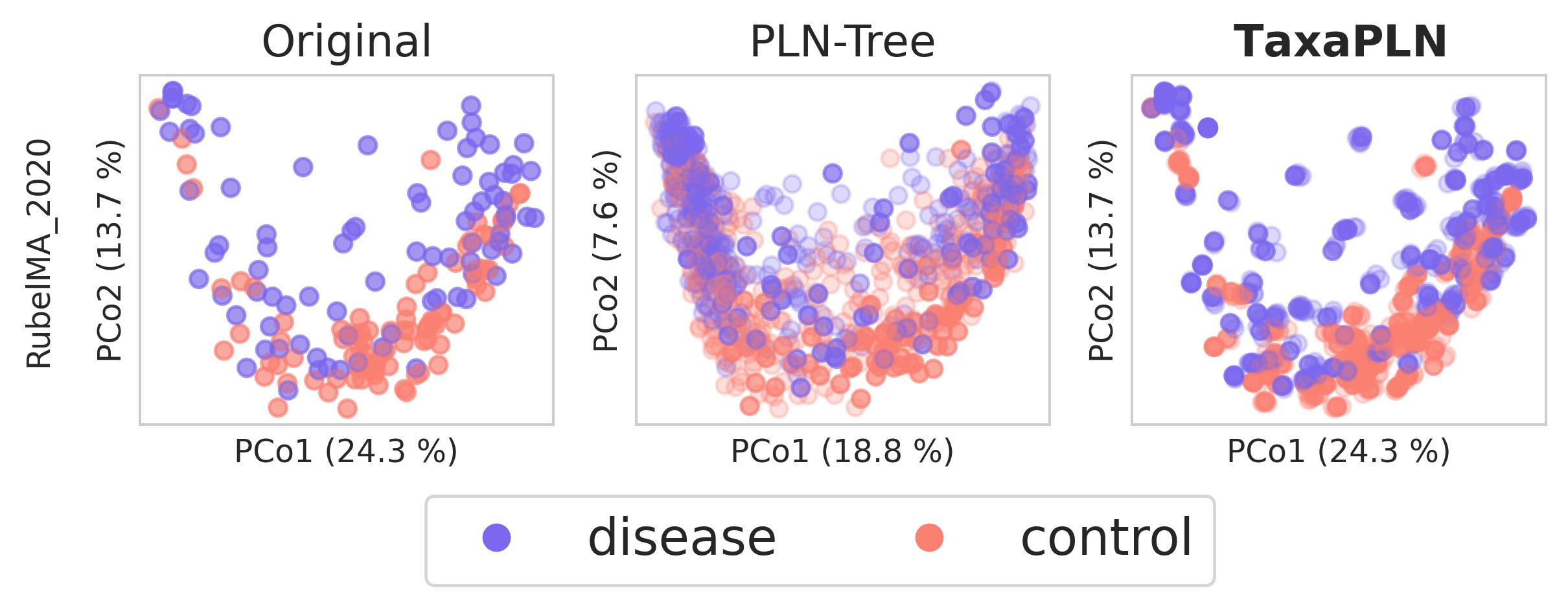}
    \end{subfigure}
    
    \begin{subfigure}[b]{0.48\linewidth}
        \centering
        \includegraphics[width=\linewidth,trim=0 40 0 22,clip]{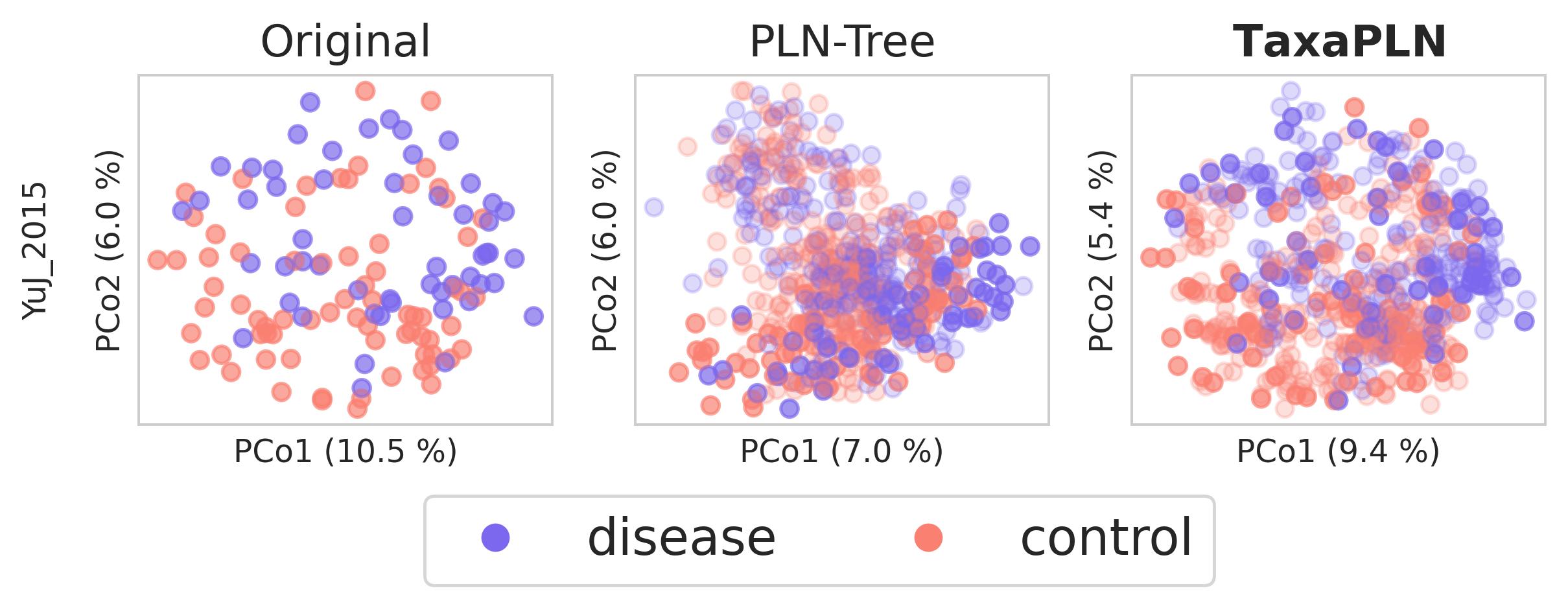}
    \end{subfigure}
    \begin{subfigure}[b]{0.48\linewidth}
        \centering
        \includegraphics[width=\linewidth,trim=0 40 0 22,clip]{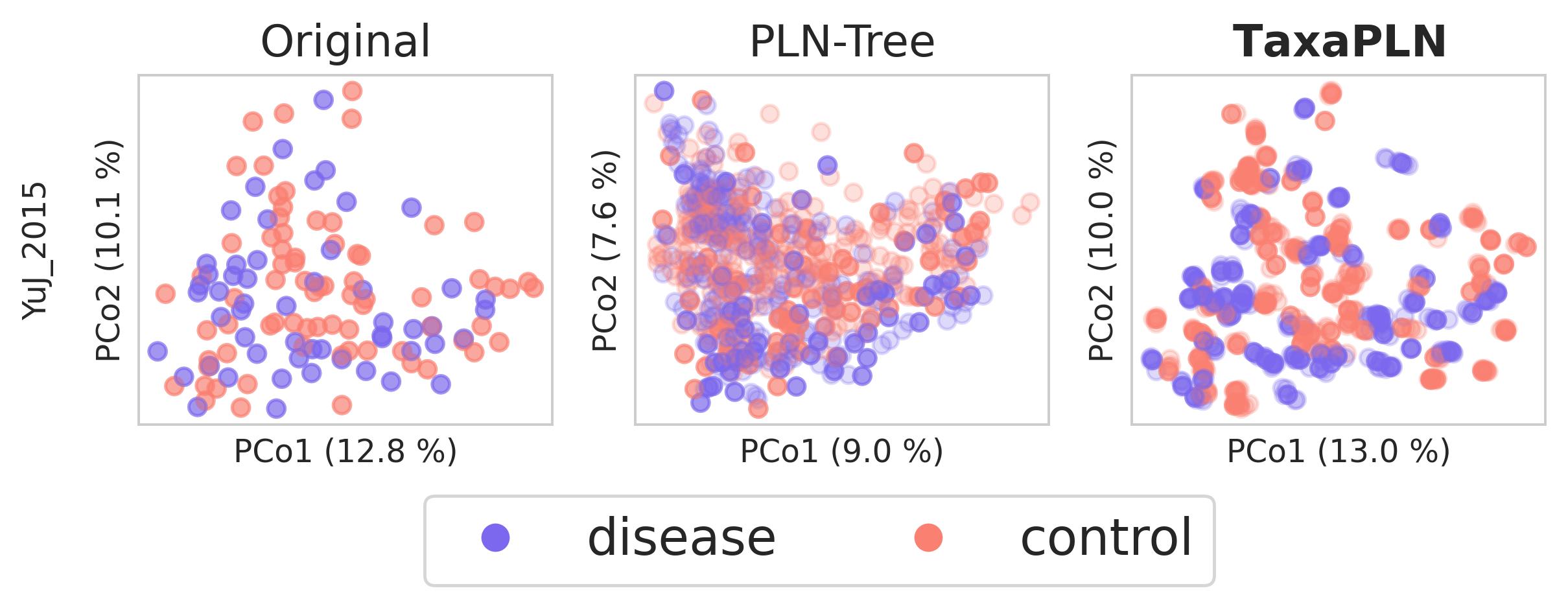}
    \end{subfigure}

    \begin{subfigure}[b]{0.48\linewidth}
        \centering
        \includegraphics[width=\linewidth,trim=0 40 0 22,clip]{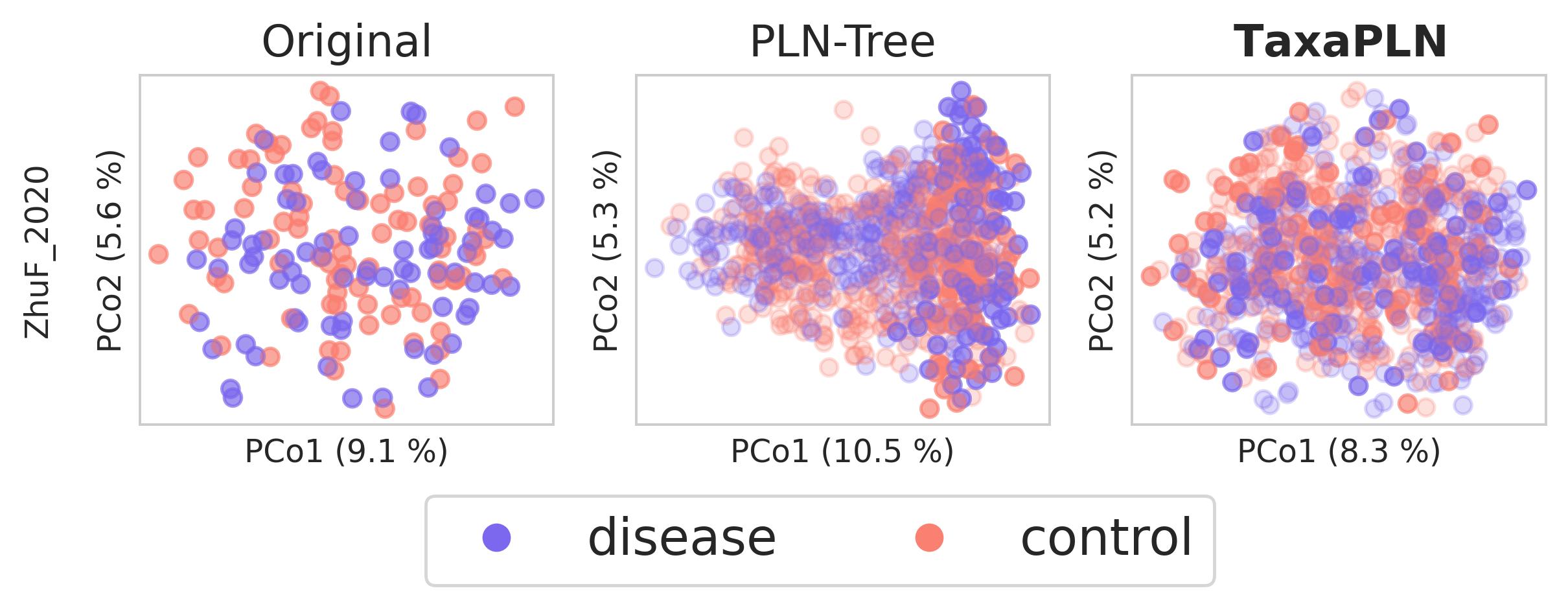}
        \caption{PCoA Aitchison distance}
    \end{subfigure}
    \begin{subfigure}[b]{0.48\linewidth}
        \centering
        \includegraphics[width=\linewidth,trim=0 40 0 22,clip]{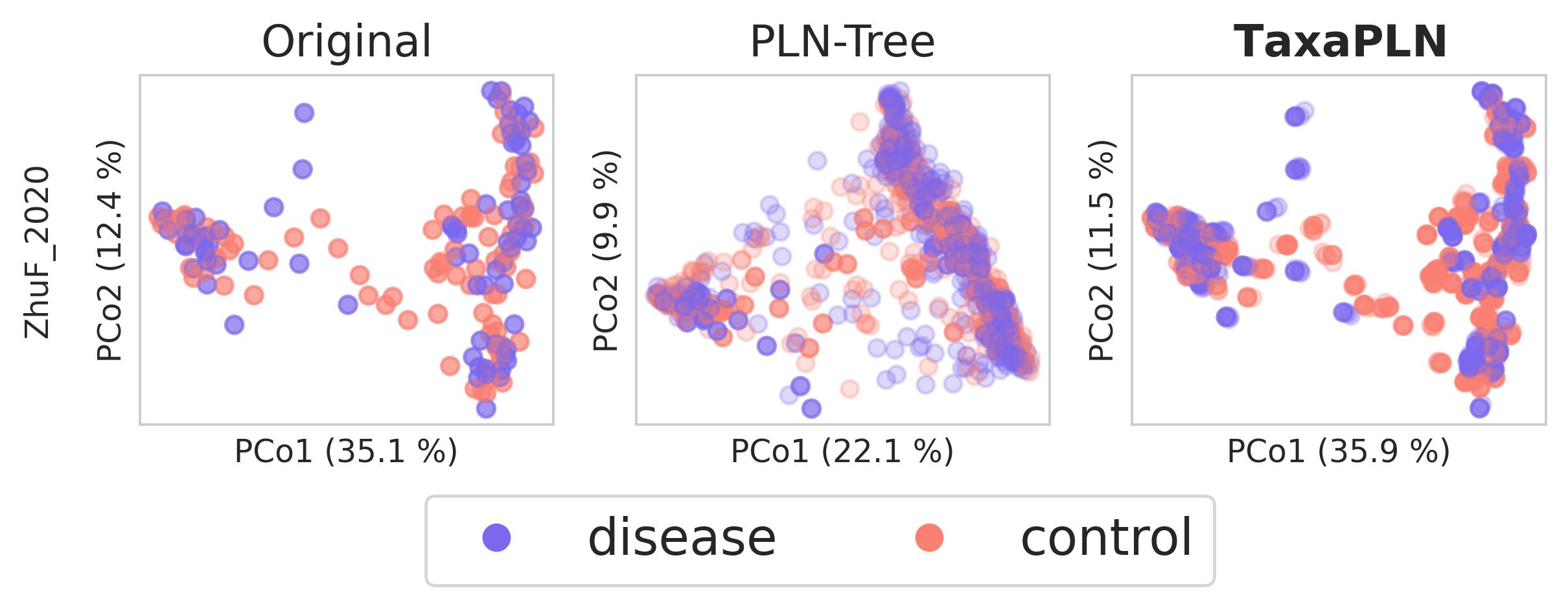}
        \caption{PCoA Bray-Curtis dissimilarity}
    \end{subfigure}

    \begin{subfigure}[b]{0.48\linewidth}
        \centering
        \includegraphics[width=\linewidth,trim=0 0 0 150,clip]{braycurtis-PCoA-WirbelJ_2018-TaxaPLNvsPLN-Tree}
    \end{subfigure}
    \caption{Aitchison and Bray–Curtis $\beta$-diversity PCoA plots of augmented datasets against the original cohort. Synthetic microbiome samples generated by TaxaPLN and PLN-Tree prior \citep{chaussard2025tree} are evaluated on their $\beta$-diversity consistency with the original microbiomes from Table~\ref{tab:curated_dataset_desc}. Each method augment the training fold with $\beta = 2$. Prior to Bray–Curtis dissimilarity computation, all data are normalized into proportions, while Aitchison distance is computed as the Euclidean distance of CLR-transformed counts. Principal Coordinates Analysis (PCoA) is used to visualize the dissimilarity structure.}
    \label{fig:taxaplnvsprior_augmentation_beta_diversity_full}
\end{figure}

In \cite{chaussard2025tree}, synthetic microbiome data are generated from the PLN-Tree model using its prior distribution $p_{\btheta}(\bZ)$, as defined in Section~\ref{sec:plntree}. While this prior-based sampling strategy significantly outperforms the non-hierarchical PLN baseline, it still produces limited numbers of realistic samples, especially when assessed at finer taxonomic resolutions. In particular, at the \textit{species} level, samples generated from the PLN-Tree prior are frequently rejected based on both $\alpha$-diversity and $\beta$-diversity metrics, as shown in Figures~\ref{fig:taxaplnvsprior_augmentation_alpha_diversity_full} and~\ref{fig:taxaplnvsprior_augmentation_beta_diversity_full}.

This limitation is consistent with known issues in variational autoencoders (VAEs), where the use of a fixed prior often results in poor sample quality \citep{tomczak2018vae,chadebec2022data}. To address this, \cite{tomczak2018vae} proposed the variational mixture of posteriors (VAMP) prior, which generates sharper and more realistic samples by relying on variational posterior distributions. While this is typically integrated into the training process, TaxaPLN instead employs a post-hoc VAMP sampler regularized by the original PLN-Tree prior.

As shown in Figures~\ref{fig:taxaplnvsprior_augmentation_alpha_diversity_full} and~\ref{fig:taxaplnvsprior_augmentation_beta_diversity_full}, this alternative sampler substantially improves the biological realism of the generated microbiome. While PCoA visualizations suggest close alignment in Bray–Curtis projected space, PCA on CLR-transformed data (Aitchison diversity) reveals that TaxaPLN explores the microbiome composition space without specifically overfitting, showing balance between exploration and biological faithfulness to the original cohort.

\section{Microbiome realism benchmarks}
\subsection{Alpha diversity}
\begin{figure}[H]
    \centering
    \begin{subfigure}[b]{0.48\linewidth}
        \centering
        \includegraphics[width=\linewidth,trim=0 30 0 0,clip]{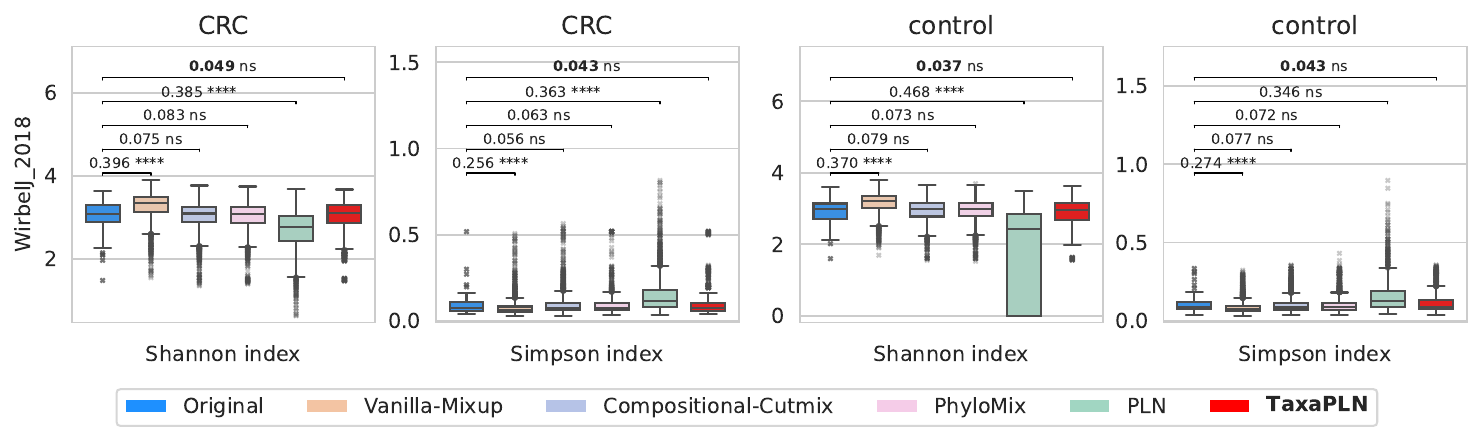}
    \end{subfigure}
    \begin{subfigure}[b]{0.48\linewidth}
        \centering
        \includegraphics[width=\linewidth,trim=0 30 0 0,clip]{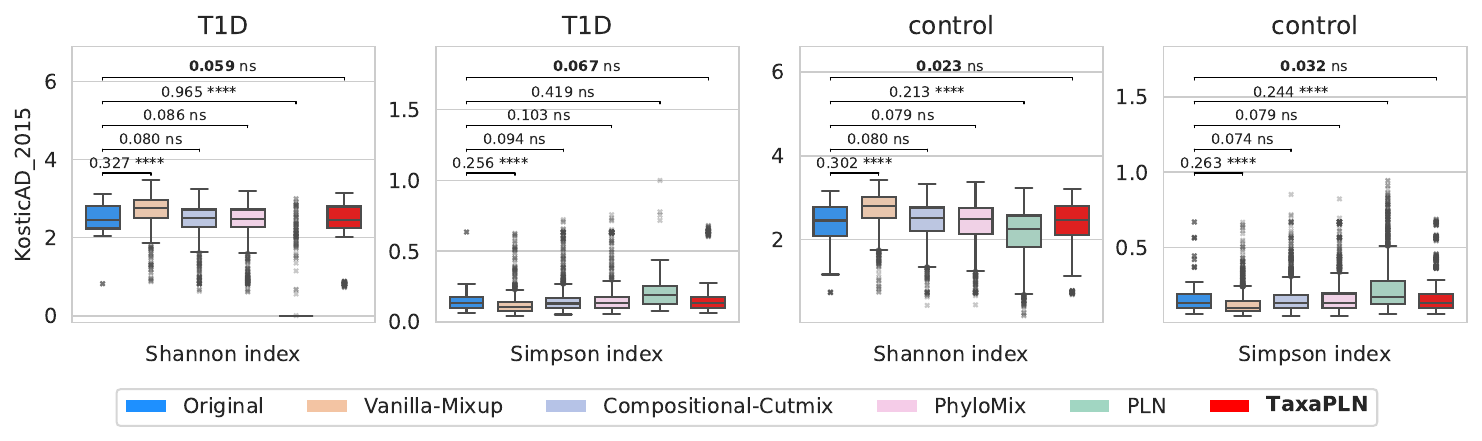}
    \end{subfigure}
    \begin{subfigure}[b]{0.48\linewidth}
        \centering
        \includegraphics[width=\linewidth,trim=0 30 0 0,clip]{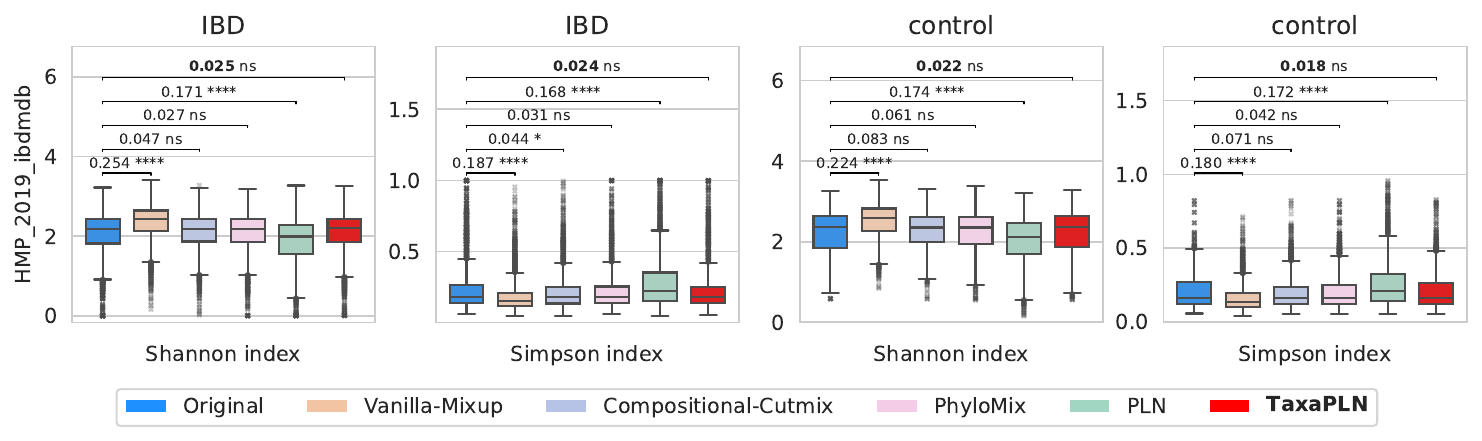}
    \end{subfigure}
    \begin{subfigure}[b]{0.48\linewidth}
        \centering
        \includegraphics[width=\linewidth,trim=0 30 0 0,clip]{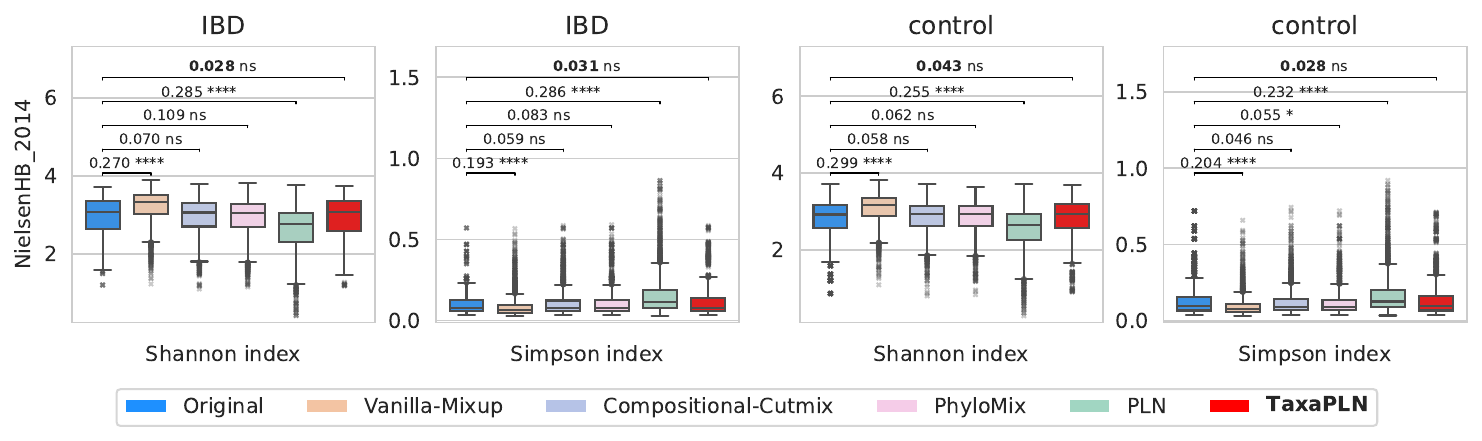}
    \end{subfigure}
    \begin{subfigure}[b]{0.48\linewidth}
        \centering
        \includegraphics[width=\linewidth,trim=0 30 0 0,clip]{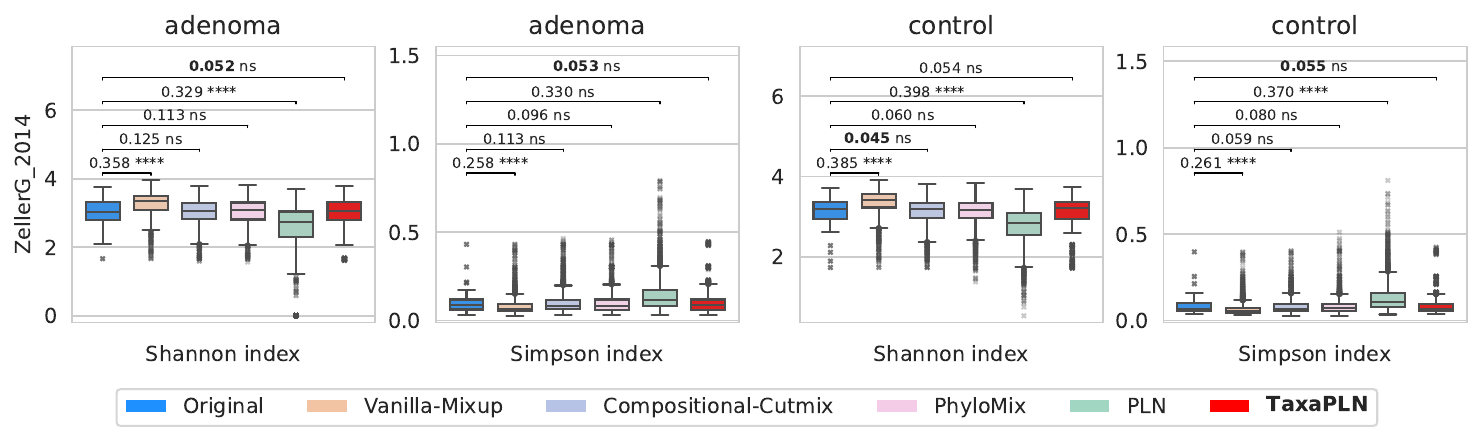}
    \end{subfigure}
    \begin{subfigure}[b]{0.48\linewidth}
        \centering
        \includegraphics[width=\linewidth,trim=0 30 0 0,clip]{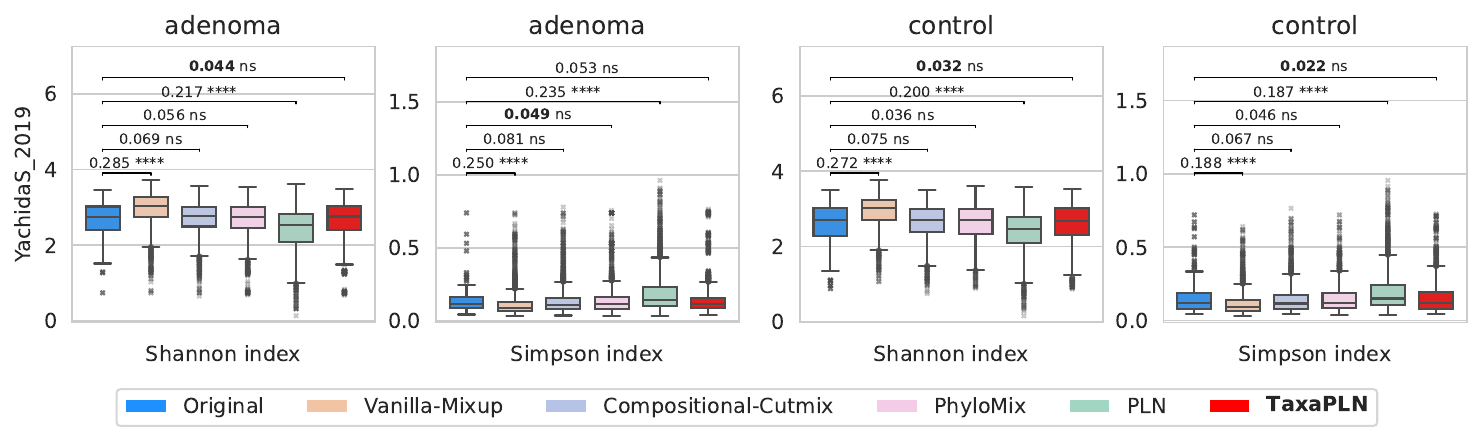}
    \end{subfigure}
    \begin{subfigure}[b]{0.48\linewidth}
        \centering
        \includegraphics[width=\linewidth,trim=0 30 0 0,clip]{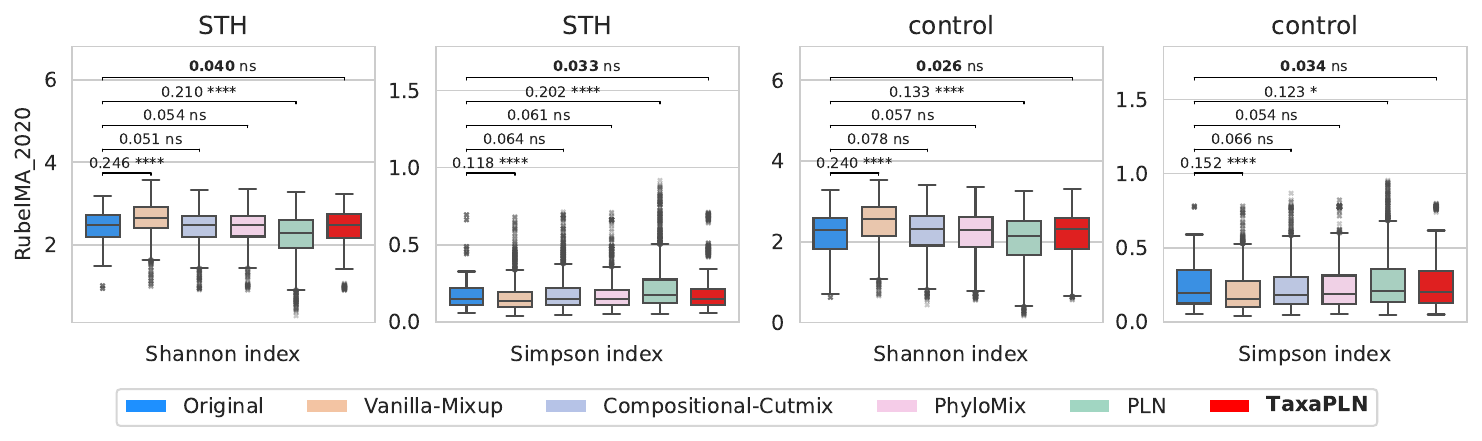}
    \end{subfigure}
    \begin{subfigure}[b]{0.48\linewidth}
        \centering
        \includegraphics[width=\linewidth,trim=0 30 0 0,clip]{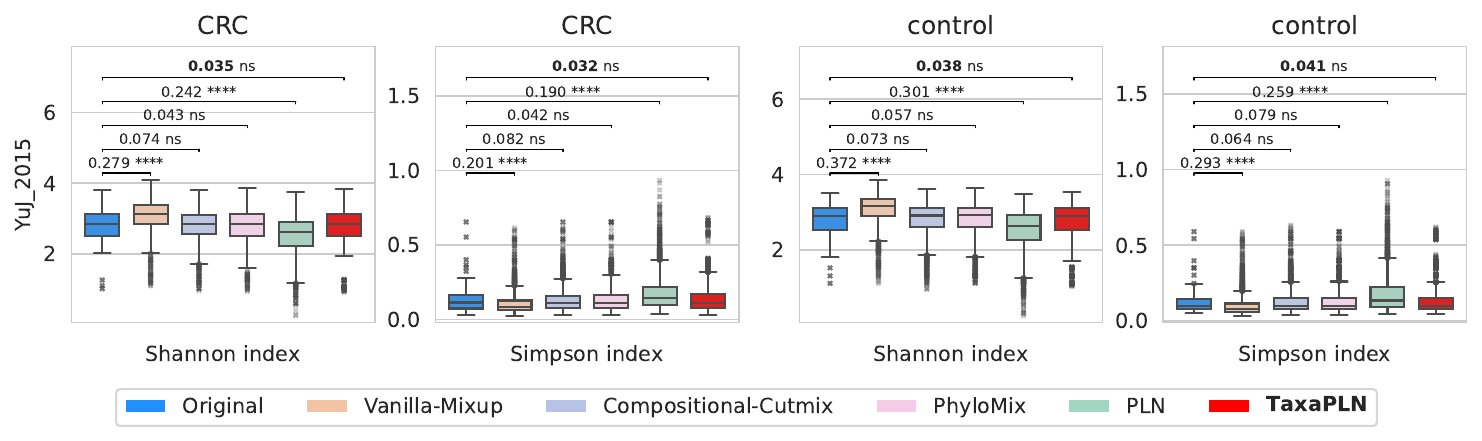}
    \end{subfigure}
    \begin{subfigure}[b]{0.98\linewidth}
        \centering
        \includegraphics[width=\linewidth,trim=0 0 0 0,clip]{alpha_diversity_benchmark_ZhuF_2020_Shannon_index_Simpson_index}
    \end{subfigure}
    \caption{Shannon and Simpson $\alpha$-diversity distributions. Synthetic microbiome samples generated by TaxaPLN and baseline augmentation methods are evaluated on their $\alpha$-diversity consistency with the original microbiomes from Table~\ref{tab:curated_dataset_desc} based on Shannon index and Simpson index. Each method generates $500$ samples. Mann-Whitney U tests are performed to assess the statistical significance of distribution differences between generated and original samples for each method. Significance $P$-values thresholds are denoted by: ****$P \leq 0.0001$; ***$P \leq 0.001$; **$P \leq 0.01$; *$P \leq 0.05$; ns: $P > 0.05$. Kolmogorov-Smirnov divergence between original samples and generated microbiomes is provided with bold value indicating the minimal distance.}
    \label{fig:vanilla_augmentation_alpha_diversity_full}
\end{figure}

\subsection{Beta diversity}
\subsubsection{Aitchison diversity}
\begin{figure}[H]
    \centering
    \begin{subfigure}[b]{0.48\linewidth}
        \centering
        \includegraphics[width=\linewidth,trim=0 40 0 0,clip]{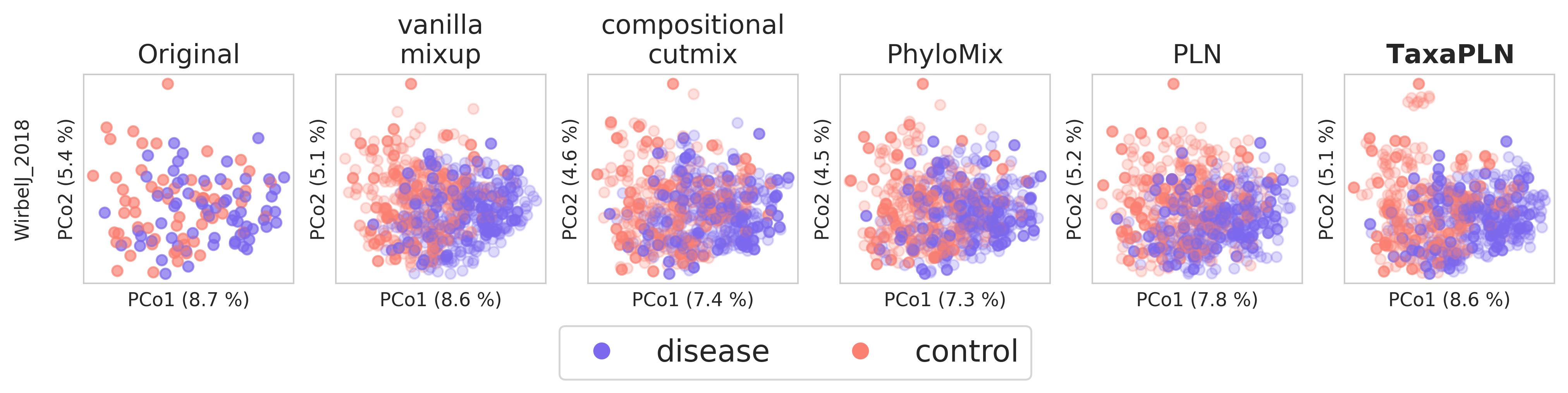}
    \end{subfigure}
    \begin{subfigure}[b]{0.48\linewidth}
        \centering
        \includegraphics[width=\linewidth,trim=0 40 0 0,clip]{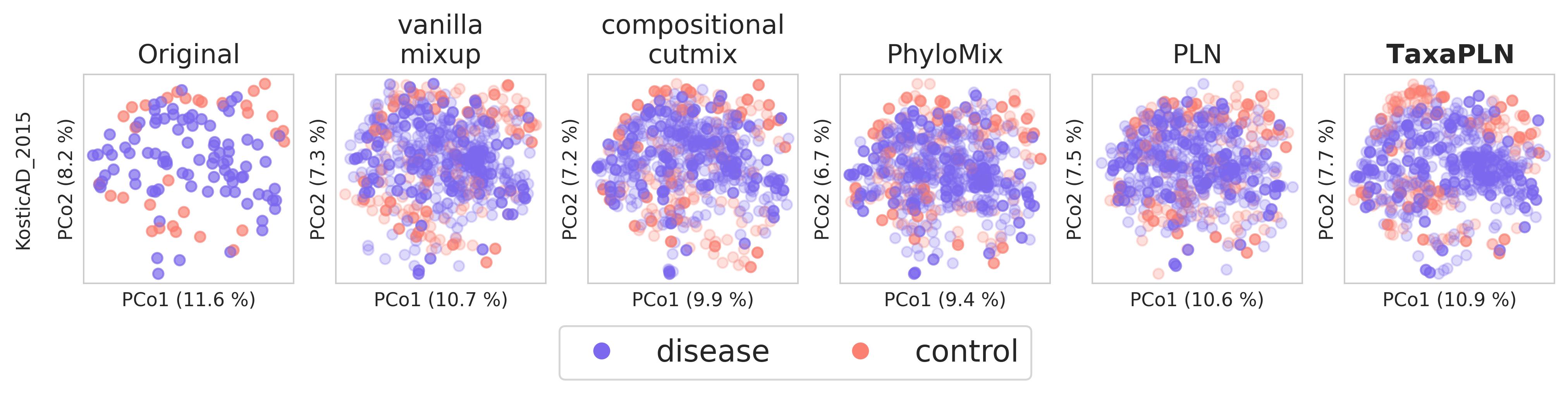}
    \end{subfigure}
    \begin{subfigure}[b]{0.48\linewidth}
        \centering
        \includegraphics[width=\linewidth,trim=0 40 0 0,clip]{aitchison-PCoA-HMP_2019_ibdmdb}
    \end{subfigure}
    \begin{subfigure}[b]{0.48\linewidth}
        \centering
        \includegraphics[width=\linewidth,trim=0 40 0 0,clip]{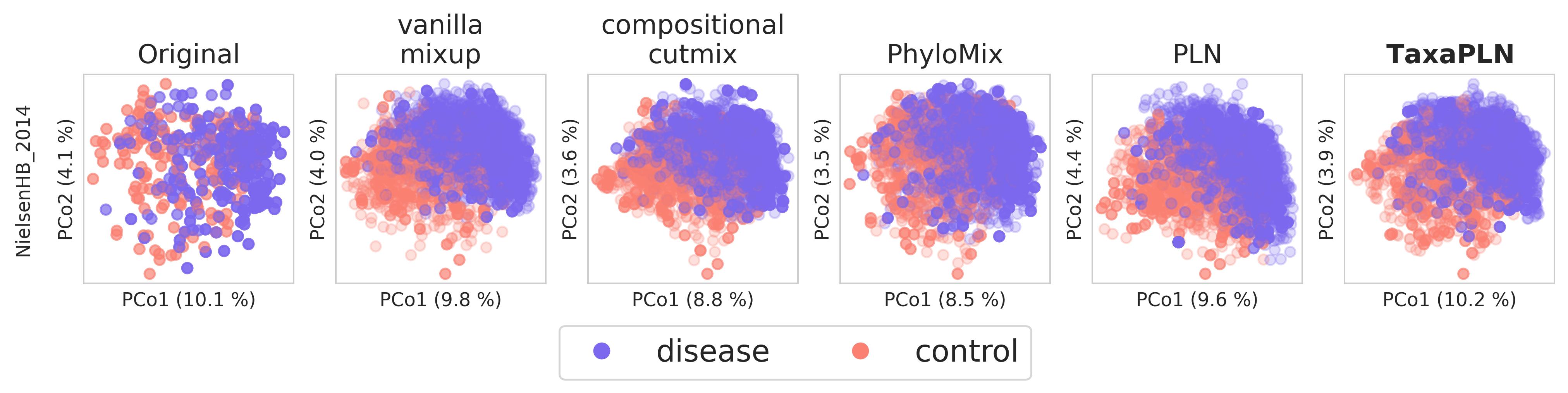}
    \end{subfigure}
    \begin{subfigure}[b]{0.48\linewidth}
        \centering
        \includegraphics[width=\linewidth,trim=0 40 0 0,clip]{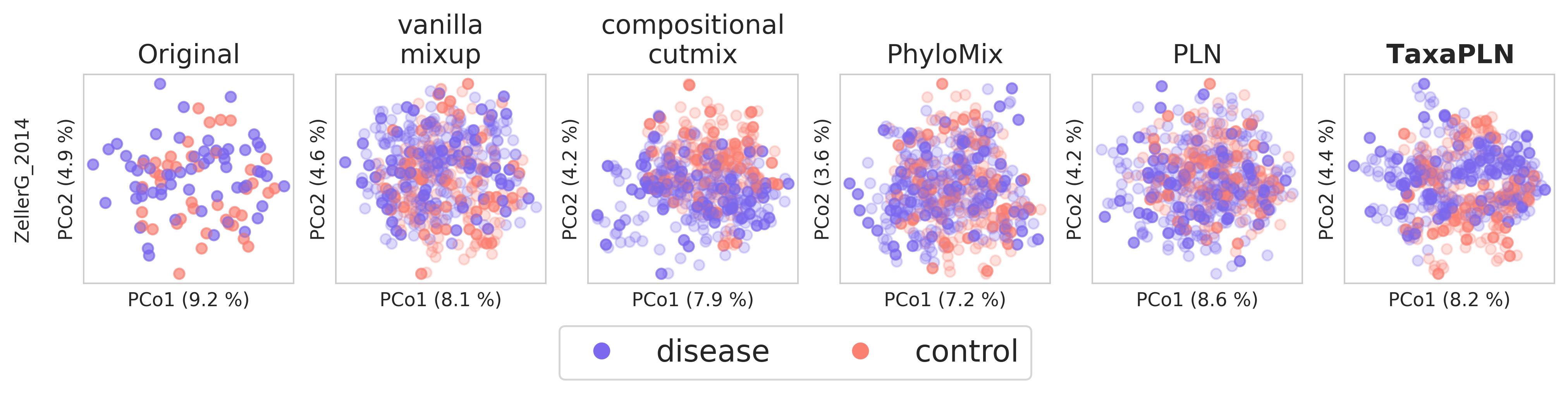}
    \end{subfigure}
    \begin{subfigure}[b]{0.48\linewidth}
        \centering
        \includegraphics[width=\linewidth,trim=0 40 0 0,clip]{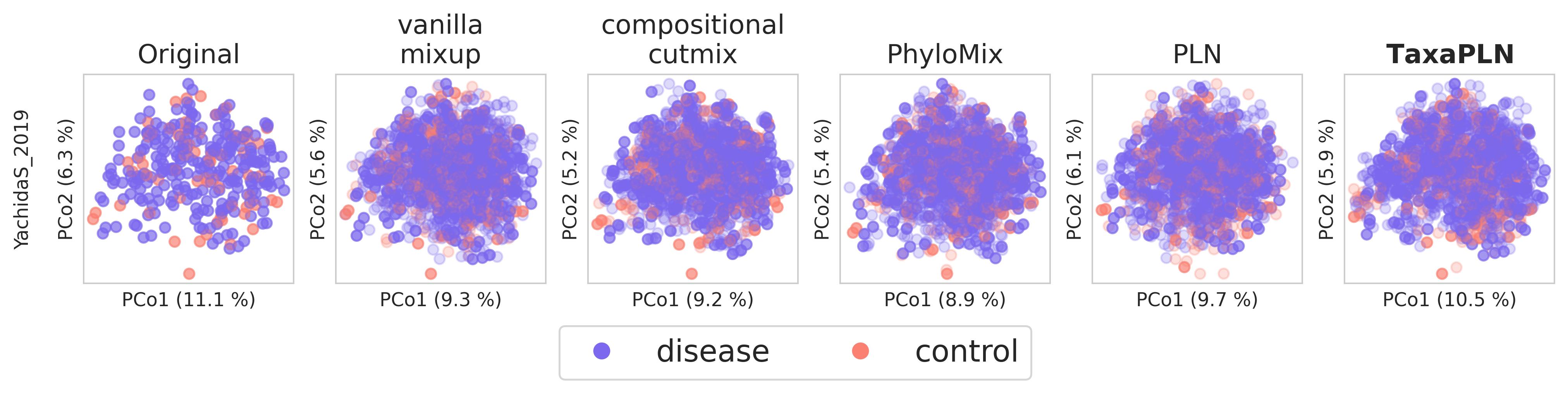}
    \end{subfigure}
    \begin{subfigure}[b]{0.48\linewidth}
        \centering
        \includegraphics[width=\linewidth,trim=0 40 0 0,clip]{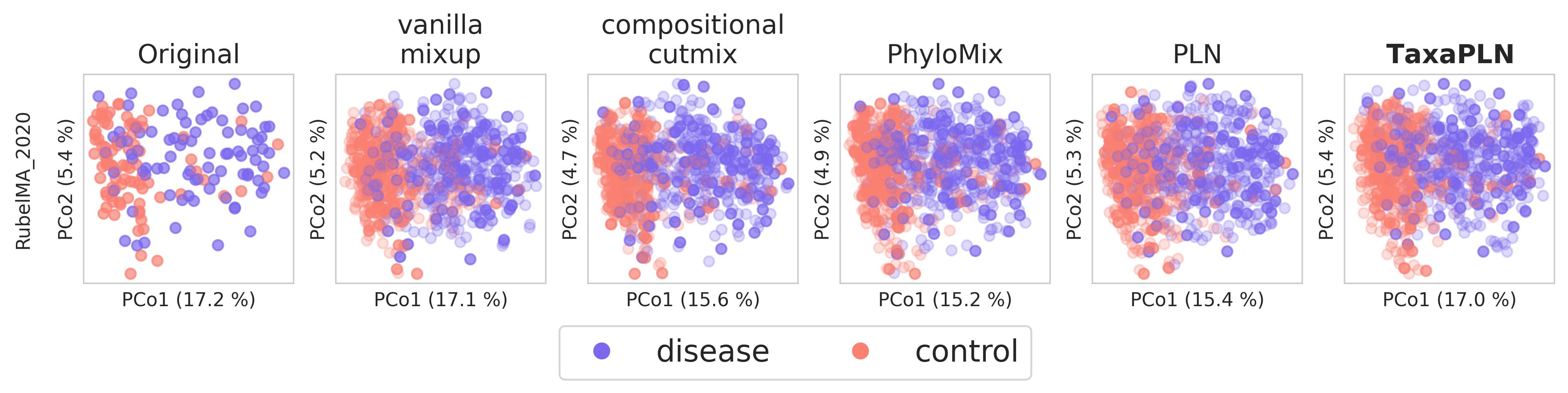}
    \end{subfigure}
    \begin{subfigure}[b]{0.48\linewidth}
        \centering
        \includegraphics[width=\linewidth,trim=0 40 0 0,clip]{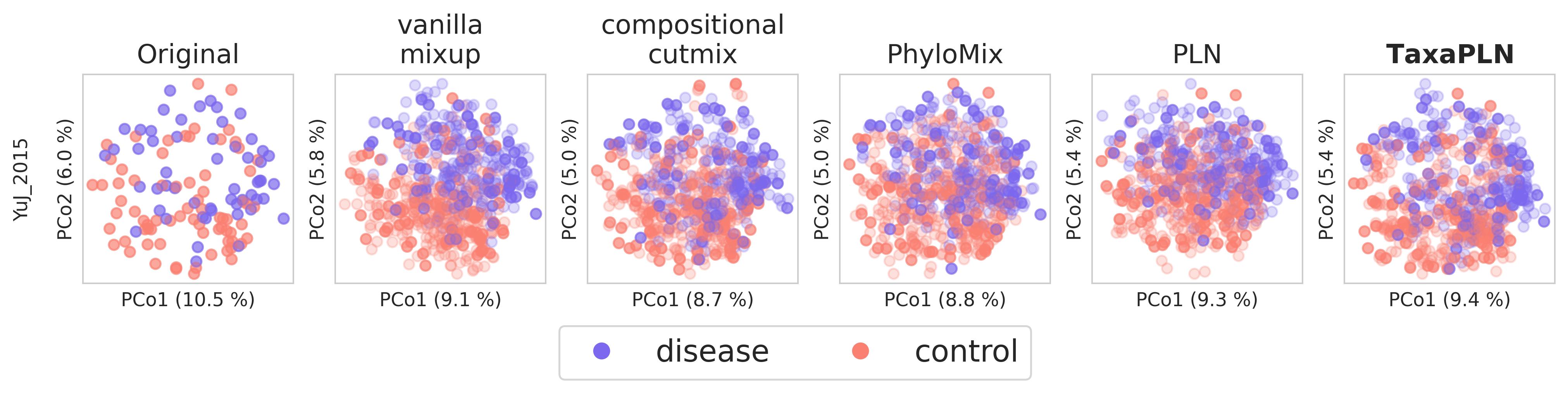}
    \end{subfigure}
    \begin{subfigure}[b]{0.98\linewidth}
        \centering
        \includegraphics[width=\linewidth,trim=0 0 0 0,clip]{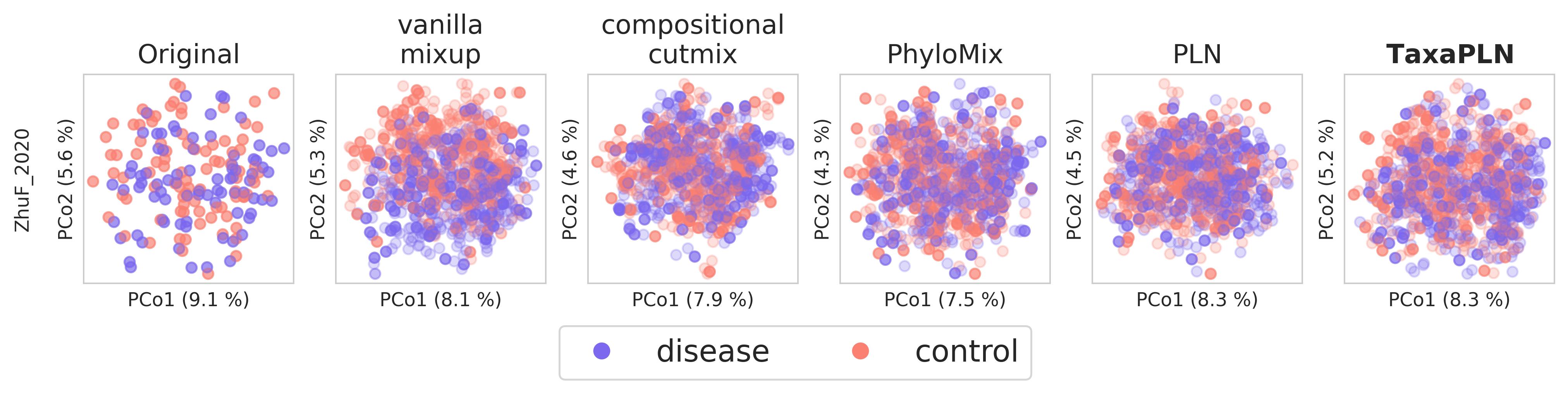}
    \end{subfigure}
    \caption{Aitchison $\beta$-diversity PCoA plots. Synthetic microbiome samples generated by TaxaPLN and baseline augmentation methods are evaluated on their Aitchison $\beta$-diversity consistency with the original microbiomes from Table~\ref{tab:curated_dataset_desc}. Each method augment the training fold with $\beta=2$. Aitchison distance is computed as the Euclidean distance of CLR-transformed counts. Principal Coordinates Analysis (PCoA) is used to visualize the dissimilarity structure.}
    \label{fig:aitchison_vanilla_augmentation_beta_diversity_full}
\end{figure}
\subsubsection{Bray-Curtis diversity}
\begin{figure}[H]
    \centering
    \begin{subfigure}[b]{0.48\linewidth}
        \centering
        \includegraphics[width=\linewidth,trim=0 40 0 0,clip]{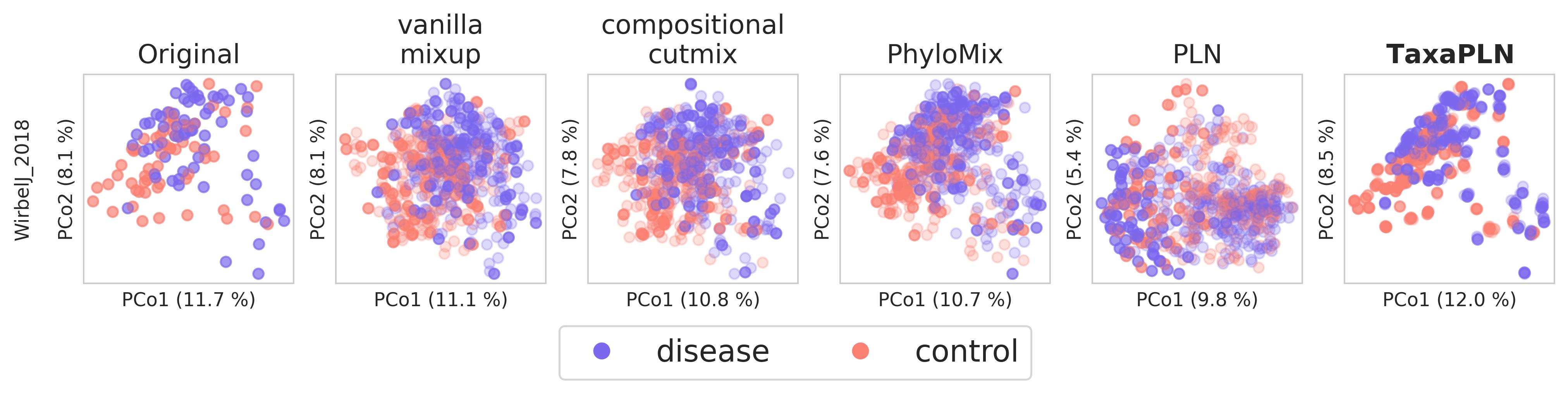}
    \end{subfigure}
    \begin{subfigure}[b]{0.48\linewidth}
        \centering
        \includegraphics[width=\linewidth,trim=0 40 0 0,clip]{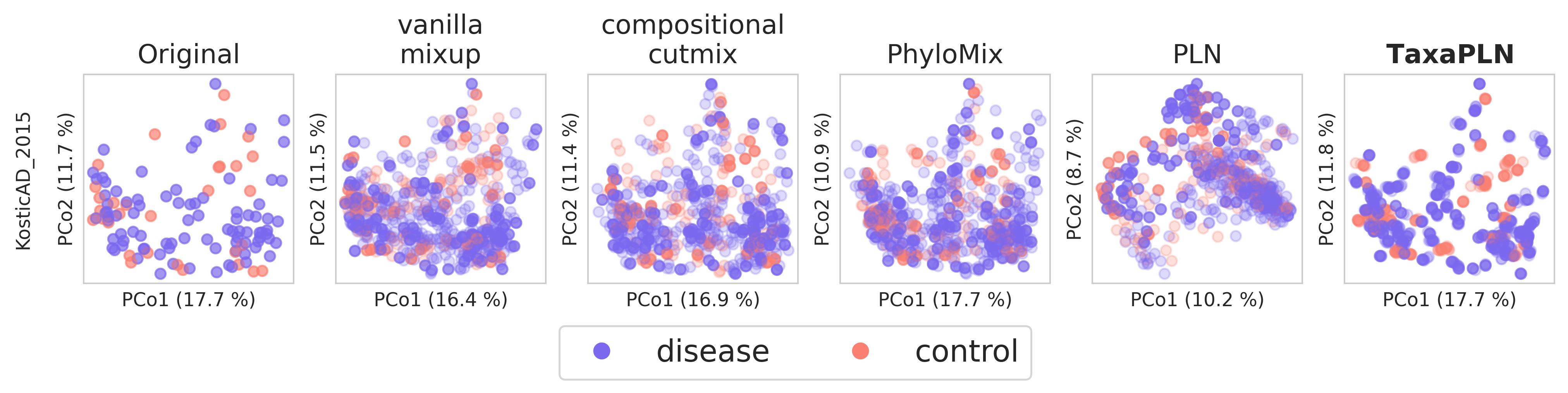}
    \end{subfigure}
    \begin{subfigure}[b]{0.48\linewidth}
        \centering
        \includegraphics[width=\linewidth,trim=0 40 0 0,clip]{braycurtis-PCoA-HMP_2019_ibdmdb}
    \end{subfigure}
    \begin{subfigure}[b]{0.48\linewidth}
        \centering
        \includegraphics[width=\linewidth,trim=0 40 0 0,clip]{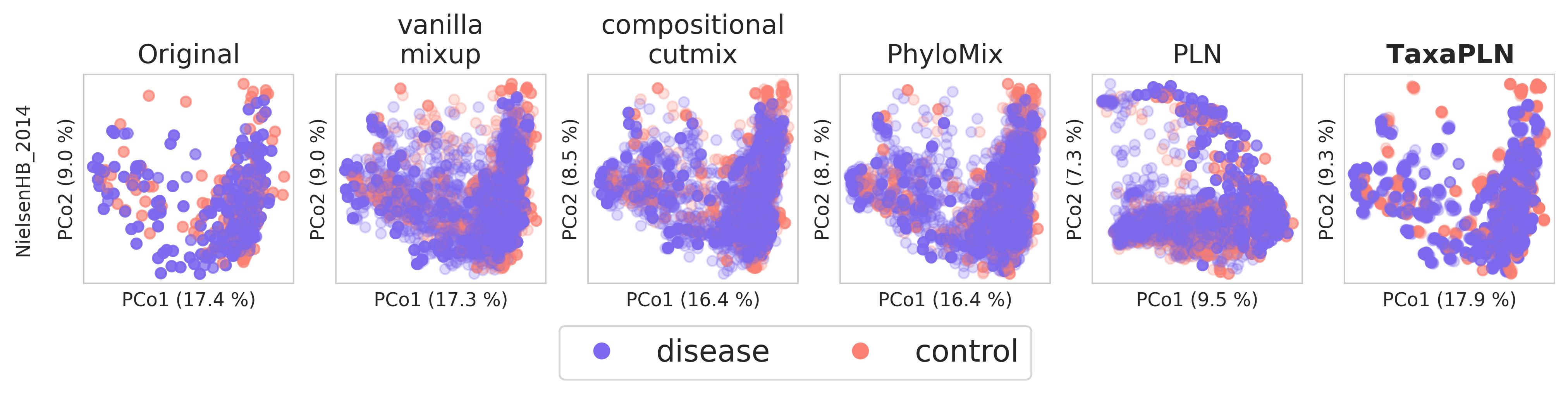}
    \end{subfigure}
    \begin{subfigure}[b]{0.48\linewidth}
        \centering
        \includegraphics[width=\linewidth,trim=0 40 0 0,clip]{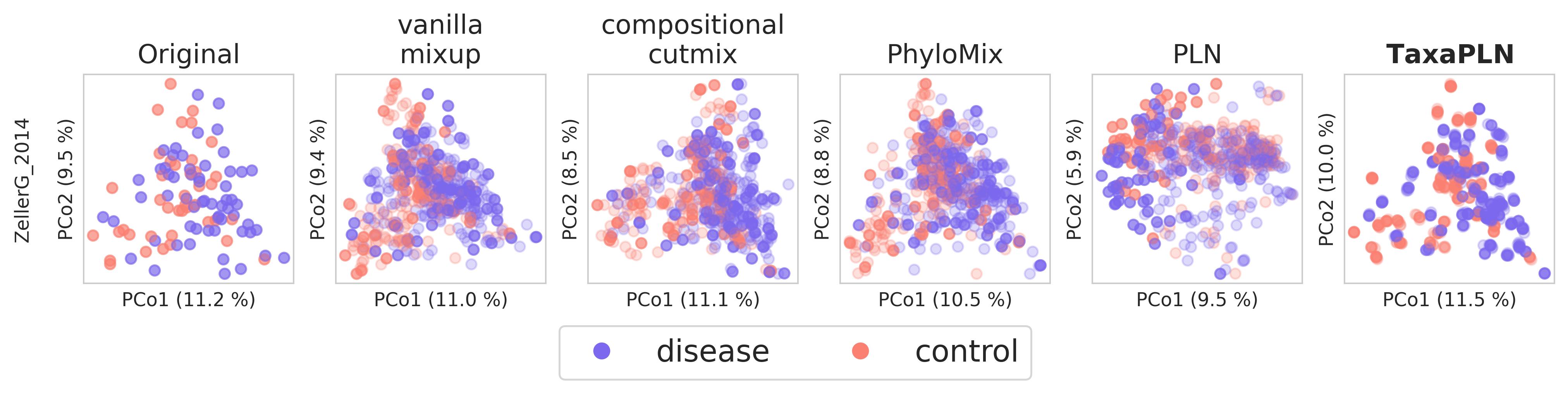}
    \end{subfigure}
    \begin{subfigure}[b]{0.48\linewidth}
        \centering
        \includegraphics[width=\linewidth,trim=0 40 0 0,clip]{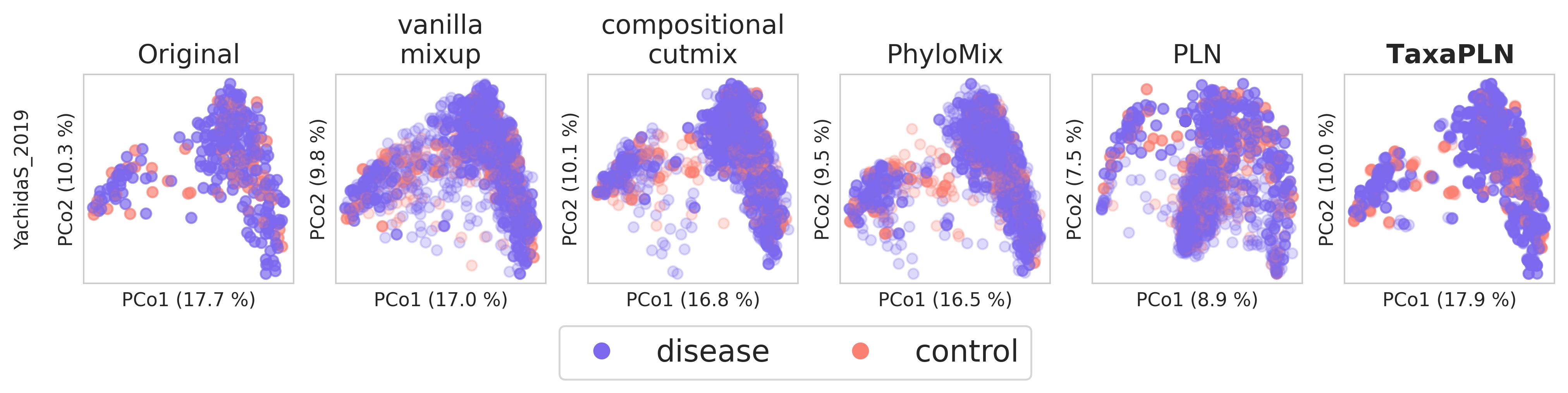}
    \end{subfigure}
    \begin{subfigure}[b]{0.48\linewidth}
        \centering
        \includegraphics[width=\linewidth,trim=0 40 0 0,clip]{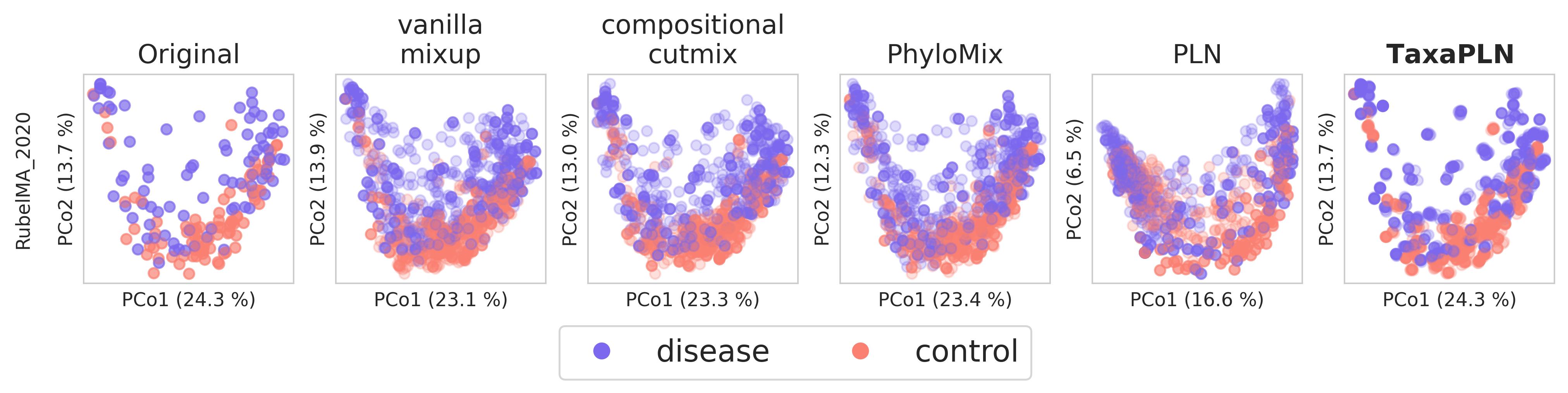}
    \end{subfigure}
    \begin{subfigure}[b]{0.48\linewidth}
        \centering
        \includegraphics[width=\linewidth,trim=0 40 0 0,clip]{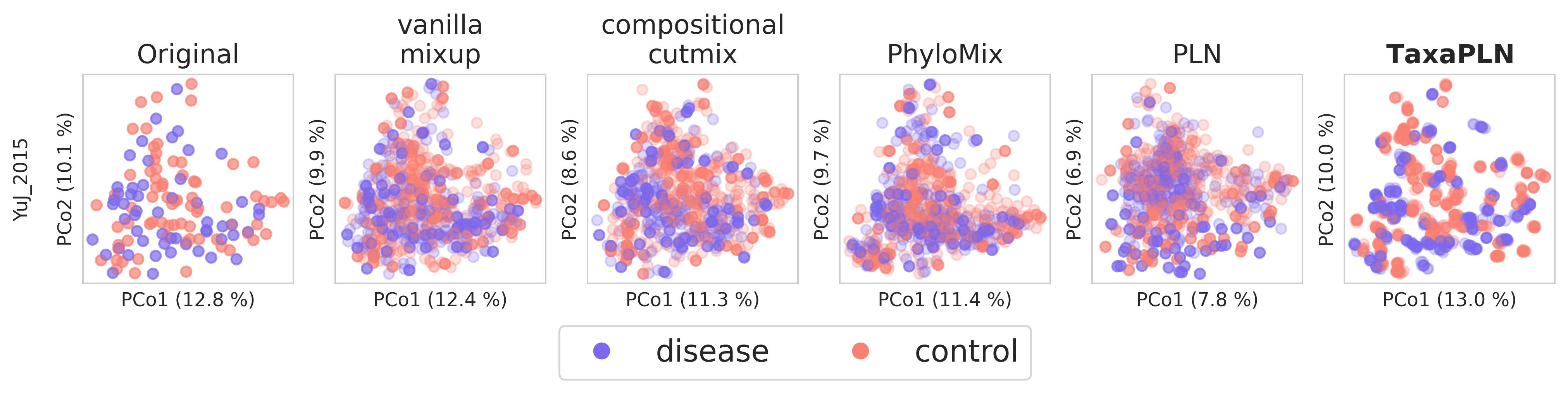}
    \end{subfigure}
    \begin{subfigure}[b]{0.98\linewidth}
        \centering
        \includegraphics[width=\linewidth,trim=0 0 0 0,clip]{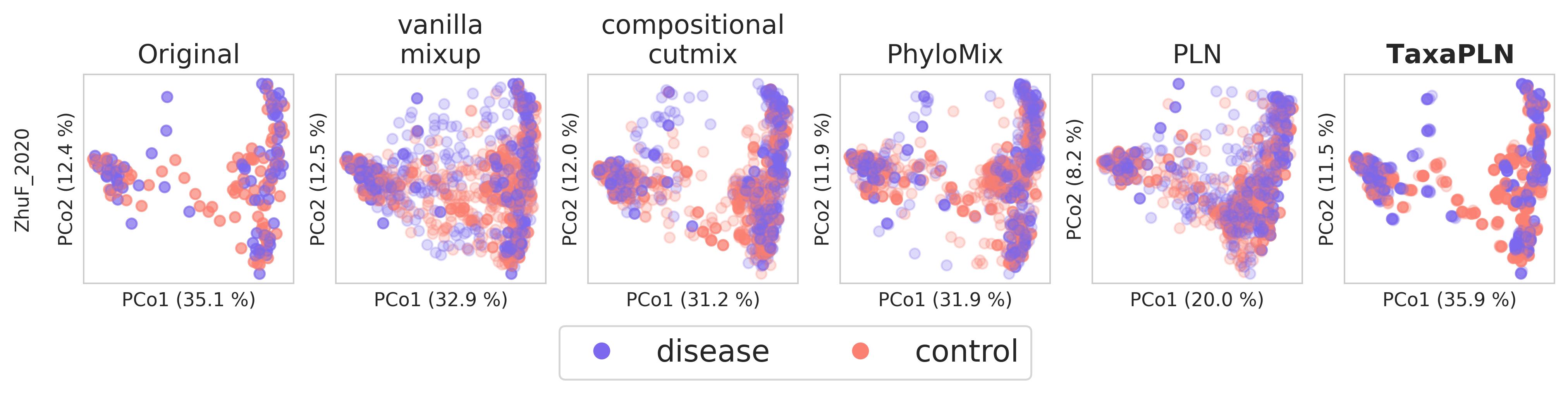}
    \end{subfigure}
    \caption{Bray-Curtis $\beta$-diversity PCoA plots. Synthetic microbiome samples generated by TaxaPLN and baseline augmentation methods are evaluated on their Bray-Curtis $\beta$-diversity consistency with the original microbiomes from Table~\ref{tab:curated_dataset_desc}. Each method augment the training fold with $\beta=2$. Bray-Curtis dissimilarity is computed on normalized counts. Principal Coordinates Analysis (PCoA) is used to visualize the dissimilarity structure.}
    \label{fig:braycurtis_vanilla_augmentation_beta_diversity_full}
\end{figure}

\clearpage
\section{Augmentation with relative abundance preprocessing}
\label{app:relative_preprocessing}
In our core experiments, data augmentation is followed by classification on microbiome compositions preprocessed using the centered log-ratio (CLR) transform \citep{aitchison1982statistical}. This preprocessing is widely employed to address the compositional nature of microbiome data which raises challenges for machine learning models. In particular, the compositional constraints introduce identifiability issues for parametric models such as logistic regression and multilayer perceptrons (MLP), which can arbitrarily hinder performances and preclude reliable downstream analysis \citep{d2022underspecification}, motivating the development of composition-aware strategies \citep{aitchison1982statistical,shi2016regression,jiang2025modeling}.

\begin{figure}[H]
    \centering
    \begin{subfigure}[b]{0.98\linewidth}
        \centering
        \includegraphics[width=\linewidth,trim=0 35 0 0,clip]{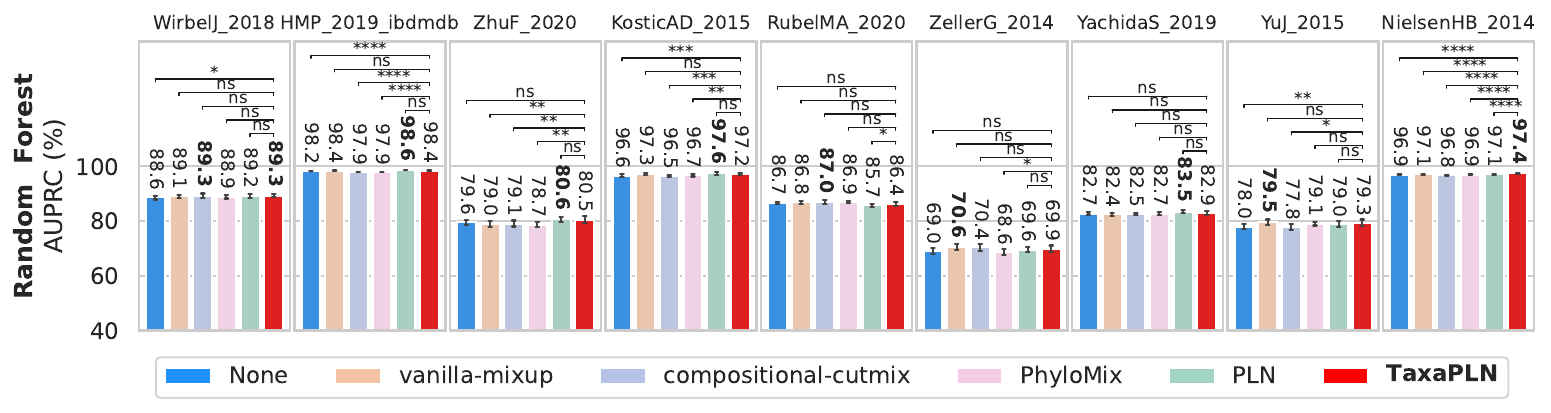}
    \end{subfigure}
    \begin{subfigure}[b]{0.98\linewidth}
        \centering
        \includegraphics[width=\linewidth,trim=0 35 0 20,clip]{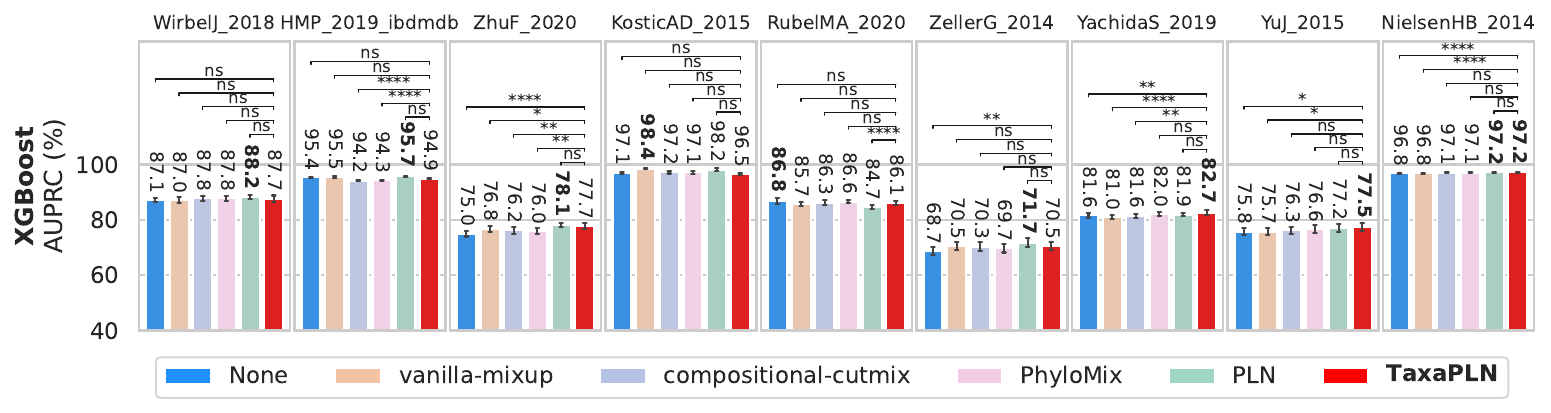}
    \end{subfigure}
    \begin{subfigure}[b]{0.98\linewidth}
        \centering
        \includegraphics[width=\linewidth,trim=0 0 0 20,clip]{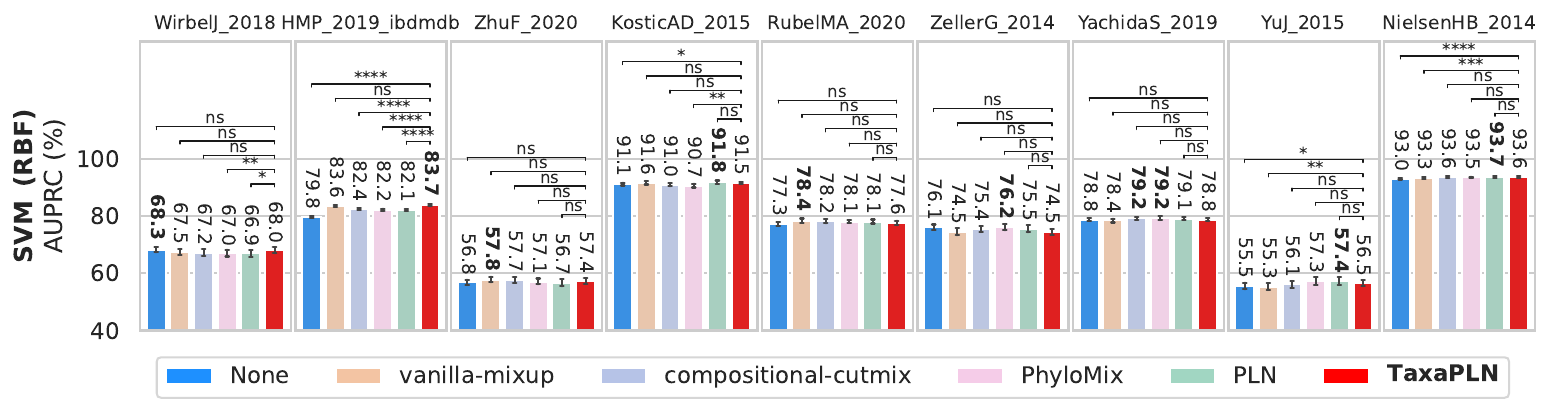}
    \end{subfigure}
    \caption{Vanilla augmentation performances on the supervised learning tasks using relative count preprocessing. The evaluation was performed on real microbiome datasets from Table \ref{tab:curated_dataset_desc} sequenced at the \textit{species} level. TaxaPLN augmentation is evaluated with five classifiers and compared against four baseline methods according to Section~\ref{sec:vanilla_augmentation}. Performance comparison between TaxaPLN and other methods is assessed using one-tailed paired $t$-tests with significance: ****$P$-value $\leq 0.0001$; ***$P$-value $\leq 0.001$; **$P$-value $\leq 0.01$; *$P$-value $\leq 0.05$; ns: $P$-value $>$ 0.05.}
    \label{fig:vanilla_augmentation_performances_relativepreproc}
\end{figure}

Despite these challenges, relative abundance representations remain prevalent in microbiome analysis, particularly when paired with non-parametric models like Random Forests, XGBoost, and SVM with RBF kernels, where they often yield competitive performance compared to CLR-transformed data. To assess the robustness of TaxaPLN across preprocessing strategies, we evaluate its augmentation performance using classifiers trained on relative abundance inputs, as reported in Figure~\ref{fig:vanilla_augmentation_performances_relativepreproc}. The parameters of the classifiers are provided in Table~\ref{tab:classifiers_parameterization}, while we follow the same 25$\times$5-Fold cross-validation protocol with augmentation ratio $\beta = 2$ as in Section~\ref{sec:training_settings}.

Results with relative abundance preprocessing are generally consistent with those obtained using CLR-transformed data. TaxaPLN achieves maximum AUPRC gains of $1.3\%$, $2.6\%$, and $4.0\%$ with Random Forest, XGBoost, and RBF SVM respectively. In comparison, PhyloMix yields maximum gains of $1.1\%$, $2.4\%$, and $1.0\%$ on the same models. On average, TaxaPLN ranks first with both Random Forest and RBF SVM, and second on XGBoost behind PLN, significantly outperforming PhyloMix in one-third of the experiments, according to paired $t$-tests.
Interestingly, the PLN baseline performs better under relative abundance preprocessing than in CLR-based experiments, narrowing the performance gap with TaxaPLN. However, PLN remains limited as a data augmentation method, since it does not preserve key biological characteristics of the data.

\section{Conditional TaxaPLN models}
\subsection{Covariate-aware PLN-Tree architecture with FiLM}
While TaxaPLN builds on the PLN-Tree framework to model microbiome count data, the original architecture does not support the incorporation of covariates. Specifically, \cite{chaussard2025tree} only define the objective for training a conditional PLN-Tree model and suggest using attention-based mechanisms to handle the heterogeneity arising from combining covariates with microbiome compositions. Although attention mechanisms offer high expressivity, they are often difficult to tune and require large datasets to achieve stable and meaningful representations \citep{vaswani2017attention}.

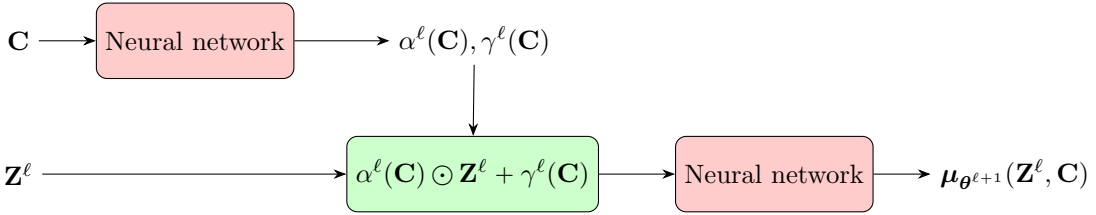
\begin{figure}[H]
\centering
    \begin{tikzpicture}[
        >=Stealth, 
        node distance=2cm,
        rnn/.style={draw, rectangle, rounded corners, fill=blue!20, minimum width=2cm, minimum height=1cm, align=center},
        ffnn/.style={draw, rectangle, rounded corners, fill=red!20, minimum width=2cm, minimum height=1cm, align=center},
        module/.style={draw, rectangle, rounded corners, fill=green!20, minimum width=2cm, minimum height=1cm, align=center},
        concat/.style={draw, circle, inner sep=0pt, minimum size=0.8cm, fill=gray!20}
    ]

    \node (start) {};
    \node[above=0cm of start] (cov) {$\bC$};
    \node[below=1cm  of start] (Z) {$\bZ^{\ell}$};

    \node[ffnn, right=0.75cm of cov] (NetFiLM) {Neural network};
    \node[right=1.25cm of NetFiLM] (modulators) {$\alpha^\ell(\bC), \gamma^\ell(\bC)$};
    \node[module, right=4cm of Z] (moduling) {$\alpha^\ell(\bC) \odot \bZ^\ell + \gamma^\ell(\bC)$};
    
    \node[ffnn, right=1.0cm of moduling] (FFNN) {Neural network};
    \node[right=0.75cm of FFNN] (output) {$\bmu_{\btheta^{\ell+1}}(\bZ^{\ell}, \bC)$};

    \draw[->] (Z) -- (moduling);
    \draw[->] (cov) -- (NetFiLM);
    \draw[->] (NetFiLM) -- (modulators);
    \draw[->] (modulators) -- (moduling);
    \draw[->] (moduling) -- (FFNN);
    \draw[->] (FFNN) -- (output);
    
    \end{tikzpicture}
\caption{Feature-wise Linear Modulation architecture applied to the PLN-Tree Markov dynamic from the prior distribution at level $\ell < L$.}
\label{fig:prior_architecture_film}
\end{figure}

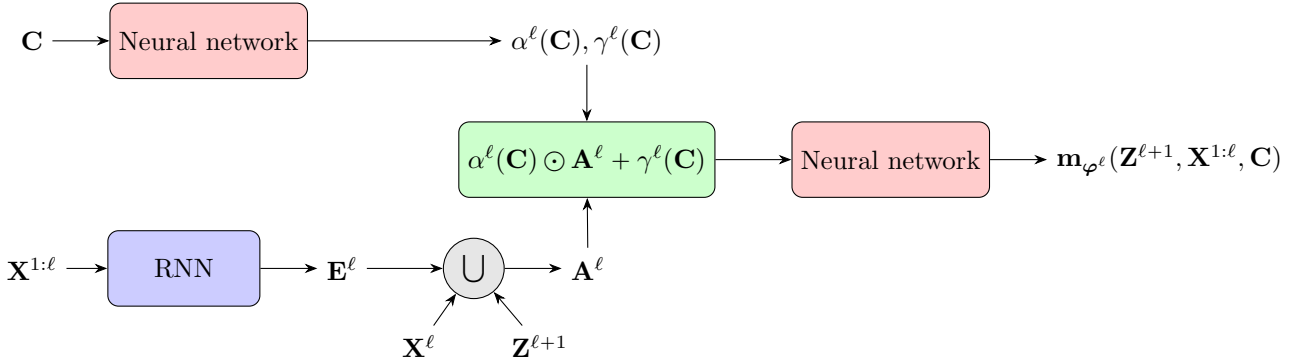
\begin{figure}[H]
\centering
    \begin{tikzpicture}[
        >=Stealth, 
        node distance=2cm,
        rnn/.style={draw, rectangle, rounded corners, fill=blue!20, minimum width=2cm, minimum height=1cm, align=center},
        ffnn/.style={draw, rectangle, rounded corners, fill=red!20, minimum width=2cm, minimum height=1cm, align=center},
        module/.style={draw, rectangle, rounded corners, fill=green!20, minimum width=2cm, minimum height=1cm, align=center},
        concat/.style={draw, circle, inner sep=0pt, minimum size=0.8cm, fill=gray!20}
    ]

    \node (start) {};
    \node[above=1.25cm of start] (cov) {$\bC$};
    \node[below=1cm  of start] (X) {$\bX^{1:\ell}$};

    \node[ffnn, right=0.75cm of cov] (NetFiLM) {Neural network};
    \node[right=2.55cm of NetFiLM] (modulators) {$\alpha^\ell(\bC), \gamma^\ell(\bC)$};
    \node[module, below=0.75cm of modulators] (moduling) {$\alpha^\ell(\bC) \odot \bA^\ell + \gamma^\ell(\bC)$};
    
    \node[rnn, right of=X] (RNN) {RNN};
    \node[right=0.75cm of RNN] (E) {$\bE^{\ell}$};
    
    \node[concat, right=1cm of E] (C) {$\bigcup$};
    \node[right=0.75cm of C] (activation) {$\bA^\ell$};
    
    \node[ffnn, right=1.0cm of moduling] (FFNN) {Neural network};
    \node[right=0.75cm of FFNN] (Z) {$\BF{m}_{\bphi^\ell}(\bZ^{\ell+1}, \bX^{1:\ell}, \bC)$};

    \node[below=0.5cm of C] (residual) {};
    \node[left=0.25cm of residual] (Res) {$\bX^{\ell}$};
    \node[right=0.25cm of residual] (Zp1) {$\bZ^{\ell+1}$};
    
    \draw[->] (X) -- (RNN);
    \draw[->] (RNN) -- (E);
    \draw[->] (E) -- (C);
    \draw[->] (cov) -- (NetFiLM);
    \draw[->] (NetFiLM) -- (modulators);
    \draw[->] (modulators) -- (moduling);
    \draw[->] (activation) -- (moduling);
    \draw[->] (Zp1) -- (C);
    \draw[->] (C) -- (activation);
    \draw[->] (moduling) -- (FFNN);
    \draw[->] (Res) -- (C);
    \draw[->] (FFNN) -- (Z);
    
    \end{tikzpicture}
\caption{Feature-wise Linear Modulation applied to the residual amortized backward architecture of PLN-Tree from \cite{chaussard2025tree}, illustrated for the variational mean at layer $\ell \leq L$. The amortizing Recurrent Neural Network is denoted by RNN, while the symbol "$\cup$" indicates a concatenation of entries. We denote the variable $\bX^{1:\ell} = (\bX^1, \dots, \bX^\ell)$ and $\bE^\ell$ the last output of the recurrent network after inputting the sequence $\bX^{1:\ell}$.}
\label{fig:res_backward_architecture_film}
\end{figure}

Given the typically limited sample sizes in microbiome studies, we propose an alternative architecture based on feature-wise linear modulation (FiLM) \citep{perez2018film}. FiLM modules are lightweight in terms of parameters and have shown competitive performance compared to attention-based models in various tasks \citep{perez2018film,dumoulin2018feature}, making them a compelling choice for modeling covariate–microbiome interactions in limited data settings.
Figures~\ref{fig:prior_architecture_film} and~\ref{fig:res_backward_architecture_film} illustrate how FiLM blocks are integrated into the PLN-Tree architecture, following the implementation of \cite{chaussard2025tree}.

\subsection{CuratedMetagenomicData available covariates summary}
The \texttt{curatedMetagenomicData} package provides a rich set of metadata associated with each patient across studies \citep{curated}. For our covariate-aware experiments, we focus on a subset of exogenous data that are both commonly available across the selected studies and known to influence microbial composition. To facilitate their integration into neural network architectures, each covariate is preprocessed according to its type, as summarized in Table~\ref{tab:covariates}.
\begin{table*}[htbp]
    \centering
    \begin{tabular}{lcccc}
    \toprule
    {\textbf{Covariate}} & \textbf{Type} & \textbf{Preprocessing}  \\
    \midrule 
    age & categorical & ordinal encoding \\
    sex & categorical & binary encoding \\
    BMI & numerical & min-max scaling \\
    country & categorical & one-hot encoding \\
    \bottomrule
    \end{tabular}
    \caption{Covariates summary and preprocessing applied. BMI: Body Mass Index.}
    \label{tab:covariates}
\end{table*}

\subsection{Diversity benchmark with conditional models}
\label{app:conditional_diversity_benchmark}
\begin{figure}[H]
    \centering
    \begin{subfigure}[b]{0.6\linewidth}
        \centering
        \includegraphics[width=\linewidth,trim=0 30 0 0,clip]{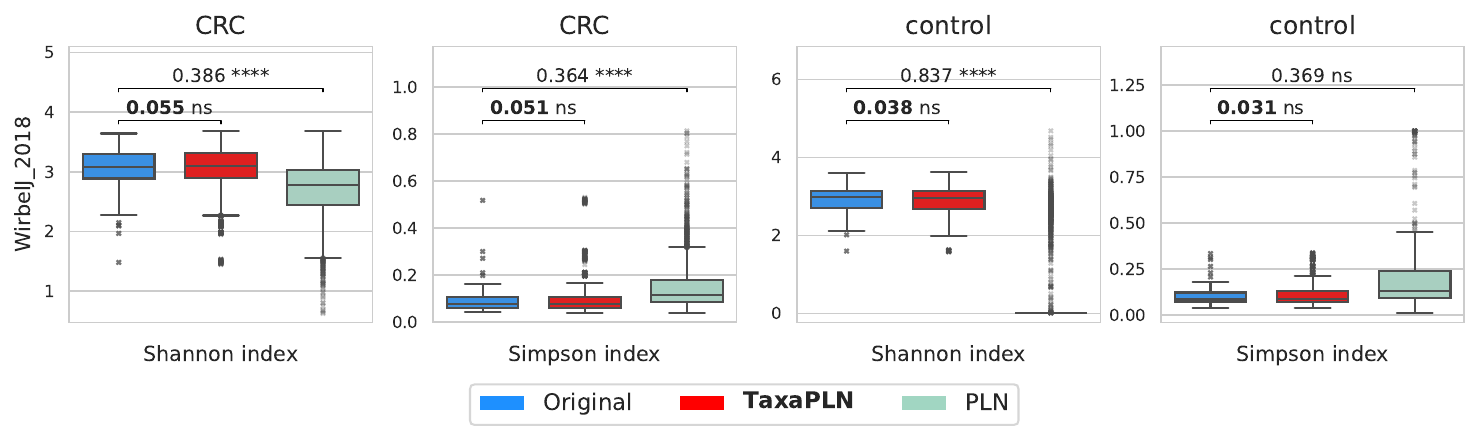}
    \end{subfigure}
    \begin{subfigure}[b]{0.6\linewidth}
        \centering
        \includegraphics[width=\linewidth,trim=0 30 0 0,clip]{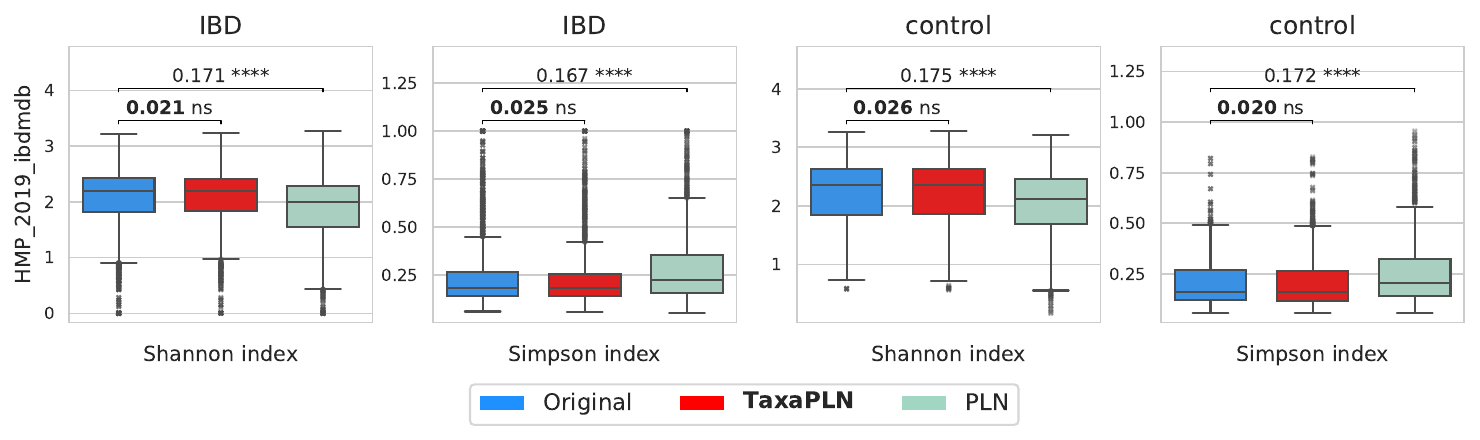}
    \end{subfigure}
    \begin{subfigure}[b]{0.6\linewidth}
        \centering
        \includegraphics[width=\linewidth,trim=0 30 0 0,clip]{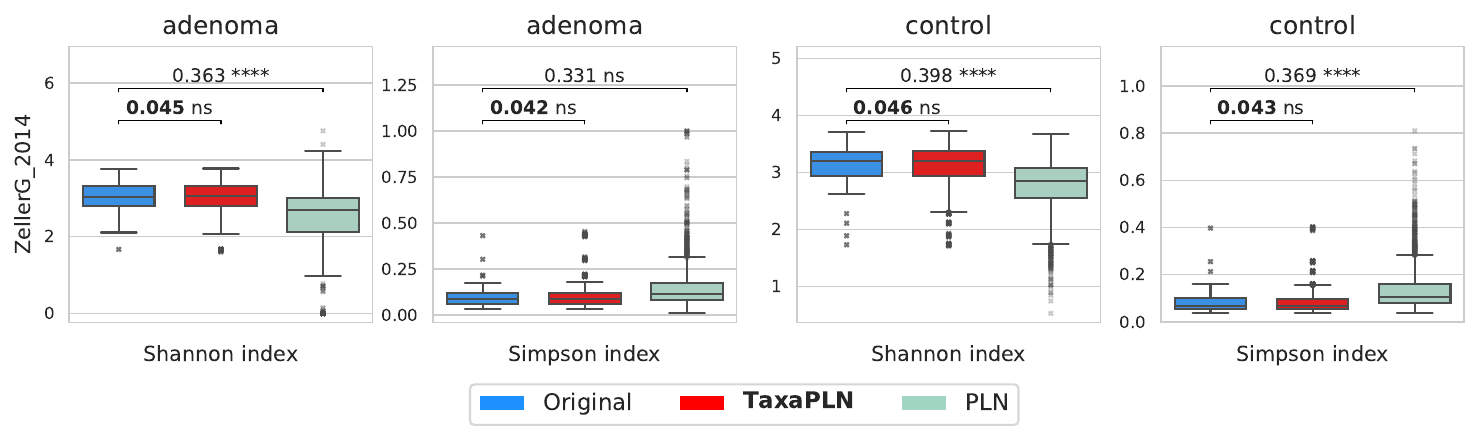}
    \end{subfigure}
    \begin{subfigure}[b]{0.6\linewidth}
        \centering
        \includegraphics[width=\linewidth,trim=0 30 0 0,clip]{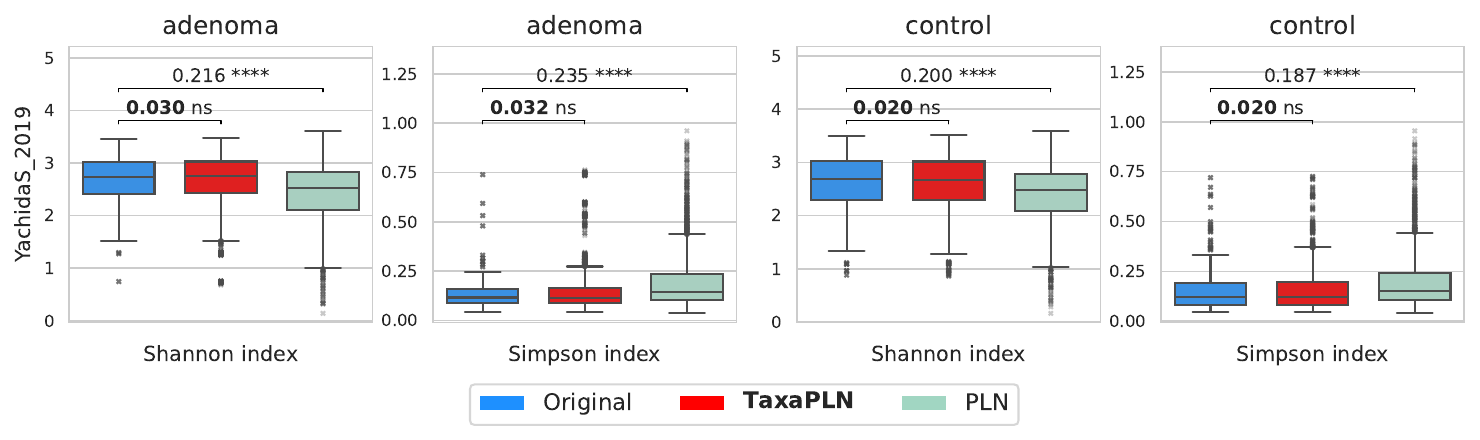}
    \end{subfigure}
    \begin{subfigure}[b]{0.6\linewidth}
        \centering
        \includegraphics[width=\linewidth,trim=0 30 0 0,clip]{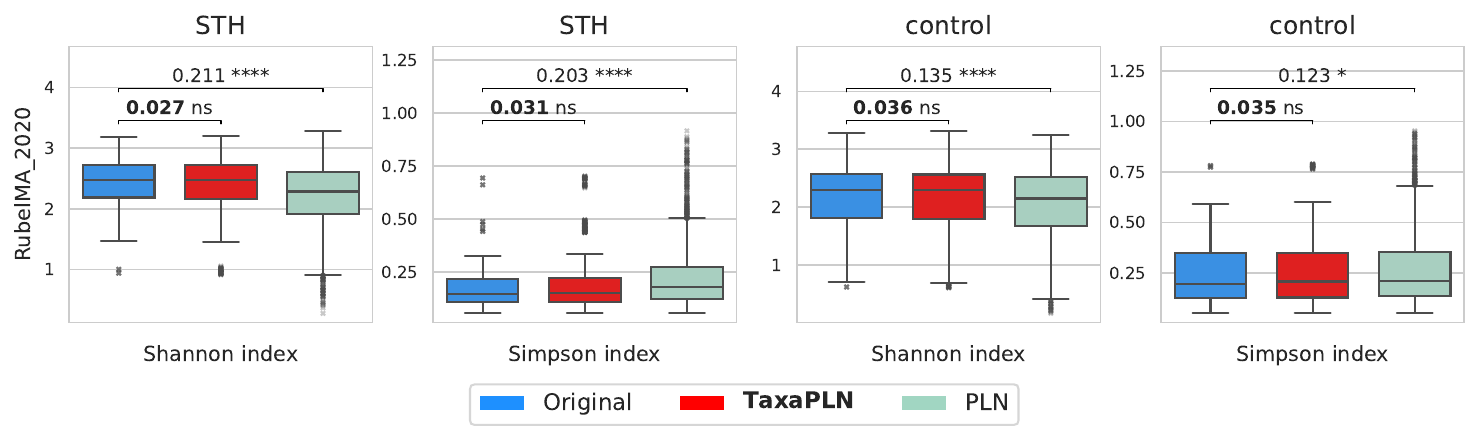}
    \end{subfigure}
    \begin{subfigure}[b]{0.6\linewidth}
        \centering
        \includegraphics[width=\linewidth,trim=0 30 0 0,clip]{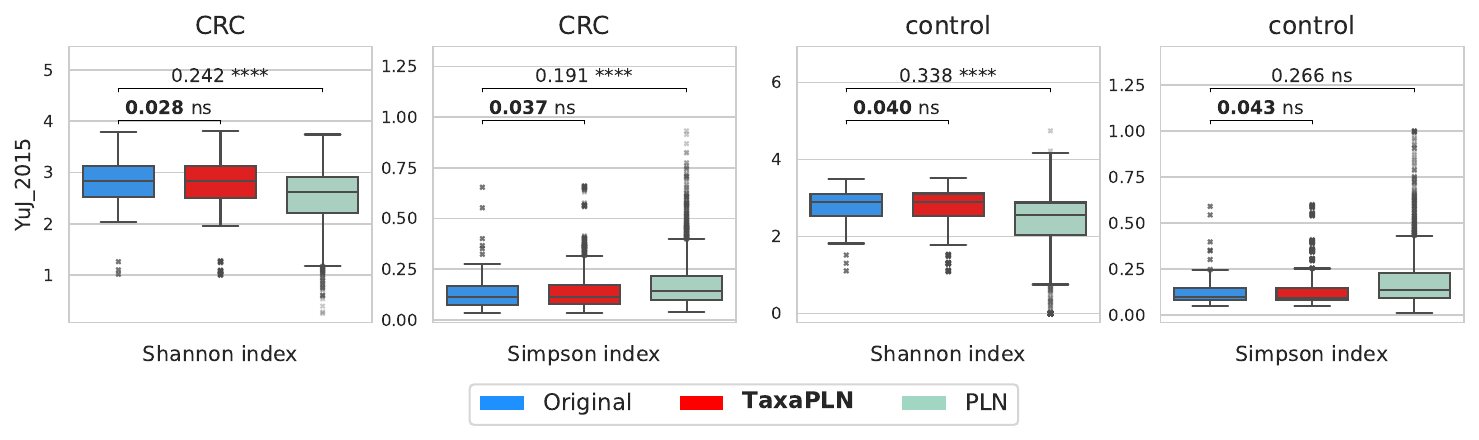}
    \end{subfigure}
    \begin{subfigure}[b]{0.6\linewidth}
        \centering
        \includegraphics[width=\linewidth,trim=0 30 0 0,clip]{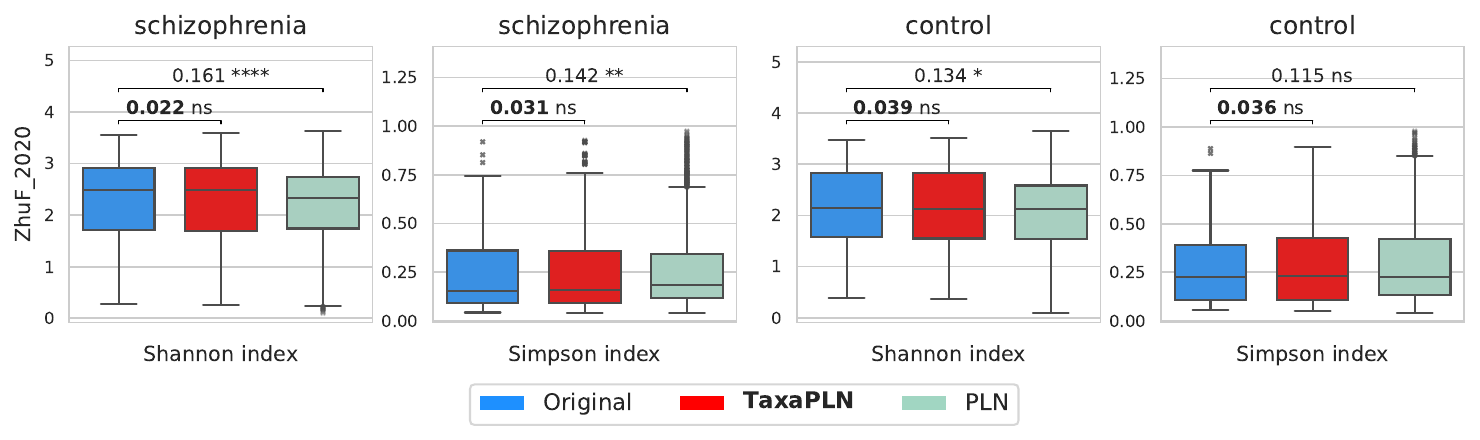}
    \end{subfigure}
    \begin{subfigure}[b]{0.98\linewidth}
        \centering
        \includegraphics[width=\linewidth,trim=0 0 0 180,clip]{conditional_alpha_diversity_benchmark_ZhuF_2020_Shannon_index_Simpson_index}
    \end{subfigure}
    \caption{Shannon and Simpson $\alpha$-diversity distributions using conditional models. Synthetic microbiome samples generated by covariate-aware TaxaPLN and conditional PLN baseline augmentation methods are evaluated on their $\alpha$-diversity consistency with the original microbiomes from Table~\ref{tab:curated_dataset_desc} based on Shannon index and Simpson index. Each method generates $500$ samples. Mann-Whitney U tests are performed to assess the statistical significance of distribution differences between generated and original samples for each method. Significance $P$-values thresholds are denoted by: ****$P \leq 0.0001$; ***$P \leq 0.001$; **$P \leq 0.01$; *$P \leq 0.05$; ns: $P > 0.05$. Kolmogorov-Smirnov divergence between original samples and generated microbiomes is provided with bold value indicating the minimal distance.}
    \label{fig:conditional_augmentation_alpha_diversity_full}
\end{figure}
\begin{figure}[H]
    \centering
    \begin{subfigure}[b]{0.48\linewidth}
        \centering
        \includegraphics[width=\linewidth,trim=0 40 0 0,clip]{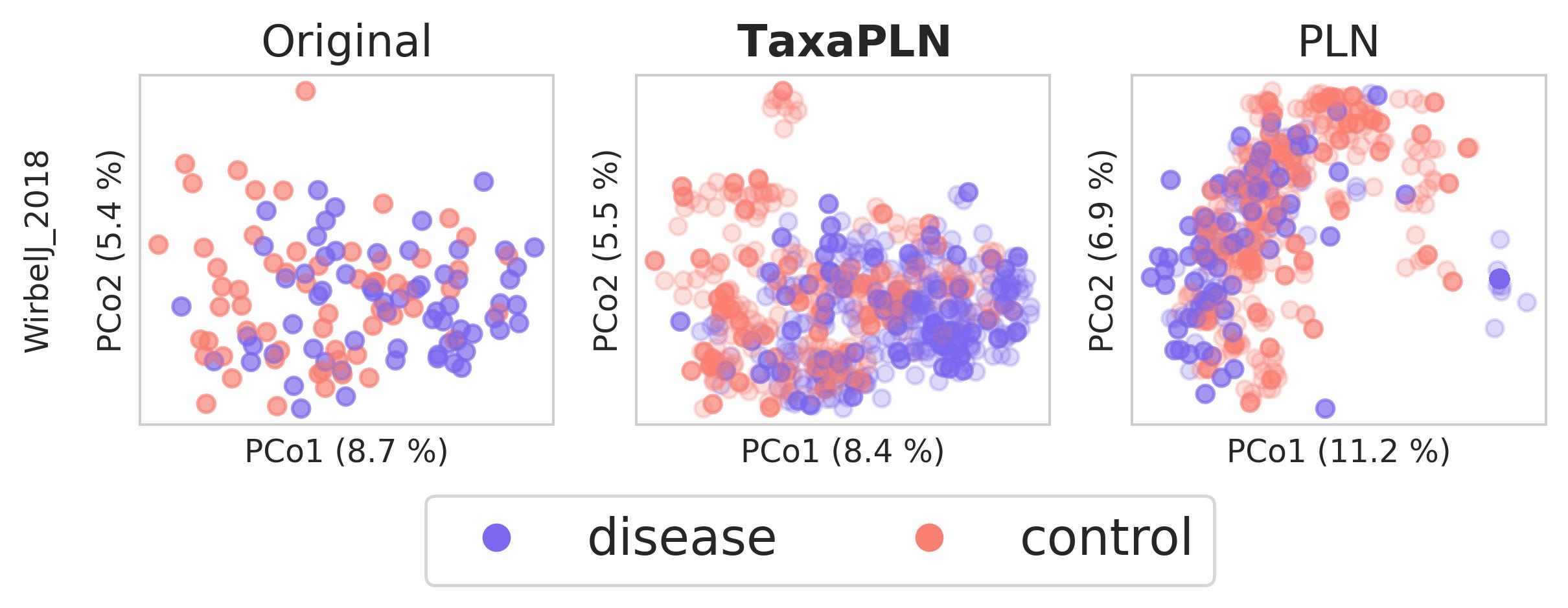}
    \end{subfigure}
    \begin{subfigure}[b]{0.48\linewidth}
        \centering
        \includegraphics[width=\linewidth,trim=0 40 0 0,clip]{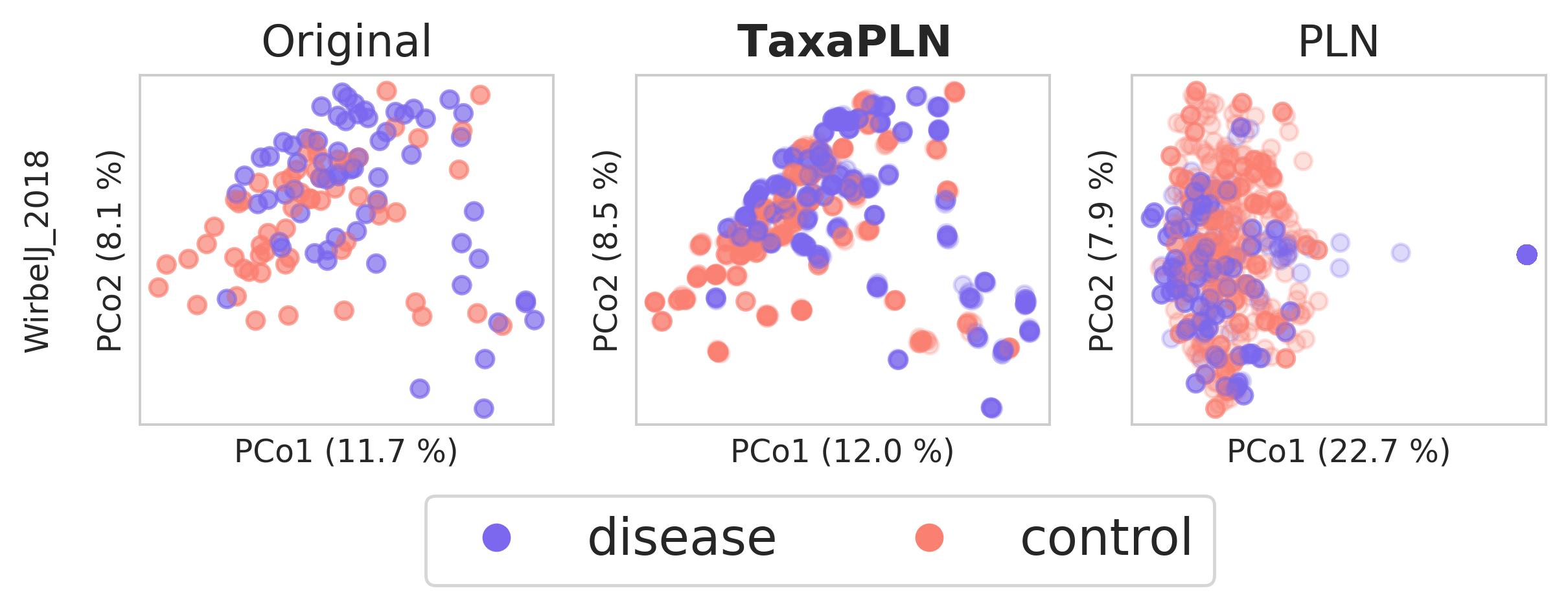}
    \end{subfigure}
    
    \begin{subfigure}[b]{0.48\linewidth}
        \centering
        \includegraphics[width=\linewidth,trim=0 40 0 23,clip]{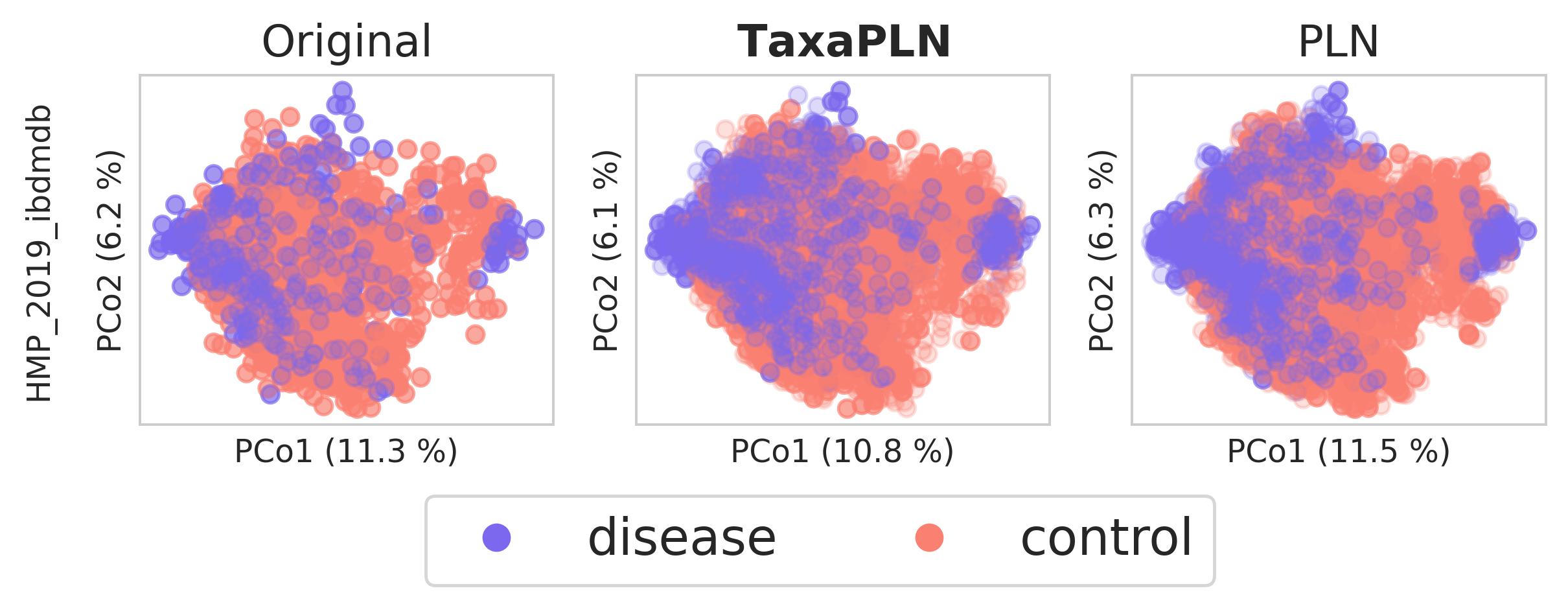}
    \end{subfigure}
    \begin{subfigure}[b]{0.48\linewidth}
        \centering
        \includegraphics[width=\linewidth,trim=0 40 0 23,clip]{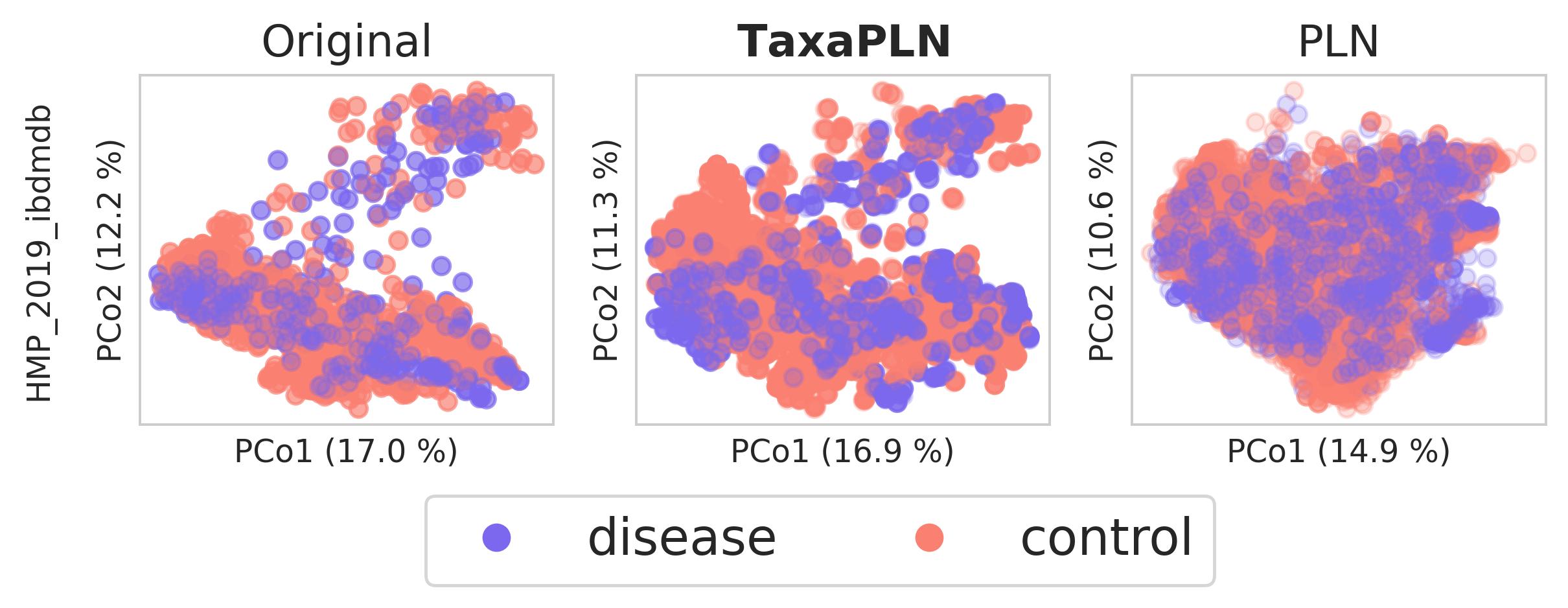}
    \end{subfigure}
    
    \begin{subfigure}[b]{0.48\linewidth}
        \centering
        \includegraphics[width=\linewidth,trim=0 40 0 23,clip]{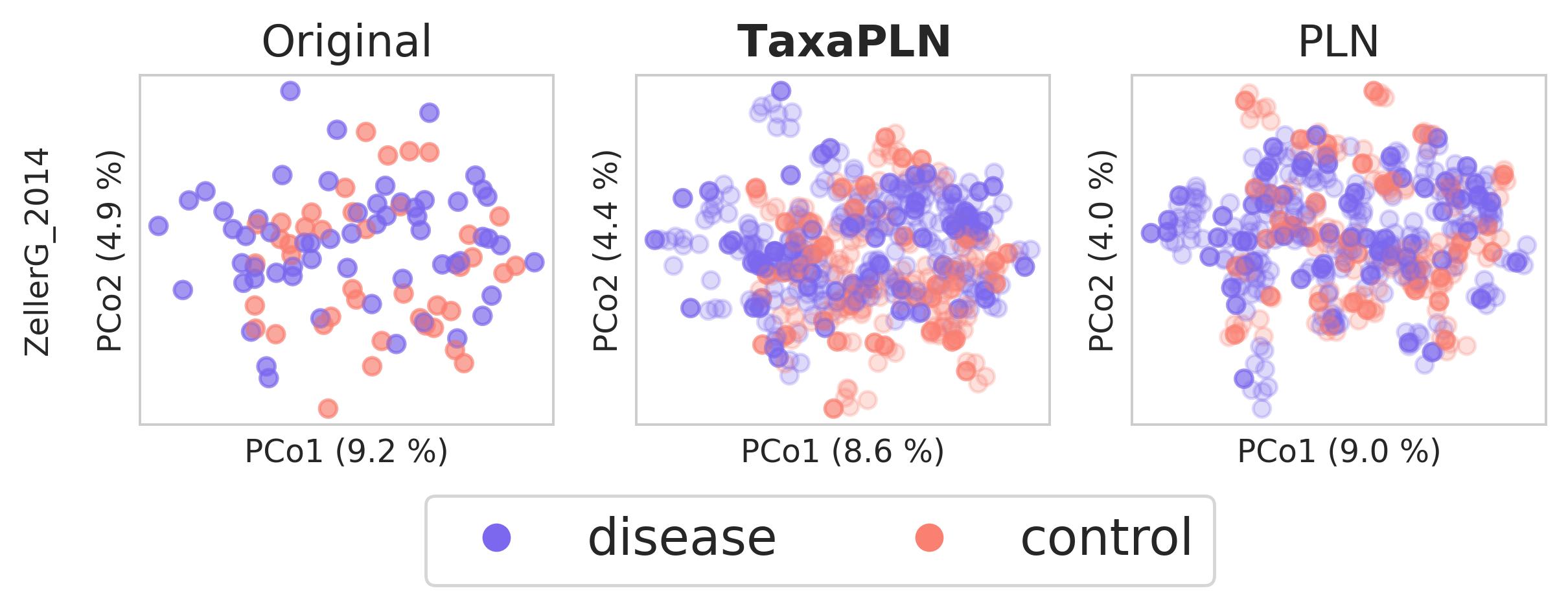}
    \end{subfigure}
    \begin{subfigure}[b]{0.48\linewidth}
        \centering
        \includegraphics[width=\linewidth,trim=0 40 0 23,clip]{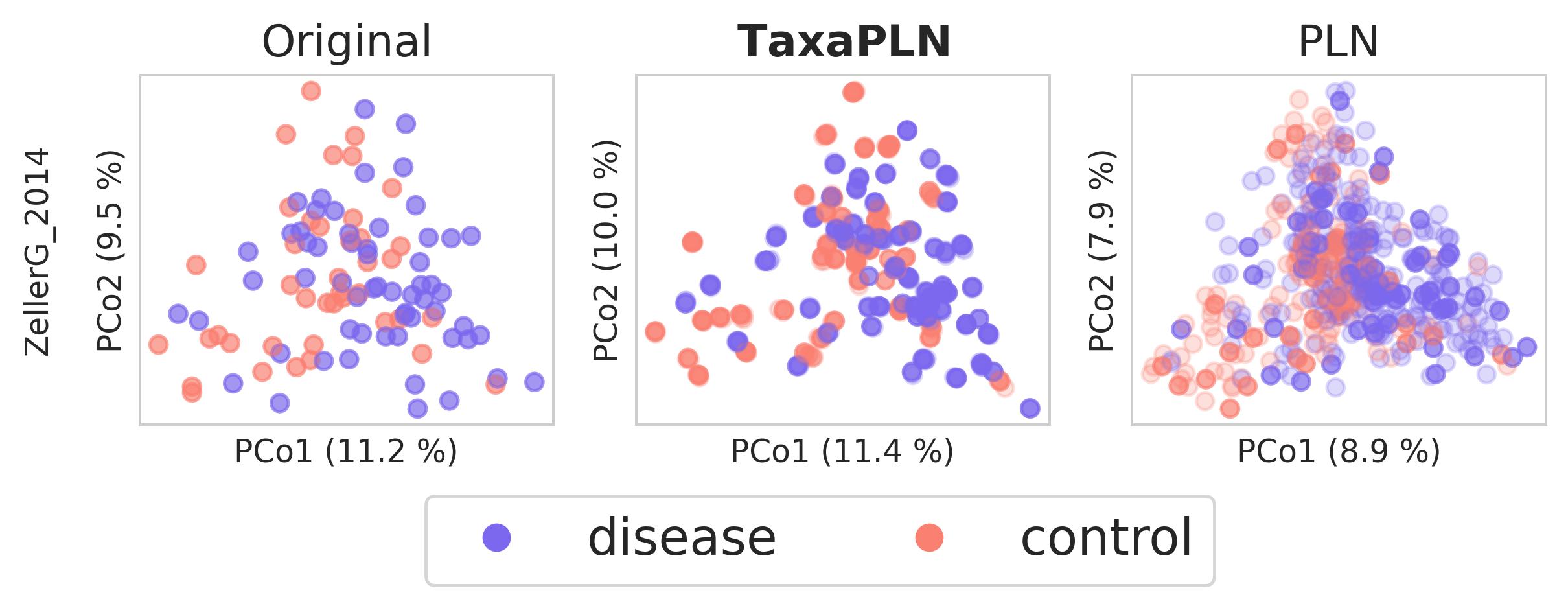}
    \end{subfigure}
    
    \begin{subfigure}[b]{0.48\linewidth}
        \centering
        \includegraphics[width=\linewidth,trim=0 40 0 23,clip]{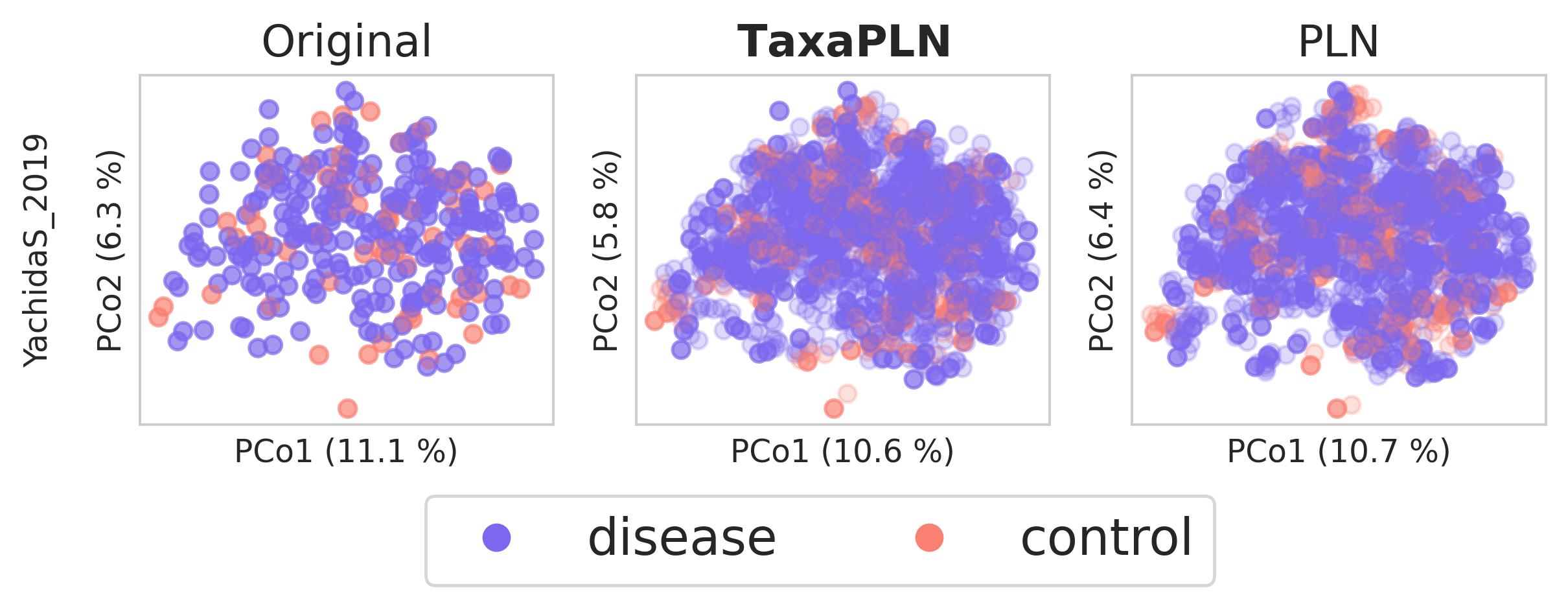}
    \end{subfigure}
    \begin{subfigure}[b]{0.48\linewidth}
        \centering
        \includegraphics[width=\linewidth,trim=0 40 0 23,clip]{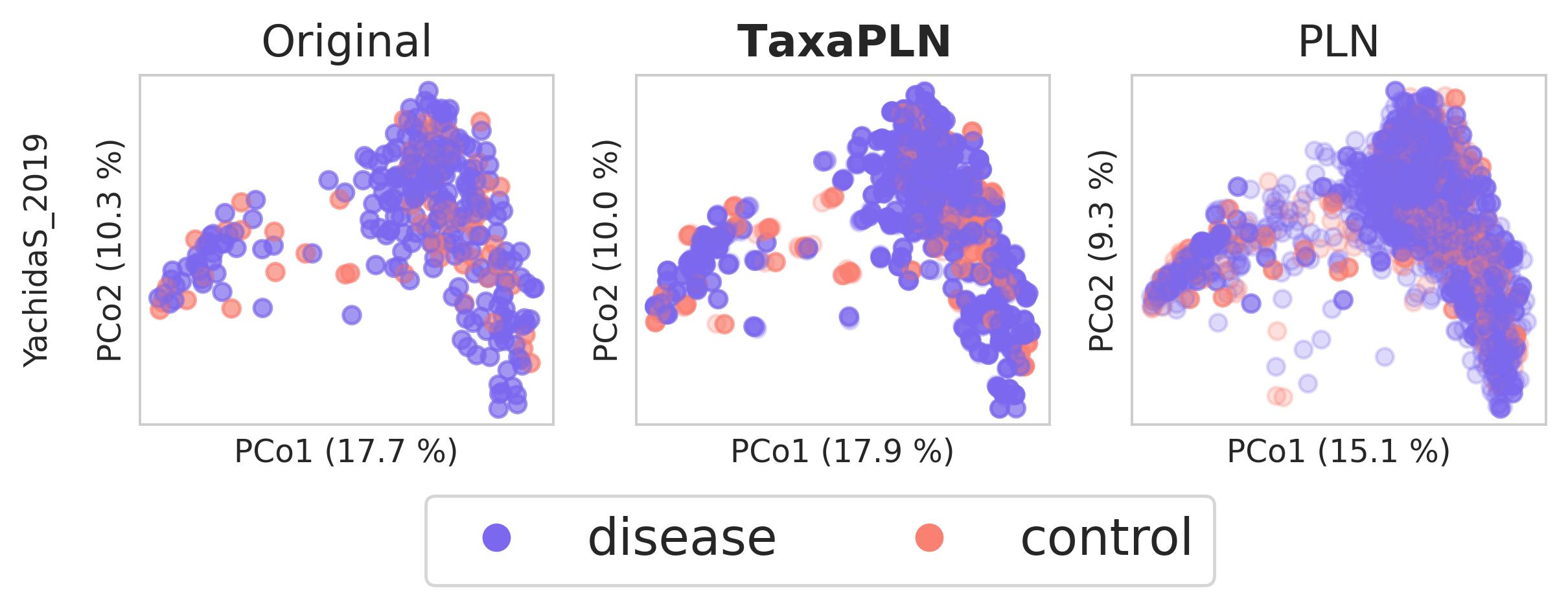}
    \end{subfigure}
    
    \begin{subfigure}[b]{0.48\linewidth}
        \centering
        \includegraphics[width=\linewidth,trim=0 40 0 23,clip]{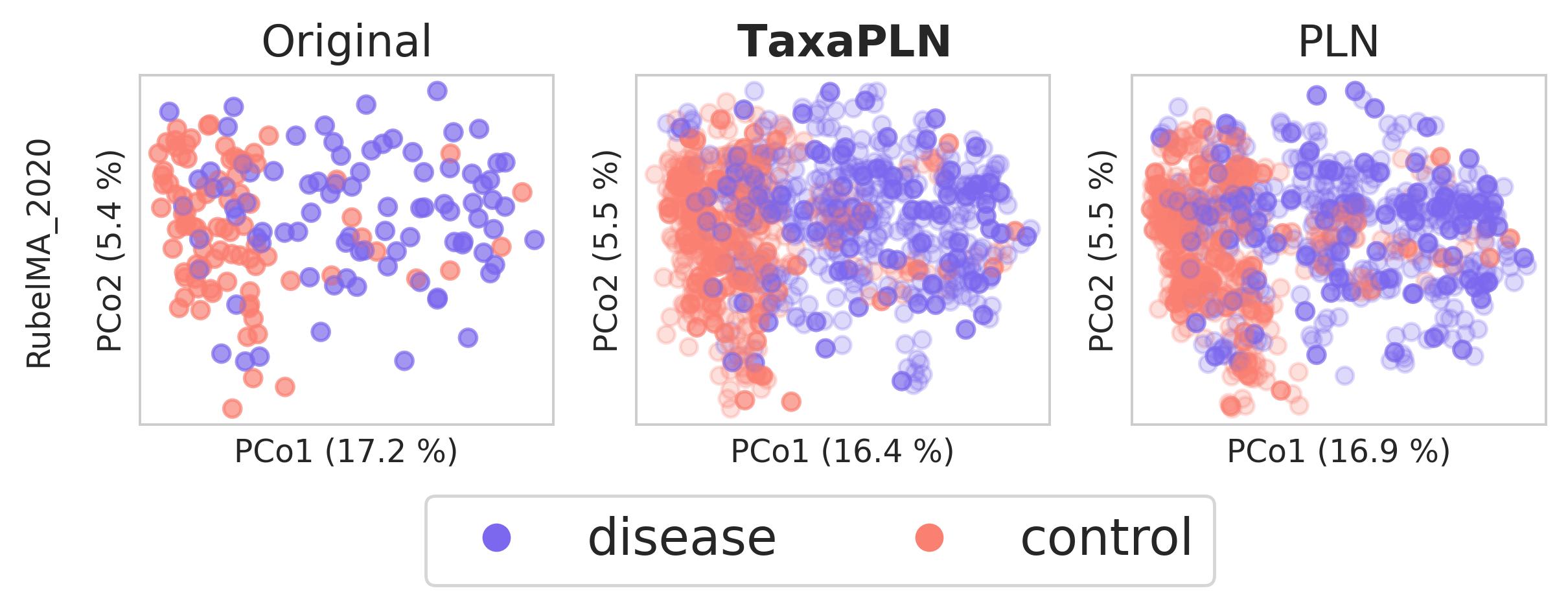}
    \end{subfigure}
    \begin{subfigure}[b]{0.48\linewidth}
        \centering
        \includegraphics[width=\linewidth,trim=0 40 0 23,clip]{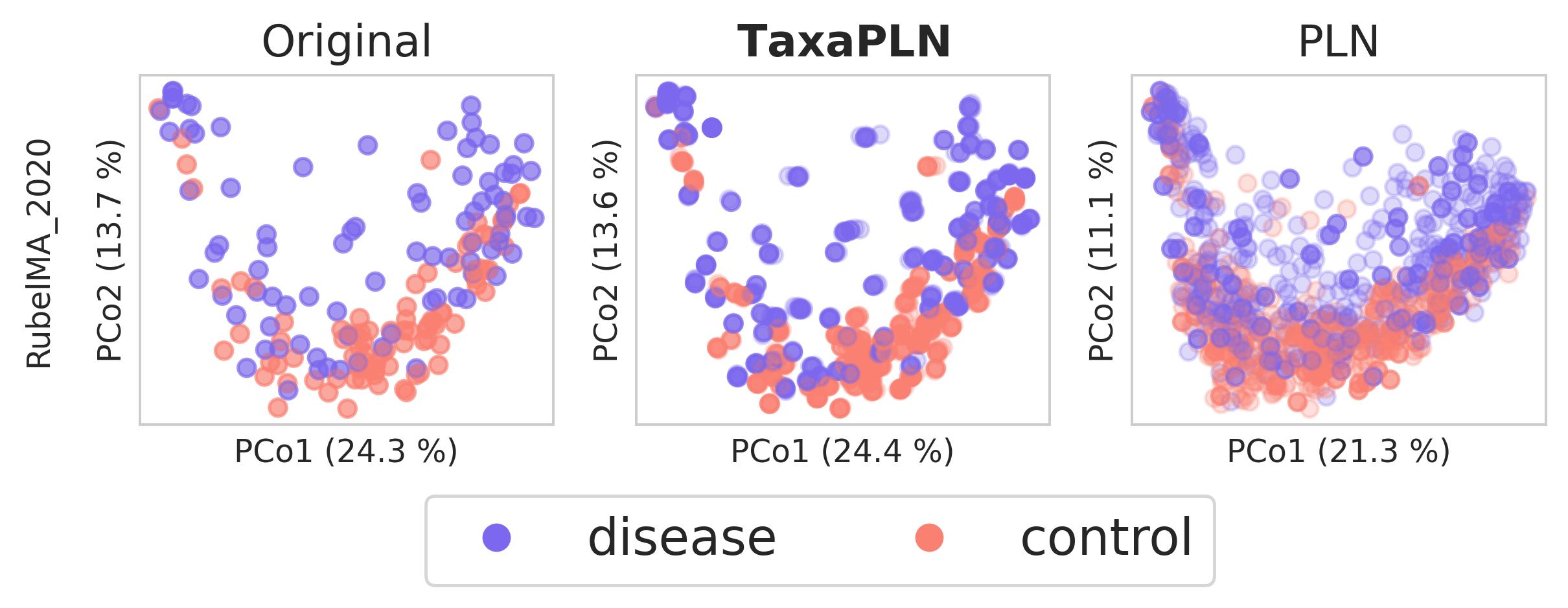}
    \end{subfigure}
    
    \begin{subfigure}[b]{0.48\linewidth}
        \centering
        \includegraphics[width=\linewidth,trim=0 40 0 23,clip]{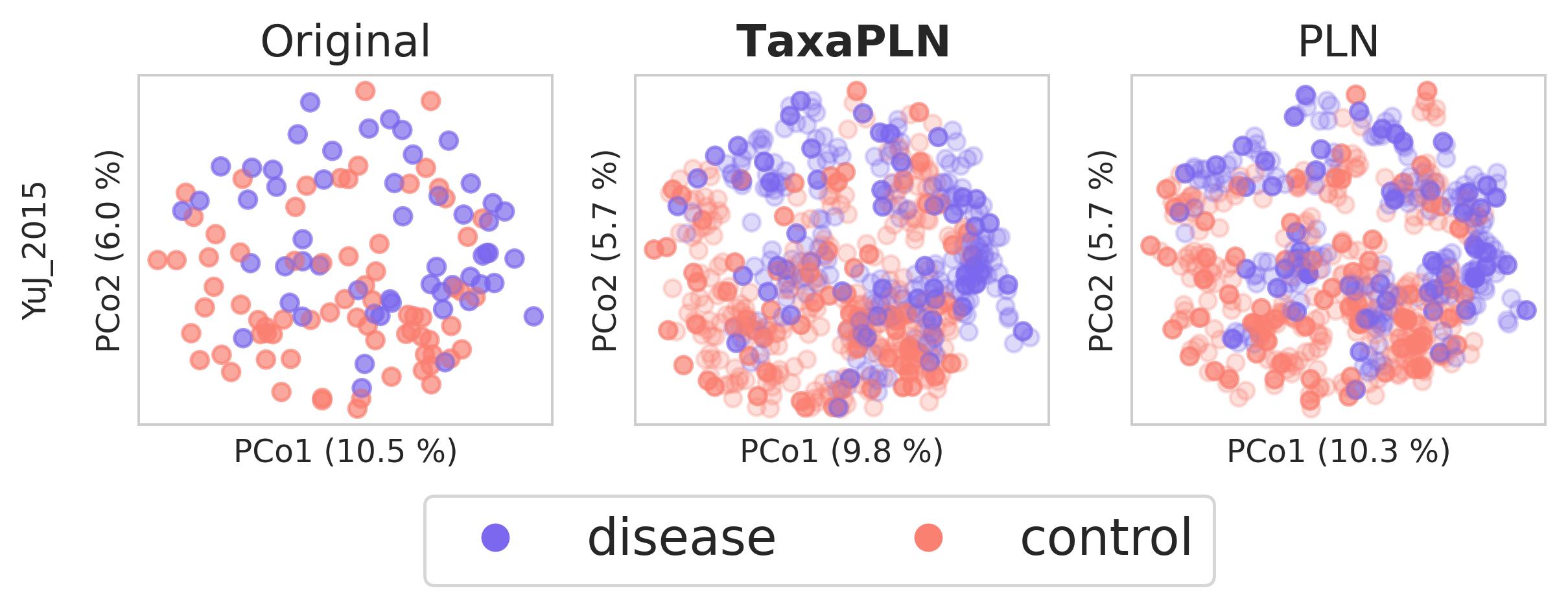}
    \end{subfigure}
    \begin{subfigure}[b]{0.48\linewidth}
        \centering
        \includegraphics[width=\linewidth,trim=0 40 0 23,clip]{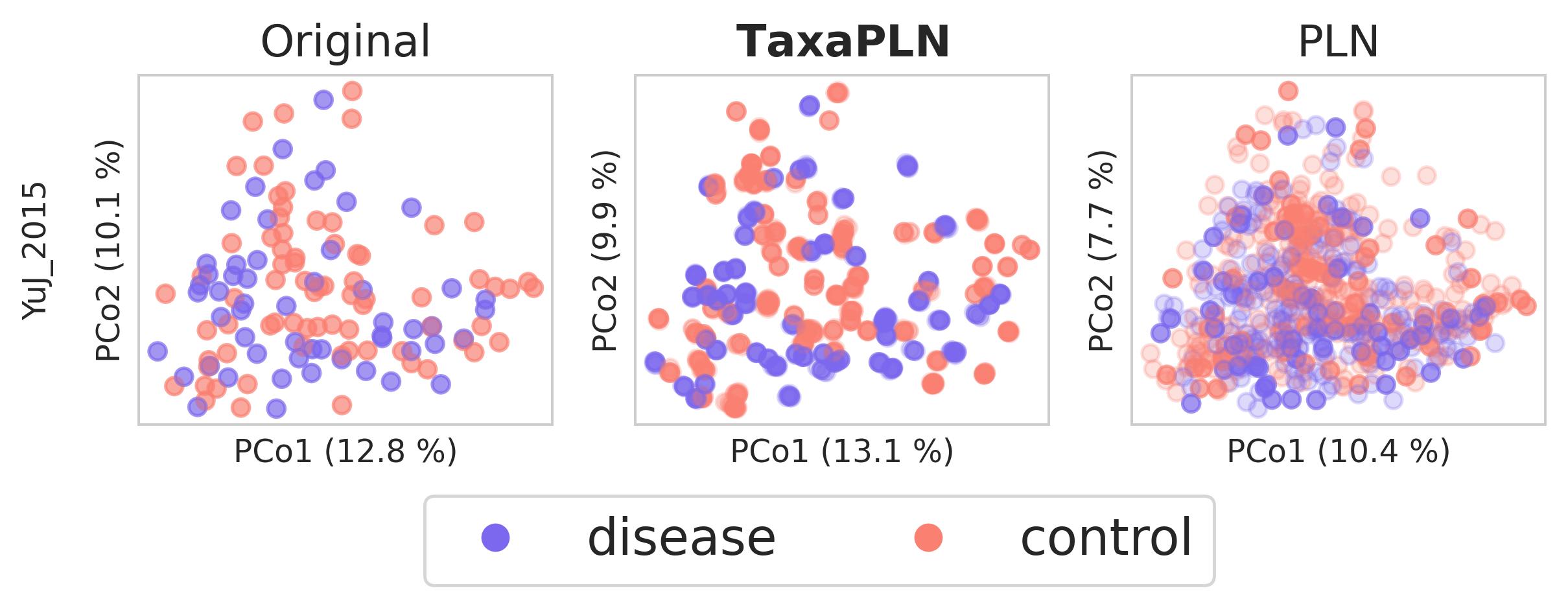}
    \end{subfigure}
    
    \begin{subfigure}[b]{0.48\linewidth}
        \centering
        \includegraphics[width=\linewidth,trim=0 40 0 23,clip]{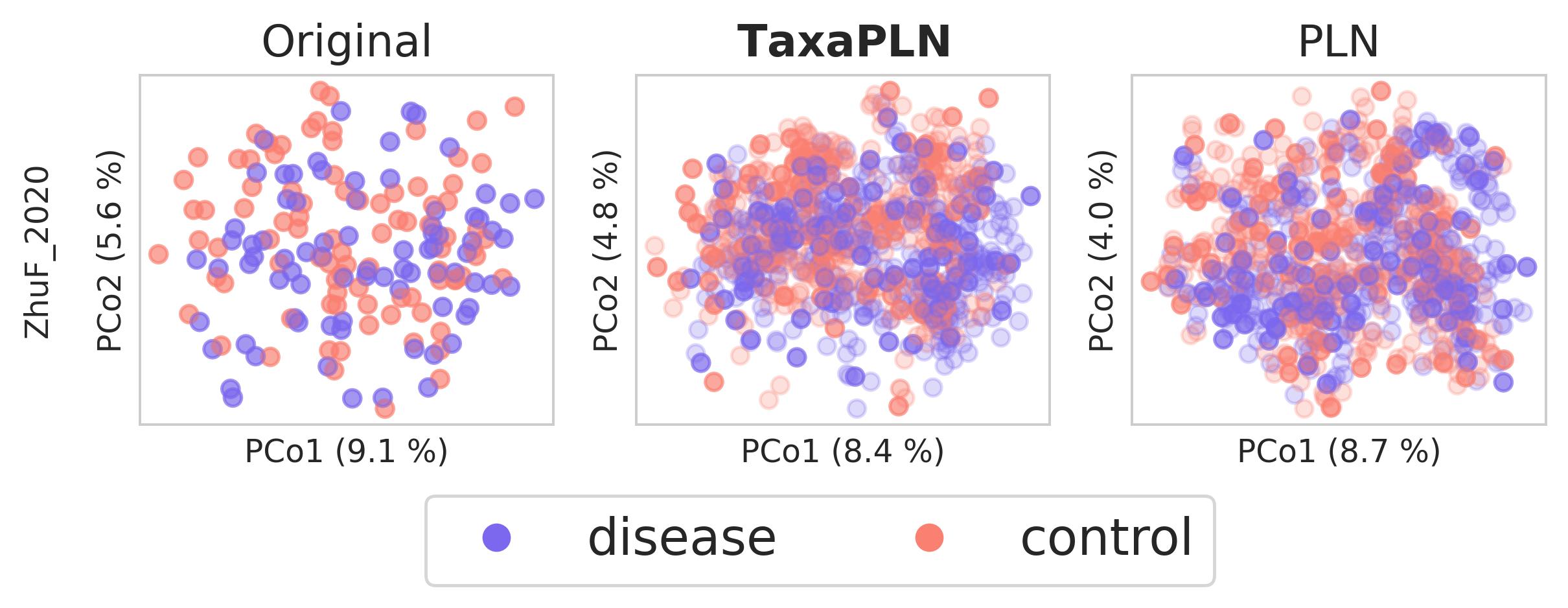}
        \caption{PCoA Aitchison distance}
    \end{subfigure}
    \begin{subfigure}[b]{0.48\linewidth}
        \centering
        \includegraphics[width=\linewidth,trim=0 40 0 23,clip]{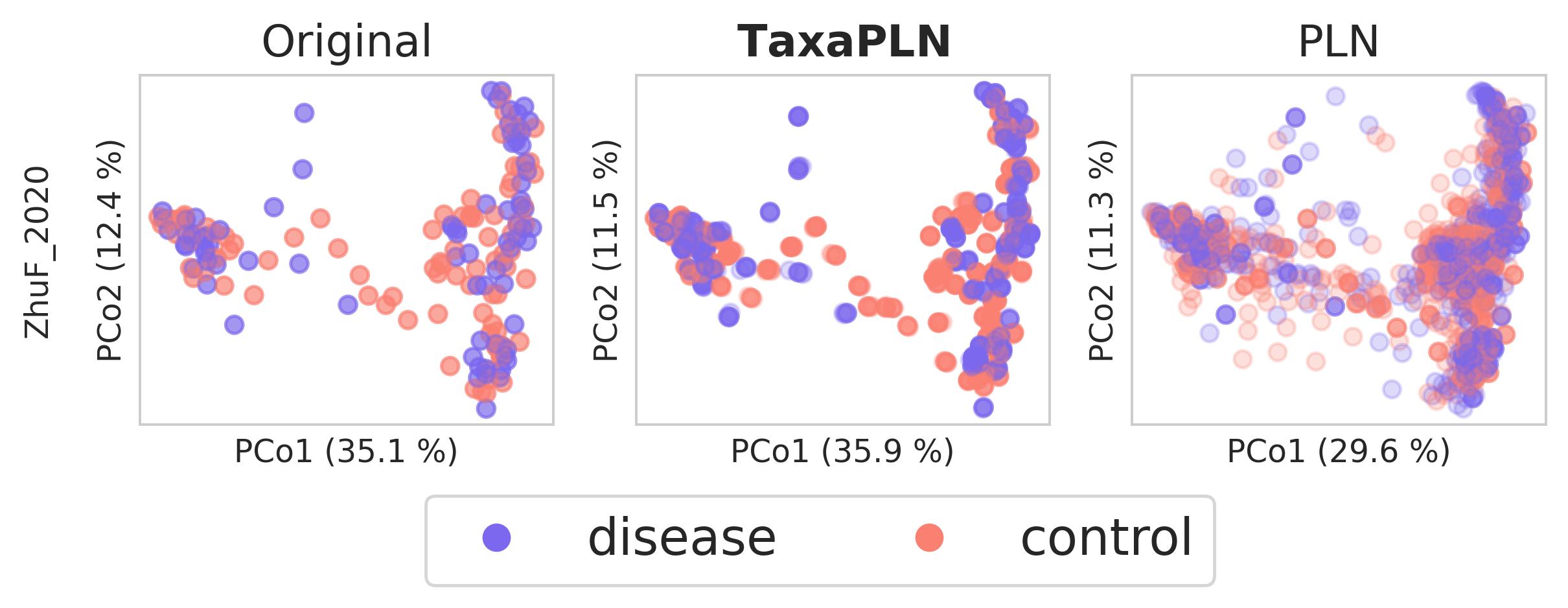}
        \caption{PCoA Bray-Curtis dissimilarity}
    \end{subfigure}
    
    \begin{subfigure}[b]{0.7\linewidth}
        \centering
        \includegraphics[width=\linewidth,trim=0 0 0 150,clip]{conditional-braycurtis-PCoA-WirbelJ_2018}
    \end{subfigure}
    \caption{Aitchison and Bray–Curtis $\beta$-diversity PCoA plots using conditional models. Synthetic microbiome samples generated by covariate-aware TaxaPLN and conditional PLN baseline augmentation methods are evaluated on their $\beta$-diversity consistency with the original microbiomes from Table~\ref{tab:curated_dataset_desc}. Each method augment the training fold with $\beta = 2$. Prior to Bray–Curtis dissimilarity computation, all data are normalized into proportions, while Aitchison distances are computed from Euclidean distance on CLR-transformed counts. Principal Coordinates Analysis (PCoA) is used to visualize the dissimilarity structure.}
    \label{fig:conditional_augmentation_beta_diversity_full}
\end{figure}

\end{appendices}
\end{document}